\documentclass[3p,12pt]{elsarticle}
\usepackage{threeparttable,booktabs,tabularx}
\usepackage[fleqn]{amsmath}
\usepackage{cases}
\usepackage{multirow}
\usepackage{xcolor}
\usepackage[normalem]{ulem}
\usepackage{tabularx}
\usepackage{siunitx}
\usepackage{comment}
\usepackage{algorithm}
\usepackage{algpseudocode}
\usepackage{algorithmicx}
\usepackage{grffile}
\usepackage{rotating}
\usepackage{array}
\usepackage{lineno}
\usepackage{hyperref}
\usepackage{makecell}
\usepackage{graphicx,enumerate}
\usepackage{amssymb}
\usepackage{bm,amsmath}
\usepackage{subfigure}
\usepackage{caption,color}
\usepackage{threeparttable}
\usepackage{subeqnarray}
\usepackage{multirow}
\usepackage{lscape}
\usepackage{rotating}
\usepackage{graphics}
\floatname{algorithm}{Algorithm}
\usepackage{epstopdf}
\journal{Journal of Computational Physics}

\biboptions{numbers,sort&compress} % 压制引用，[1,2,3]--> [1-3]

\begin{document}
	
\begin{frontmatter}

\title{Multiscale simulation of interacting turbulent and rarefied gas flows in the DSMC framework}
%{Particle-based general synthetic iterative scheme for rarefield gas flow}
\author{Liyan Luo}
\author{Songyan Tian}  
\author{Lei Wu\corref{mycorrespondingauthor1}}
\ead{wul@sustech.edu.cn}
\cortext[mycorrespondingauthor1]{Corresponding author}
\address{Department of Mechanics and Aerospace Engineering,
Southern University of Science and Technology, 518055 Shenzhen, China}

\begin{abstract}
A multiscale stochastic-deterministic coupling method is proposed to investigate the complex interactions between turbulent and rarefied gas flows within a unified framework. This method intermittently integrates the general synthetic iterative scheme with the shear stress transport turbulence model into the direct simulation Monte Carlo (DSMC) approach, enabling the simulation of gas flows across the free-molecular, transition, slip, and turbulent regimes. 
First, the macroscopic synthetic equations, derived directly from DSMC, are coupled with the turbulence model to establish a constitutive relation that incorporates not only turbulent and laminar transport coefficients but also higher-order terms accounting for rarefaction effects. Second, the macroscopic properties, statistically sampled over specific time intervals in DSMC, along with the turbulent properties provided by the turbulence model, serve as initial conditions for solving the macroscopic synthetic equations. Finally, the simulation particles in DSMC are updated based on the macroscopic properties obtained from the synthetic equations. % ensuring that the particle distribution evolves consistently with both turbulence and rarefaction effects.
Numerical simulations demonstrate that the proposed method asymptotically converges to either the turbulence model or DSMC results, adaptively adjusting to different flow regimes. Then,  this coupling method is applied to simulate an opposing jet surrounded by hypersonic rarefied gas flows, revealing significant variations in surface properties due to the interplay of turbulent and rarefied effects. This study presents an efficient methodology for simulating the complex interplay between rarefied and turbulent flows, establishing a foundational framework for investigating the coupled effects of turbulence, hypersonic conditions, and chemical reactions in rarefied gas dynamics in the future.
\end{abstract}

\begin{keyword}
Rarefied gas dynamics; turbulent modeling; multiscale numerical method; direct simulation Monte Carlo.
\end{keyword}

\end{frontmatter}
\section{Introduction}\label{sec:1}

With the rapid progress of modern industry, rarefied gas flows have become increasingly common in critical applications, particularly in the development of near-space hypersonic vehicles with jet flows operating at altitudes of 20 to 100 kilometers. These vehicles encounter not only conventional aerodynamic processes, but also complex rarefied gas dynamics and turbulence. Due to the substantial differences in flow characteristics and applicable conditions, turbulence and rarefied flows are typically studied independently, with turbulence described by the Navier-Stokes (NS) equations and rarefied flows governed by the Boltzmann equation. From a physical perspective, however, the dynamics of dilute gases, whether in rarefied or turbulent flow regimes, are governed by the Boltzmann equation. Only when the Knudsen number (defined as the ratio of the molecular mean free path $\lambda$ to the characteristic system length $L$) is small can the NS equations be derived from the Boltzmann equation~\cite{chapman-1990}. 

In the simulation of turbulent flows, the NS equations are typically solved directly or in conjunction with turbulence models. However, since the turbulent flow has many length scales, if the characteristic length $L$ is chosen as the smallest Kolmogorov length scale, the local Knudsen number can be large, which might cause the NS equations to fail in capturing rarefaction effects.
Although Tennekes and Lumley estimated that this is unlikely under most practical conditions~\cite{tennekes-1972}, the advancement of computational capabilities has spurred efforts to identify rarefaction effects in turbulent flows.  The most reliable large scale simulation is conducted by McMullen \textit{et al.} who, however, demonstrated that the energy spectrum obtained from the direct simulation Monte Carlo (DSMC\footnote{Since  DSMC is primarily designed for simulating laminar rarefied flows, its computational cost is prohibitive in near-continuum regimes due to its stringent constraints on grid size and time step; in this case the DSMC simulation costed over 500 hours on half a million CPU cores.}, a numerical solver for the Boltzmann equation)~\cite{bird-1994} and NS simulations agree well with each other, except that discrepancy occurs at large wave numbers, where the thermal fluctuations (in DSMC, as well as in real gas) cause the spectrum to scale with the square of the wave number~\cite{mcmullen-2022}. Nevertheless, they emphasized that ``thermal fluctuations have little impact on the large-scale evolution of the flow".

Recently, a GSIS-SST coupling solver has been developed to investigate the possibility of coexistence and interaction of turbulent and rarefied flows~\cite{tian-2024}. This approach integrates the general synthetic iterative scheme (GSIS) for solving the Boltzmann equation~\cite{su-2020,su-2020-can} with the $k$-$\omega$ shear stress transport (SST) model for turbulent flows~\cite{menter-1994}. By combining these methods, the constitutive relations incorporate both turbulent and laminar viscosities, as well as higher-order terms that capture rarefaction effects. Consequently, the GSIS-SST method is capable of describing gas flows from highly rarefied conditions to fully turbulent continuum flows. Interestingly, in studies of supersonic rarefied gas flows interacting with turbulent jets, the GSIS-SST solver predicts surface heat flux and pitch moment that differ significantly from those predicted by the pure SST model or the pure Boltzmann equation~\cite{tian-2024,tian-2025}. This highlights the critical importance of simultaneously accounting for turbulence and rarefied gas dynamics in such scenarios.

However, since the GSIS-SST is a deterministic solver for the Boltzmann equation where both physical and velocity space are discretized, its application to high Mach number cases requires large computational memory. 
To address this challenge, we aim to develop a particle version of the GSIS-SST within the standard DSMC framework. This approach will build on our recent advances in the Direct Intermittent GSIS-DSMC coupling method (DIG)~\cite{luo-2024}, which solves the Boltzmann equation for laminar flows across multiscale regimes, with fast convergence and less numerical dissipation. 
Inspired by GSIS-SST, the DIG method will be extended to couple with the $k$-$\omega$ SST model to simulate the high-density jet flows in higher-altitude and high-Mach-number rarefied flows, in the DSMC framework.  %Generally speaking, the proposed algorithm has the so-called asymptotic preserving property, since it preserves to the original $k$-$\omega$ SST model for turbulent flows when Reynolds number is high. And for large Knudsen numbers, where rarefaction effects dominate, the results produced by the present coupling algorithm are consistent with those obtained from the DSMC method. Moreover, due to the adaptive nature of the stochastic method in velocity space, the DIG-SST method exhibits inherent advantages of computational storage over GSIS-SST in such scenario.

% In our previous work, the fast convergence and asymptotic-preserving properties have been demonstrated. For instance, 

% \leir{we will develop the particle version of the GSIS-SST. .., on top of the DSMC method.} \lei{This is based on our recent advance of the DIG method, which ... solves the Boltzmann equation for laminar flows.} \leir{then briefly explain the DIG method}
% GSIS has been employed to couple with other stochastic particle methods~\cite{luo-2023,Luo2024arXiv}, among which the direct intermittent GSIS-DSMC coupling method (DIG) stands out for its simplicity in incorporating other complex physical processes~\cite{luo-2024}. Other numerous strategies have also been developed to enhance the efficiency of the DSMC method in near-continuum regimes~\cite{pareschi-2001,fei-2020,fei-2023,ren-2014}, but they still encounter difficulties in effectively solving turbulent flows. 

% And the interaction between the turbulent and rarefied flows is reflected in the constitutive relations within the synthetic equations, as these relations account for both laminar and turbulent viscosities.

The remainder of this paper is organized as follows. Section \ref{sec:2} provides a detailed description of the DSMC method for polyatomic gases and the DIG method. In Section \ref{sec:3}, the DIG and SST models are coupled to enable the DSMC method to accurately describe rarefied and turbulent flows using coarse grids. The algorithm is validated under both turbulent and rarefied conditions through simulations of hypersonic flow over a wedge and a blunt body in Section~\ref{sec:4}. Section~\ref{sec:5} investigates the interaction between turbulent and rarefied flows based on simulations of an opposing jet. Finally, Section~\ref{sec:6} presents concluding remarks and outlines future research directions.

%%%%%%%%%%%%%%%%%%%%%%%%%%%%%%%%%%%%%%%%%%%%%%%%%%%%%%%%%%%%%%%
%  And as Kn increases, the constitutive relations underlying the NS equations become invalid. Since the Boltzmann equation defines the distribution function within a six-dimensional phase space, its high-dimensional nature presents considerable difficulties for both analytical and numerical solutions.
% Therefore, if the pure turbulent flow is present, then the NS equations are directly solved or solved with some turbulence models.

\section{The DSMC method for polyatomic gas flow and its acceleration by DIG}\label{sec:2}

Throughout this paper, all macroscopic variables are non-dimensionalized using the reference length $L_0$, reference density $\rho_0$, reference temperature $T_0$, and most probable speed $c_0=\sqrt{k_BT_0/m}$, with $k_B$ and $m$ being the Boltzmann constant and molecular mass, respectively. In this section, we first review the governing equations and the DSMC method for simulating rarefied polyatomic gases. Then we introduce the DIG method, which enables efficient simulation of polyatomic gas flows across a wide range of Knudsen numbers, from the laminar continuum to the rarefied regime, but without turbulence.

\subsection{Kinetics of polyatomic gas}

According to quantum mechanics, polyatomic gases possess internal energy levels. In the gas kinetic theory, the state of polyatomic gas flows at the $i$-th level of internal energy is determined by the probability distribution function $f_i(t,\bm{x},\bm{v})$, where $\bm{v}$ and $\bm{x}$ represent the molecular velocity and position, respectively, while $t$ denotes time. The evolution of the distribution function is governed by the Wang-Chang-Uhlenbeck (WCU) equation:
\begin{equation}
\frac{\partial f_i}{\partial t}+\bm{v}\cdot\frac{\partial f_i}{\partial\bm{x}}=\sum_{i'j'}\sum_j\int_{-\infty}^{\infty}\int_{4\pi}\left(\frac{g_ig_j}{g_{i'}g_{j'}}f_{i'}f_{j'}-f_if_j\right)c_r\sigma_{ij}^{i'j'} d\Omega d\bm{v}_*,
\label{eq:boltzmann_equation}
\end{equation}
where $c_r=|\bm{v}-\bm{v}_*|$ is the relative translational velocity of the collision pair, and $\bm{v}$ and $\bm{v}_*$ represent the pre-collision velocities of the two molecules with internal states $i$ and $j$, respectively. The superscript $'$ indicates the states after collision; $g_i$ is the degeneracy of the $i$-th internal energy level;  $\Omega$ is the solid angle; $\sigma_{ij}^{i'j'}$ is the differential cross-section of the binary collision normalized by $\pi d^2$, where $d$ is the effective molecular diameter. 
% Particularly, $\sigma_{ij}^{i'j'}$ is determined by the intermolecular potential and the variable hard sphere (VHS) model \leir{in DSMC} is employed in the following paper~\cite{bird-1994}. 

For simplicity, we consider only the rotational internal energy, although the method can be readily extended to include vibrational and electronic energy levels. Due to the small energy gap between successive rotational levels, the rotational energy can be approximately treated as a continuous variable. Accordingly, the distribution function $f_i$ can be expressed as $f_i(t,\bm{x},\bm{v},I_r)$, where $I_r$ denotes the molecular rotational energy (normalized by $k_BT_0$). By taking moments of distribution function, the macroscopic properties, mass density $\rho$, velocity $\bm{u}$,  translational temperature $T_t$, rotational temperature $T_r$, shear stress $\bm{\sigma}$, translational heat flux $\bm{q}_t$ and rotational heat flux $\bm{q}_r$, can be obtained:
\begin{equation}
\begin{aligned}
\left(\rho,\rho\bm{u},\bm{\sigma}\right)&=\iint \left(1,\bm{v},\bm{c}\bm{c}-\frac{c^2}{3}\textbf{\text{I}}\right)fd\bm{v}dI_r,\\
\left(3\rho T_t,d_r\rho T_r\right)&=\iint \left(\frac{c^2}{2},I_r\right)fd\bm{v}dI_r,\\
\left(\bm{q}_t,\bm{q}_r\right)&=\iint \bm{c}\left(\frac{c^2}{2},I_r\right)fd\bm{v}dI_r,\\
\end{aligned}
\label{eq:macroscopic_properties}
\end{equation}
where \textbf{I} is a $3\times3$ identity matrix and the $\bm{c}=\bm{v}-\bm{u}$ is the peculiar velocity. Note that the shear stress and heat flux are normalized by $\rho_0RT_0$ and $\rho_0 RT_0c_0$, respectively. 

For polyatomic gases, the exchange of energy among different degrees of freedom are important in the binary collision. When there is no exchange of energy between the translational and rotational modes, the collision is considered as elastic; otherwise it is inelastic. For inelastic collisions, the rotational energy does not instantaneously reach equilibrium but instead undergoes a relaxation process over multiple collisions. This relaxation behavior is often described using the Jeans-Landau equation, which characterizes the time evolution of the rotational temperature $T_r$ as it approaches its equilibrium value:
\begin{equation}
    \frac{dT_r}{dt}=\frac{T-T_r}{Z\tau_c},
\label{eq:relaxation_Tr}
\end{equation}
where $\tau_c=\mu/p$ is the instantaneous collision time and $Z$ is the rotational collision number. The total temperature $T$ is defined as $T=(3T_t+d_r T_r)/(5+d_r)$ with $d_r$ being the constant rotational degrees of freedom (The corresponding normalized total pressure is defined as $p=\rho T$). The ratio between the bulk viscosity $\mu_b$ and shear viscosity is given by 
\begin{equation}
    \frac{\mu_b}{\mu}=\frac{2d_rZ}{3(d_r+3)}.
\end{equation}

Furthermore, since the exchange of thermal energy among different degrees of freedom influences the heat transport, the relaxation of heat fluxes must account for their mutual coupling, which can be written as~\cite{mason-1962,li-2021},
\begin{equation}
    \begin{bmatrix} \partial \bm{q}_t/\partial t  \\ \partial \bm{q}_r/\partial t \end{bmatrix} = 
    -\frac{1}{\tau_c}
    \begin{bmatrix} A_{tt} & A_{tr}  \\ A_{rt} &A_{rr} \end{bmatrix}
    \begin{bmatrix} \bm{q}_t \\ \bm{q}_r \end{bmatrix},
    \label{eq:relaxation_heatflux}
\end{equation}
where $\bm{A}$ is a matrix characterizing the thermal relaxation rates associated with the translational and internal thermal conductivities of the polyatomic gas. In general, since Eqs.~\eqref{eq:relaxation_Tr} and \eqref{eq:relaxation_heatflux} describe distinct relaxation processes, the relaxation parameter $Z$ and the matrix $\bm{A}$ are not necessarily correlated and may vary independently in practical applications.

\subsection{The DSMC method with Borgnakke-Larsen model}

In the DSMC method, each simulation particle represents a large number of gas molecules; each carries information of its own velocity $\bm{v}$, location $\bm{x}$, and internal energy $I_{r}$. Thus, the macroscopic properties in Eq.~\eqref{eq:macroscopic_properties} can be sampled and non-dimensionalized within each computational cell accordingly, see Ref.~\cite{luo-2023}. Moreover, the fundamental principle of the standard DSMC method lies in the decoupling of particle advection and collisions:
\begin{equation}
\begin{aligned}
&\text{Advection:}\qquad \frac{\partial f}{\partial t}+\bm{v}\cdot\frac{\partial f}{\partial \bm{x}}=0.\\
&\text{Collision:}\qquad \left[\frac{\partial f}{\partial t}\right]_{\text{coll}}=Q(f_i),
\label{eq:splitting_eq}
\end{aligned}
\end{equation}
where $Q(f_i)$ is the collision operator. During the advection step, the position of each simulation particle is updated, while its velocity remains unchanged. In the subsequent collision step, however, directly employing Eq.~\eqref{eq:boltzmann_equation} for polyatomic gas requires the determination of the cross-section for all possible transitions between different energy levels, which imposes a significant computational burden in practical applications. To address this challenge, numerous phenomenological models have been developed. These models integrate the energy exchange between different modes into the existing monatomic DSMC collision framework, offering a more computationally efficient alternative. In our work, we focus solely on the rotational internal energy and employ the Borgnakke-Larsen model~\cite{borgnakke-1975}. For inelastic collisions, the total energy of the two colliding particles, comprising the relative translational energy of the pair and the internal energy of each particle, is recorded and subsequently redistributed between the translational and internal modes. This redistribution ensures energy conservation while accounting for the interplay between translational and internal energy. The permitting double-relaxation particle selection method are employed as proposed by Hass \textit{et al} \cite{haas-1994}, where the fraction of inelastic collisions $P_{inelastic}$ is the inverse of the relaxation collision number in DSMC, $Z_{DSMC}$:
\begin{equation}
    P_{inelastic}=\frac{1}{Z_{DSMC}}=\frac{\alpha(5-2\omega)(7-2\omega)}{5(\alpha+1)(\alpha+2)Z},
\end{equation}
where $\omega$ is the shear stress viscosity index such that $\mu(T)=\mu_{0}(T_{0})T^\omega$ and the parameter $\alpha$ indicates the scattering angle for binary collision (e.g., $\alpha = 1$ for the variable hard sphere  model in binary collisions).

The DSMC method is highly efficient for simulations flows with  large Knudsen numbers. However, the strict requirement that the grid size (time step) must be smaller than one-third of the local mean free path (mean collision time) leads to an extremely high computational cost in the near-continuum regime. To address this challenge, the next section introduces the recently developed DIG method, which eliminates these limitations on grid size and time step, thereby significantly enhancing convergence in near-continuum flow simulations.

\subsection{The DIG method for multiscale simulation}

To facilitate the fast convergence and reduce the numerical dissipation, it is necessary to use the macroscopic equations to guide the evolution of simulation particles in DSMC. By taking moments of Eq.~\eqref{eq:boltzmann_equation}, the governing equations for macroscopic properties $\rho, \bm{u}, T_t, T_r$ can be obtained as follows:
\begin{equation}
\begin{aligned}
\frac{\partial \rho}{\partial t}+\nabla\cdot(\rho \bm{u})=&0, \\
\frac{\partial \rho\bm{u}}{\partial t}+\nabla\cdot(\rho\bm{u}\bm{u})+\nabla\cdot(\rho T_t\bm{I}+\bm{\sigma})=&0,\\
\frac{\partial \rho E}{\partial t}+\nabla\cdot(\rho E\bm{u})+\nabla\cdot\left(\rho T_t\bm{u}+\bm{u}\cdot\bm{\sigma}+\bm{q}_t+\bm{q}_r \right)=&0, \\
\frac{\partial \rho E_r}{\partial t} + \nabla\cdot(\rho E_r\bm{u}) + \nabla\cdot\bm{q}_r = &\frac{d_r\rho}{2}\frac{T-T_r}{Z\tau},
\end{aligned}
\label{eq:Navior-Stokes}
\end{equation}
where $E=(3T_t+d_rT_r)/2+u^2/2$ and $E_r=d_rT_r/2$ are the specific total and rotational energies, respectively. In Eq.~\eqref{eq:Navior-Stokes}, the first three equations represent the conservation of mass, momentum and energy, while the last equation describes the exchange between the translational and rotational energy. 

However, Eq.~\eqref{eq:Navior-Stokes} is not closed since the shear stress $\bm{\sigma}$ and heat fluxes $\bm{q}_t,\,\bm{q}_r$ cannot be expressed in terms of low order moments. According to the Chapman-Enskog theory~\cite{chapman-1990}, the NS constitutive relations can be obtained by the first-order expansion of the kinetic equation, which are
\begin{equation}
\begin{aligned}
    &\bm{\sigma}^{\text{NS}}=-\mu\left(\nabla\bm{u}+\nabla\bm{u}^T-\frac{2}{3}\nabla\cdot\bm{uI}\right),\\
    &\bm{q}_t^{\text{NS}}=-\kappa_t \nabla T_t, \\ & \bm{q}_r^{\text{NS}}=-\kappa_r\nabla  T_r,
\end{aligned}
\label{eq:NS_constitutive}
\end{equation}
where the translational and rotational thermal conductivities $\kappa_t,\,\kappa_r$ are determined according to the thermal relaxation rates $\bm{A}$ in~\cite{li-2021}:
\begin{equation}
    \begin{bmatrix} \kappa_{t}  \\ \kappa_{r} \end{bmatrix} = 
    \frac{\mu}{2}
    \begin{bmatrix} A_{tt} & A_{tr}  \\ A_{rt} &A_{rr} \end{bmatrix} ^{-1}
    \begin{bmatrix} 5  \\ d_{r} \end{bmatrix}.
    \label{eq:kappat_kappar}
\end{equation}

For highly non-equilibrium gas flow, the NS constitutive relations are not accurate; instead, the shear stress and the heat fluxes should be extracted from the DSMC without any truncation. In DIG, the constitutive relations are constructed as:
\begin{equation}
\begin{aligned}
    &\bm{\sigma}=-\mu\left(\nabla\bm{u}+\nabla\bm{u}^T-\frac{2}{3}\nabla\cdot\bm{uI}\right)+\text{HoT}_{\bm{\sigma}},\\
    &\bm{q}_t=-\kappa_t \nabla T_t+\text{HoT}_{\bm{q}_t}, \\
    &\bm{q}_r=-\kappa_r\nabla T_r+\text{HoT}_{\bm{q}_r},
\end{aligned}
\label{eq:GSIS_constitutive}
\end{equation}
and the high-order terms (HoTs) representing the rarefaction effects are directly constructed according to the definition of the shear stress and heat flux:
\begin{equation}
\begin{aligned}
% &\text{HoT}_{\sigma_{ij}}=\iint f^{*}c_{(i}^{*}c_{j)}^{*}d\bm{v}dI_r-\sigma_{ij}^{\text{NS}*}, \\
% &\text{HoT}_{\bm{q}_t}={\frac{1}{2}}\iint f^{*}\bm{c}^{*}\left(c^{*}\right)^{2}d\bm{v}dI_r-\bm{q}_{t}^{\text{NS}*},
% \\
% &\text{HoT}_{\bm{q}_r}=\iint f^{*}\bm{c}^{*}I_r^*d\bm{v}dI_r-\bm{q}_{r}^{\text{NS}*},
&\text{HoT}_{\bm{\sigma}}=\bm{\sigma}^{\text{DSMC}}-\bm{\sigma}^{\text{NS}*}, \\
&\text{HoT}_{\bm{q}_t}=\bm{q}_{t}^{\text{DSMC}}-\bm{q}_{t}^{\text{NS}*},
\\
&\text{HoT}_{\bm{q}_r}=\bm{q}_{r}^{\text{DSMC}}-\bm{q}_{r}^{\text{NS}*},
\end{aligned}
\label{eq:higher-oder_terms}
\end{equation}
where $\bm{\sigma}^{\text{DSMC}}$, $\bm{q}_t^{\text{DSMC}}$ and $\bm{q}_r^{\text{DSMC}}$ are sampled statistically based on preceding DSMC steps, and $\bm{\sigma}^{\text{NS}*}$, $\bm{q}_t^{\text{NS}*}$ and $\bm{q}_r^{\text{NS}*}$ are NSF constitutive relations based on Eq.~\eqref{eq:NS_constitutive} with the macroscopic properties in these relations sampled from DSMC as well. As demonstrated in Ref.~\cite{luo-2024}, iteratively solving both the mesoscopic equation~\eqref{eq:boltzmann_equation} and the macroscopic synthetic equations~\eqref{eq:Navior-Stokes} and \eqref{eq:GSIS_constitutive} guarantees the continuous update of macroscopic properties until the steady state is achieved. This treatment not only facilitates fast convergence but also maintains asymptotic-preserving properties in the continuum regimes.
% The macroscopic synthetic equations Eqs.~\eqref{eq:Navior-Stokes} and \eqref{eq:GSIS_constitutive}, when solved to the steady state, help to guide the DSMC quickly towards convergence.

A common characteristic shared by both the DIG~\cite{luo-2024} and GSIS~\cite{su-2020,su-2020-can} methods is the iterative coupling of mesoscopic and macroscopic equations. However, unlike GSIS, which deterministically obtains the distribution function in the discrete velocity space, DIG utilizes higher-order terms and macroscopic properties extracted from DSMC statistically, which inherently exhibits significant fluctuations. To circumvent this, the synthetic equations in DIG are not solved at every step of the mesoscopic solver but are instead solved intermittently every $N_s$ DSMC steps. Moreover, $n_a$ preceding weighted moving-time averaged samples are applied to further reduce the fluctuations, ensuring a more stable and efficient computation. In practice, the optimal value of these physical quantities are case-independent and required careful consideration. In our previous work, we empirically employ $N_s=50$ and $n_a=100$ with the time step of the DSMC simulation being roughly equal to the mean collision time, and then the fast convergence and asymptotic-preserving properties have been demonstrated. For instance, in the case of a hypersonic argon gas passing over a cylinder, only 40,000 computational cells are employed in DIG, compared to over 2 million cells in standard DSMC simulations. And the CPU time of DIG is smaller than DSMC by nearly two orders of magnitude. 

\section{A multiscale stochastic method from turbulent to the rarefied flows}\label{sec:3}
So far, the DSMC method has primarily been used for simulating laminar rarefied flows. However, directly solving the Boltzmann equation for continuum flows, especially in the presence of turbulence, remains computationally prohibitive. For example, simulating the turbulent Taylor-Green vortex using DSMC required over half a million CPU cores for 500 hours~\cite{gallis-2017}. Even with advancements in computational resources, a 2D hypersonic side-jet flow simulation still demands approximately half a million CPU hours~\cite{Karpuzcu2023Study}. Given these substantial computational costs, incorporating turbulence models is essential for practical engineering applications. In this section, we first introduce the $k\text{-}\omega$ SST turbulence model in the context of continuum flows, and then integrate it into the DIG framework for simulations ranging from free-molecular to turbulent regimes.

\subsection{The $k$-$\omega$ turbulence model}

Although the NS equations adequately describe gas dynamics in the continuum flow regime, incorporating turbulence effects becomes crucial at extremely high Reynolds numbers. Typically, these effects are captured by directly solving the NS equations. However, the prohibitive computational cost of direct numerical simulation has spurred the development of turbulence models. Despite being derived under multiple simplifying assumptions, these models are widely adopted in industrial applications due to their computational efficiency and reliable accuracy. Here the popular $k\text{-}\omega$ SST model is applied \cite{menter-1994}, where the NS constitutive relations in~\eqref{eq:NS_constitutive} are rewritten as:
\begin{equation}
    \begin{aligned}
        \mu &= \mu_{lam}+\mu_{turb},\\
        \kappa_t &= \kappa_{t,lam}+\kappa_{t,turb},\\
        \kappa_r &= \kappa_{r,lam}+\kappa_{r,turb},
    \end{aligned}
    \label{eq:mukappa_lam_turb}
\end{equation}
where the subscript $lam$ and $turb$ denote the laminar and turbulent parts respectively. The dimensionless form of turbulent viscosity is given by
\begin{equation}
    \mu_{turb}=a_R\rho k/\max\left(a_R\omega,\frac{\lVert\Omega\rVert F_\mu}{\text{Re}_{\text{ref}}}\right),
\label{eq:muturb}
\end{equation}
where $k$ represents the turbulent kinetic energy, $\omega=\rho k/\mu_{turb}$ is the turbulent dissipation frequency with $\lVert\Omega\rVert$ being the vorticity magnitude. In Eq.~\eqref{eq:muturb}, $a_R$ is a constant based on the Bradshaw assumption, and $F_\mu$ is a blending function with $F_\mu=1$ for boundary layers and $F_\mu=0$ for free shear flows. Note that the turbulent heat conductivity, regardless of whether it is translational or rotational, is given by:
\begin{equation}
    \kappa_{turb}=\frac{\mu_{turb}\text{Pr}_{lam}}{\text{Pr}_{turb}}\frac{c_p}{\text{Pr}_{lam}},
\label{eq:kappaturb}
\end{equation}
where the laminar to turbulent Prandtl number ratio is set to 0.8, with the value of $c_p/\text{Pr}_{lam}$ for $\kappa_t$ and $\kappa_r$ which is determined according to Eq.~\eqref{eq:kappat_kappar}.

The $k\text{-}\omega$ SST model is a widely applied turbulence model that effectively combines the strengths of both low- and high-Reynolds number turbulence models. Specifically, it blends the $k\text{-}\omega$ model in the near-wall region, where its formulation excels at capturing boundary layer phenomena, with the $k\text{-}\epsilon$ model in the freestream, which is well-suited for regions away from solid boundaries. This hybrid combination enables the SST model to accurately predict key aerodynamics behaviors, such as boundary layer separation and flow response under adverse pressure gradients, making it particularly effective for simulating complex aerodynamic flows. Furthermore, it incorporates Bradshaw’s assumption, linking turbulent shear stress to turbulent kinetic energy and improving the predictions of turbulent boundary layers~\cite{menter-1994}. While the model was originally developed for incompressible flows, its adaptability has allowed extensions to compressible and even hypersonic flow scenarios. According to Sarkar~\cite{sarkar-1992}, additional terms, such as pressure-dilatation and dilatation dissipation, are included to account for compressibility effects. And these terms are essential for reducing the turbulent kinetic energy for flows with Mach number (Ma) larger than 5. The final form of the implemented $k\text{-}\omega$ SST model are written as:
\begin{equation}
\begin{aligned}
\frac{\partial\rho k}{\partial t}+\frac{\partial\rho u_{j}k}{\partial x_{j}}= & \frac{1}{\text{Re}_{\text{ref}}}\frac{\partial}{\partial x_{j}}\left[(\mu_{lam}+\sigma_{k}\mu_{turb})\frac{\partial k}{\partial x_{j}}\right] 
+\text{Prod}-\underbrace{\text{Re}_{\text{ref}}\beta^{*}\rho\omega k}_{Diss_{k}} \\
 & -\underbrace{\text{Re}_{\text{ref}}\beta^{*}\xi^{*}M_{t}^{2}\rho\omega k}_{\text{Dilatation dissipation}} 
 +\underbrace{\alpha_{2}\tau_{turb,ij}\frac{\partial u_{i}}{\partial x_{j}}M_{t}+\mathrm{Re}_{\mathrm{ref}}\alpha_{3}\beta^{*}\rho\omega kM_{t}^{2}}_{\text{pressure dilatation}},
\end{aligned}
\label{eq:Eqfork}
\end{equation}
\begin{equation}
\begin{aligned}
\frac{\partial\rho\omega}{\partial t}+\frac{\partial\rho u_{j}\omega}{\partial x_{j}}= & \frac{1}{\text{Re}_{\text{ref}}}\frac{\partial}{\partial x_{j}}\left[(\mu_{lam}+\sigma_{\omega}\mu_{turb})\frac{\partial\omega}{\partial x_{j}}\right] 
  +\gamma\frac{\rho}{\mu_{turb}}\tau_{turb,ij}\frac{\partial u_{i}}{\partial x_{j}}-\text{Re}_{\text{ref}}\beta\rho\omega^{2} \\
 & +\frac{2(1-F_{\phi})}{\text{Re}_{\text{ref}}}\frac{\rho\sigma_{\omega2}}{\omega}\frac{\partial k}{\partial x_{j}}\frac{\partial\omega}{\partial x_{j}} 
  +\underbrace{\text{Re}_{\text{ref}}\beta^{*}\xi^{*}M_{t}^{2}\rho\omega^{2}}_{\text{Dilatation dissipation}},
\end{aligned}
\label{eq:Eqforw}
\end{equation}
where $M_t=\sqrt{2k}/\sqrt{\gamma RT}$ is the turbulent Mach number. Note that $\tau_{turb}$ is the Favre-averaged Reynolds-stress computed under the Boussinesq eddy-viscosity hypothesis, which is given by:
\begin{equation}
    \tau_{turb}=\mu_{turb}\left(\nabla\bm{u}+\nabla\bm{u}^{\mathrm{T}}-\frac{2}{3}\nabla\cdot\bm{u}\bm{I}\right)-\frac{2}{3}\rho k\bm{I}.
\end{equation}
And to eliminate certain erroneous spikes of $\mu_{turb}$ encountered in two-equation turbulence models~\cite{menter1993}, the production term in Eq.~\eqref{eq:Eqfork} is designed as
\begin{equation}
    \text{Prod}=\min\left(\tau_{turb,ij}\frac{\partial u_i}{\partial x_j},20Diss_k\right).
\end{equation}
Additionally, $k=0$ and $\omega=(60\mu_{lam})/(\text{Re}_\text{ref}\beta_1\rho D_1^2)$ are specified at the solid surface, with $D_1$ being the distance of the first cell center to the wall. For other coefficients in the $k\text{-}\omega$ SST model such as $(\alpha_2,\alpha_3,\xi^*)$, see Refs. \cite{wilcox-1993,tian-2024}.

\subsection{The DIG-SST coupling for coexisting turbulent and rarefied flows}

% To accurately capture rarefaction effects, the cell size and time step in the DSMC method must be set smaller than one-third of the mean free path and mean collision time, respectively. However, these stringent requirements result in a prohibitive computational burden, particularly in the near-continuum regime. Given that practical engineering problems predominantly focus on steady-state solutions, the implicit iteration method for solving the NS equations can be employed to accelerate the convergence of DSMC simulations toward steady states. In DIG, such coupling of mesoscopic and macroscopic solvers has been demonstrated to be significantly more efficient than the standard DSMC approach. Here, we first review the recently developed DIG method, which enables efficient simulation of gas flows across the entire flow regime, and then propose the DIG-SST algorithm to simulate problems involving the coexistence of rarefied and turbulent flows.

As stated in previous section, although the DIG is efficient in capturing the flow properties in the multiscale problems, the macroscopic solver is still based on the NS equation, rendering it incapable of efficiently describing high Reynolds number flows with turbulence effects. Since the NS constitutive relations for heat fluxes and stress in Eq.~\eqref{eq:GSIS_constitutive} are explicitly incorporated, the eddy-viscosity assumption employed in Reynolds-averaged Navier-Stokes (RANS) models can be readily embedded in the general DIG framework. That is, the turbulence viscosity determined according to the $k$-$\omega$ SST model is directly employed as the physical viscosity in the macroscopic synthetic equations, while the mesoscopic part remains the same. Consequently, the constitutive relations of heat fluxes and stress are reconstructed as~\cite{tian-2024,tian-2025}:
\begin{equation}
\begin{aligned}
    &\bm{\sigma}=-(\mu_{lam}+\mu_{turb})\left(\nabla\bm{u}+\nabla\bm{u}^T-\frac{2}{3}\nabla\cdot\bm{uI}\right)+\text{HoT}_{\bm{\sigma}},\\
    &\bm{q}_t=-(\kappa_{t,lam}+\kappa_{t,turb}) \nabla T_t+\text{HoT}_{\bm{q}_t}, \\
    &\bm{q}_r=-(\kappa_{r,lam}+\kappa_{r,turb})\nabla T_r+\text{HoT}_{\bm{q}_r},
\end{aligned}
\label{eq:GSIS_SST_constitutive}
\end{equation}
where the turbulence viscosity is determined from the $k$-$\omega$ SST model.

As demonstrated in Refs. \cite{su-2020,su-2020-can}, the HoTs derived from the Chapman-Enskog expansion of the kinetic model are proportional to $\text{Kn}^2$ in the continuum limit. Showing from Eq.~\eqref{eq:GSIS_SST_constitutive}, when the Knudsen number is small and the HoTs are negligible, the DIG-SST coupling method asymptotically preserves to the original $k$-$\omega$ SST model for turbulent flows. Conversely, for large Knudsen numbers, where rarefaction effects dominates, the turbulent viscosity $\mu_{turb}$, driven by density and shear motion, becomes negligible in comparison to the laminar viscosity $\mu_{lam}$. As a results, the outcomes of the proposed coupling algorithm align with those obtained from the DSMC method.

\subsection{General algorithm of the DIG-SST method}
\begin{figure}[t]
	\centering
	\includegraphics[width=0.8\textwidth]{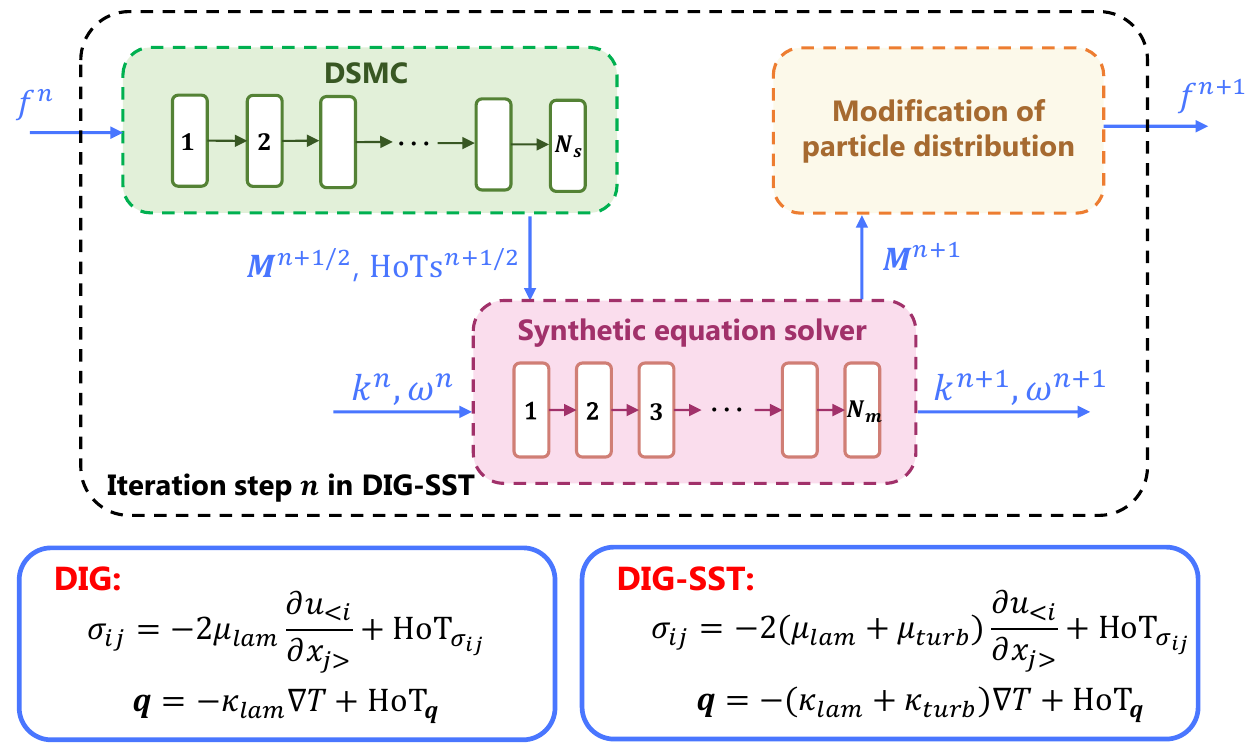}
	\caption{
 Flowchart of the DIG-SST coupling algorithm in a unit cycle in the $n$-th step. In each iteration step, the synthetic equations together with $k$-$\omega$ SST turbulent model are solved for $N_m=500\text{-}2000$ iterations, or until the relative error of the conservative variables between two successive steps falls below $10^{-5}$. Different from the DIG method, the additional turbulent viscosity and heat conductivities are employed in DIG-SST method.
	}
	\label{fig:flowchart}
\end{figure}

Since the DIG-SST method is generally an extension of the DIG method, it naturally inherits the properties of the DIG method described in previous sections. The sole distinction in the DIG-SST method lies in the additional solving of the $k$-$\omega$ SST equation alongside the synthetic equations. When solving these macroscopic synthetic equations, the time-implicit finite volume scheme is employed, with the standard Lower-Upper Symmetric Gauss-Seidel applied for the iterative process. Detailed process for solving synthetic equations can be found in Ref.~\cite{zeng-2023}. Moreover, the time-averaged macroscopic properties $\bm{M}^{n+1/2}=[\rho^{n+1/2},\bm{u}^{n+1/2},T_t^{n+1/2},T_r^{n+1/2}]$ should be extracted from preceding DSMC steps and then obtain higher-order terms $\text{HoT}^{n+1/2}$ according to Eq.~\eqref{eq:higher-oder_terms}. However, since the turbulent kinetic energy $k$ and turbulent dissipation rate $\omega$ are derived from the SST turbulent model, they cannot be determined directly from the mesoscopic equations. Initially, the values of $k$ and $\omega$ are set based on the specified initial turbulence intensity $I_t$ and the turbulent-to-laminar viscosity ratio $\mu_r=\mu_{turb}/\mu_{lam}$, which are defined at different boundaries. Subsequently, during the $n$-th iteration step in DIG-SST, the values of $k^n$ and $\omega^n$, which are determined after the $(n-1)$-th iteration step, serve as the initial inputs for solving the synthetic equations in the $n$-th step. The general algorithm of DIG-SST method illustrated in Fig.~\ref{fig:flowchart}, is briefly outlined in Algorithm \ref{algo:DIG-SST}.

\begin{algorithm}[!h]
    \caption{Overall algorithm of DIG-SST coupling method} 
    \label{algo:DIG-SST}
    \begin{algorithmic}[1]
        \Require
            Initial distribution of macroscopic properties $\bm{M}$ and turbulent properties $k$, $\omega$;
        \Ensure
            Time-averaged macroscopic properties $\bm{M}$ after steady state;
        \State Solve classic NS equations with $k$-$\omega$ SST model to obtain updated macroscopic properties $\bm{M}$ and turbulent properties $k$, $\omega$;
        \State Draw $N_p$ simulation particles within each computational cell according to the updated $\bm{M}$ and Maxwellian distribution;
        \State Run standard DSMC for $n_a$ time steps to obtain sufficient samples;
        \State Set iteration step $n = 1$;
        \While {$n \le \text{MaxSteps}$}
            \State Run standard DSMC for $N_s$ time steps;
            \State Calculate the time-averaged macroscopic properties $\bm{M}^{n+1/2}$, see Ref.~\cite{luo-2023};
            \State Obtain the higher-order terms $\text{HoT}^{n+1/2}$ based on Eq.~\eqref{eq:higher-oder_terms};
            \State Use $k^n$ and $\omega^n$ obtained from the $(n-1)$-th step as initial input;
            \State Solve Eqs.~\eqref{eq:Navior-Stokes}, \eqref{eq:Eqfork} and \eqref{eq:Eqforw} by $N_m$ steps to obtain $\bm{M}^{n+1}$, $k^{n+1}$ and $\omega^{n+1}$;
            \State Modify the distribution of particles to match $\bm{M}^{n+1}$, see Ref.~\cite{luo-2023};
            \State $n ++$;
        \EndWhile
    \end{algorithmic}
\end{algorithm}

\section{Validation of the DIG-SST method}\label{sec:4}%%%%%%%%%%%%%%%%%%%%

In this section, the DIG-SST method is validated in the simulation of hypersonic flow passing over a wedge and a blunt body. All simulations are conducted with nitrogen gas, employing the variable hard sphere model for binary collisions and the Maxwellian diffuse boundary condition at the solid surface. The definitions of the global Knudsen number and Reynolds number, along with the physical parameters of nitrogen gas employed in this study, are given as:
\begin{equation}
\text{Kn} = \frac{\mu_0}{p_{0}L_{0}}\sqrt{\frac{\pi k_B T_{0}}{2m}},\,\,
\text{Re} = \frac{\rho u L_{0}}{\mu_{0}}, \,\,
\omega=0.74,\,\,
Z=2.59,\,\,
\bm{A}=\left[\begin{matrix}0.786&-0.201\\ -0.059 & 0.842 \end{matrix}\right]
\label{eq:KnRe}
\end{equation}
As discussed in previous section, both the DIG-SST and DIG methods necessitate a sufficient sample size of $n_a=1000$ to effectively suppress fluctuations in macroscopic properties, with the macroscopic synthetic equations being solved every $N_s=100$ steps. Furthermore, the number of simulation particles per cell plays a critical role in determining the stability and accuracy of the macroscopic equations. While increasing the particle count reduces statistical fluctuations, it also imposes a greater computational cost. Conversely, a reduced particle count leads to pronounced fluctuations, which may compromise the stability of the macroscopic solver. Here, an initial particle count of $N_p=200$ per cell is empirically chosen to achieve an optimal balance between computational efficiency and solution stability.

\subsection{Large Reynolds number}

The hypersonic flow passing over a wedge is a classic problem included in the NASA Turbulence Test Cases database, serving as a standard benchmark for turbulence model development and validation. The Mach number of the free stream is $\text{Ma}=9.22$ with the reference temperature $T_0=64.5$K, while the temperature of the solid wall is $T_{\text{wall}}=295$K. The reference length is taken as $L = 1~$m while the global Reynolds number is $4.7\times10^7$ and Knudsen number is $2.91\times10^7$. The computational domain is discretized using a total of 31,284 non-uniform spatial grid cells, with the first cell layer exhibiting a height of $10^{-6}~$m.

\begin{figure}[!t]
    \centering
    \includegraphics[width=0.49\textwidth,trim=10pt 10pt 10pt 10pt, clip]{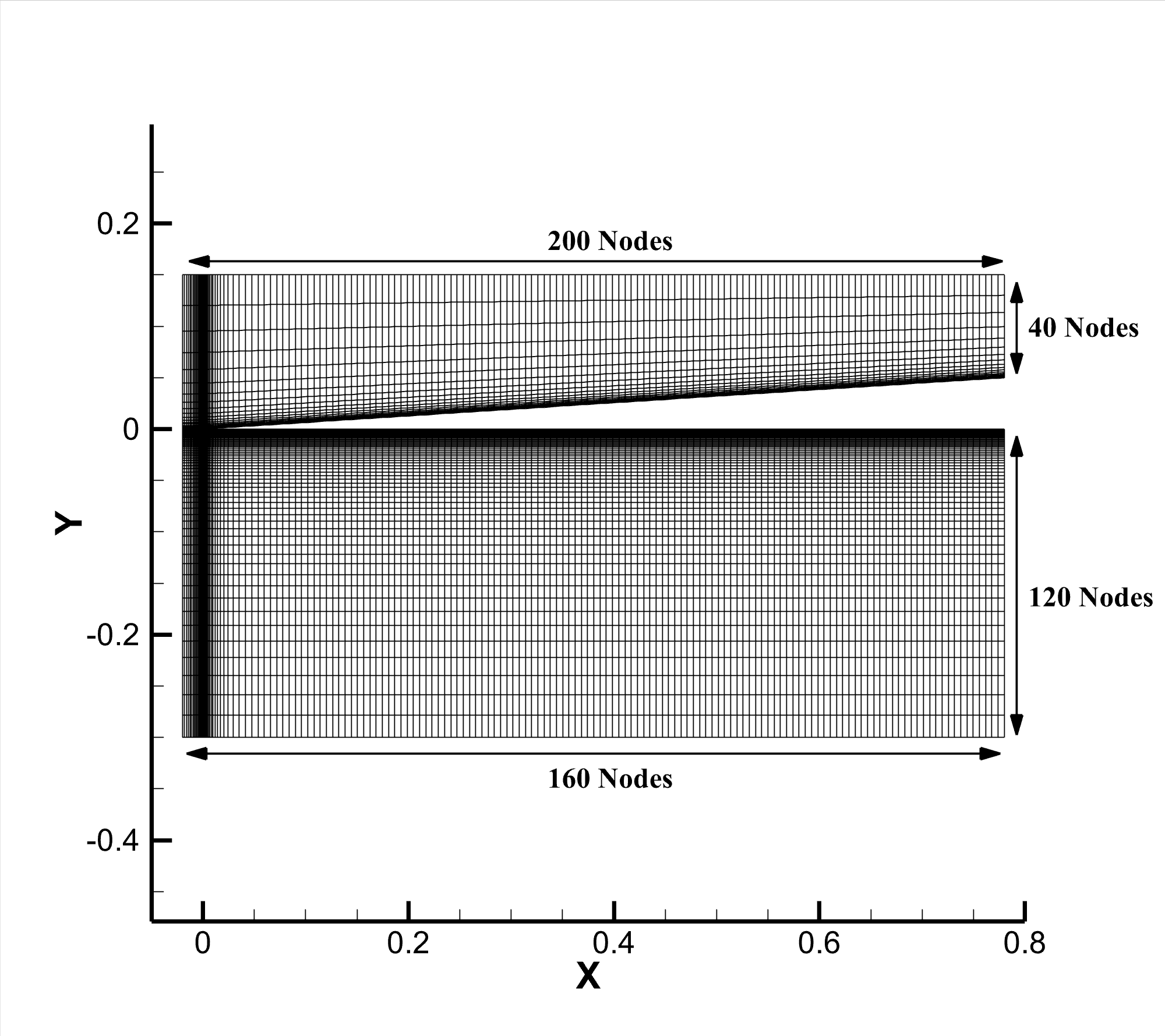}
    \includegraphics[width=0.49\textwidth]{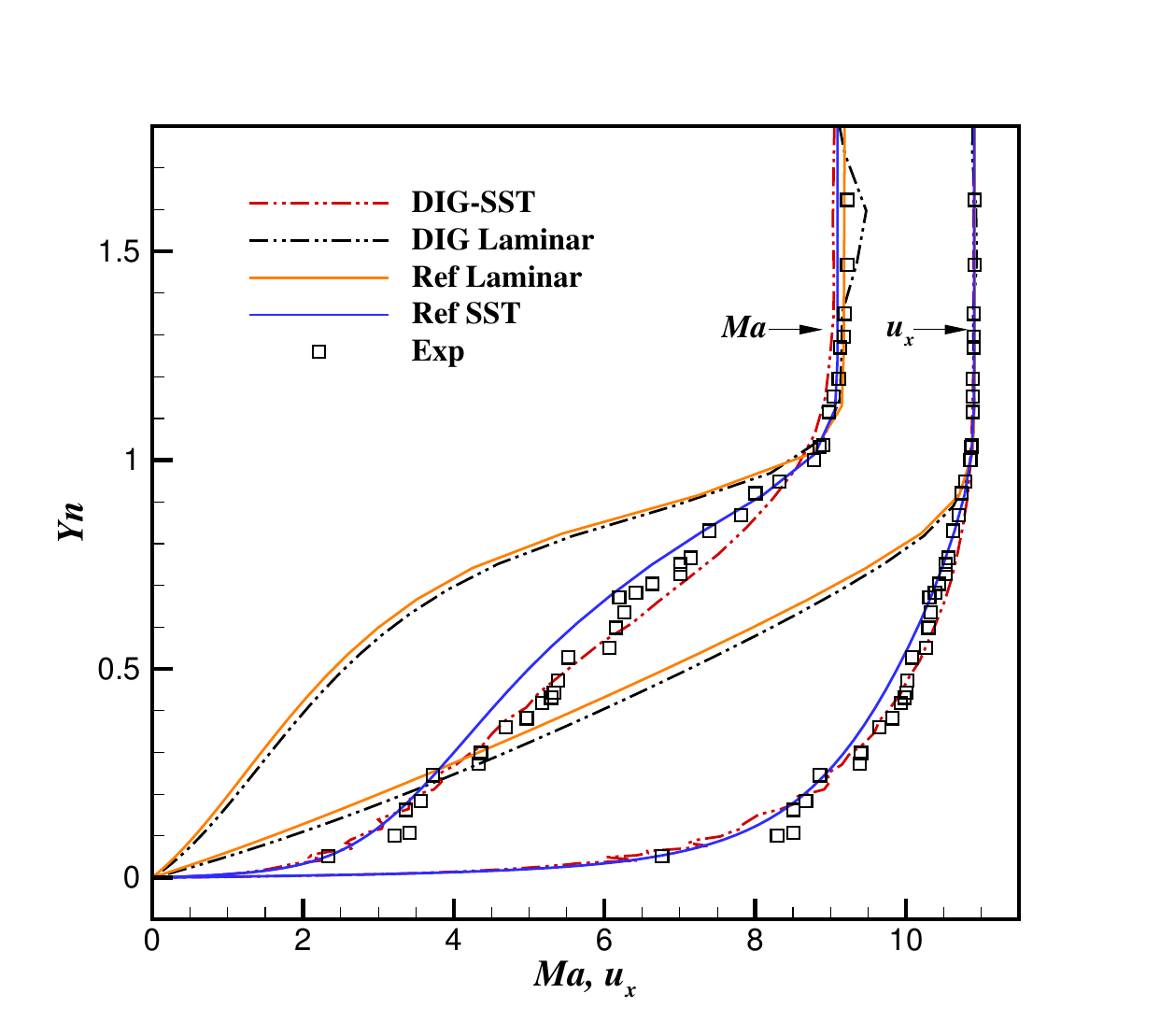}\\
    \caption{(Left) The computational domain for the hypersonic flow passing over a wedge. (Right) The comparison of the boundary layer velocity as well as the local Mach number between the experimental data and different methods. The reference CFD solution is obtained by CFD++.  }
    \label{fig:flate_plate}
\end{figure}

As shown in Fig.~\ref{fig:flate_plate}, the experimental data are extracted at $X=0.076~$m on the surface of the lower flat plate \cite{NASA_Database}, where $Y_n$ represents the vertical distance normalized by the boundary layer thickness, defined at 99.5\% of the free-stream velocity. The boundary layer velocity and local Mach number predicted by the DIG-SST method show excellent agreement with the experimental data. Additionally, the original $k$-$\omega$ SST model, solved using a conventional CFD solver, also aligns closely with the DIG-SST results. Thus, these solutions validate the capability of the DIG-SST approach to accurately simulate problems in the continuum regime with turbulence modeling.

\subsection{Large Knudsen number}
\begin{figure}[!t]
    \centering
    \subfigure[]{\includegraphics[width=0.49\textwidth]{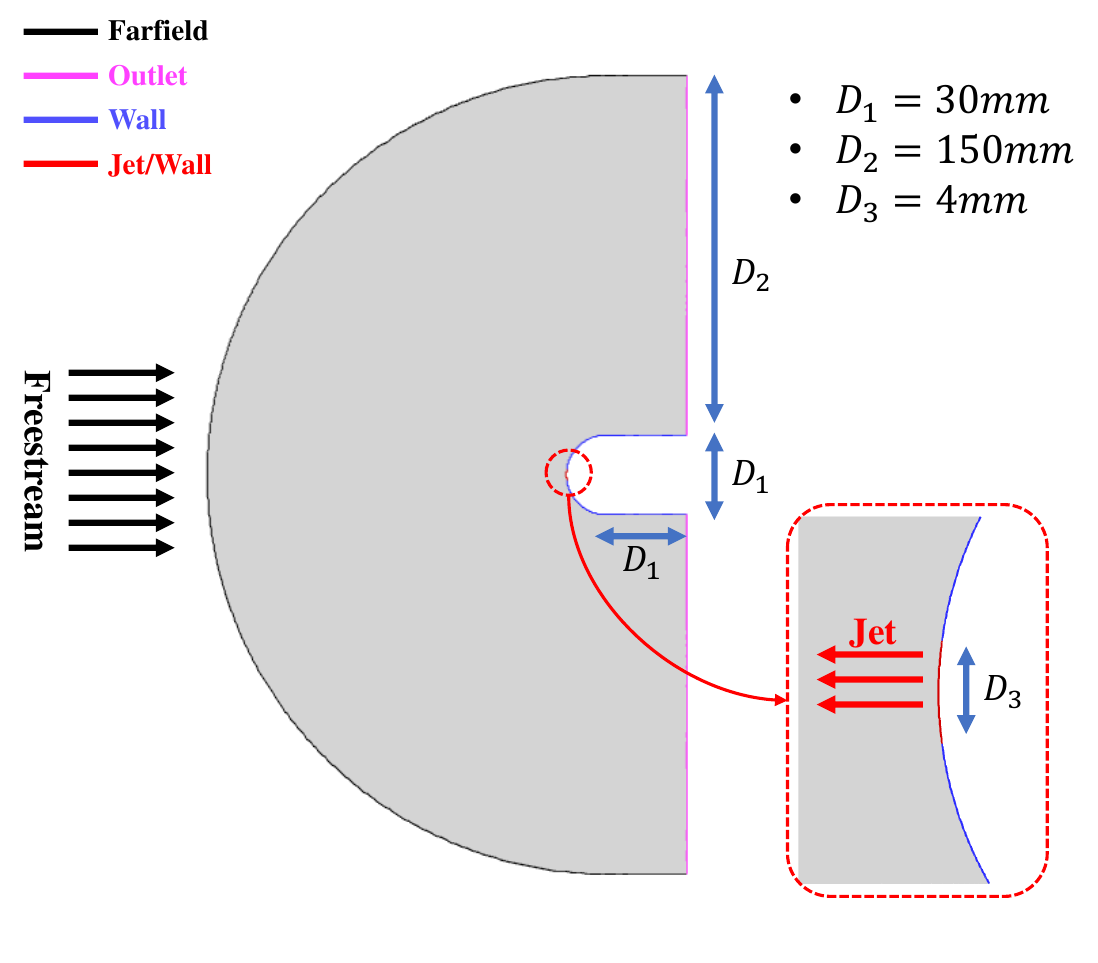}}
    \subfigure[]{\includegraphics[width=0.49\textwidth,trim=10pt 10pt 10pt 10pt,clip]{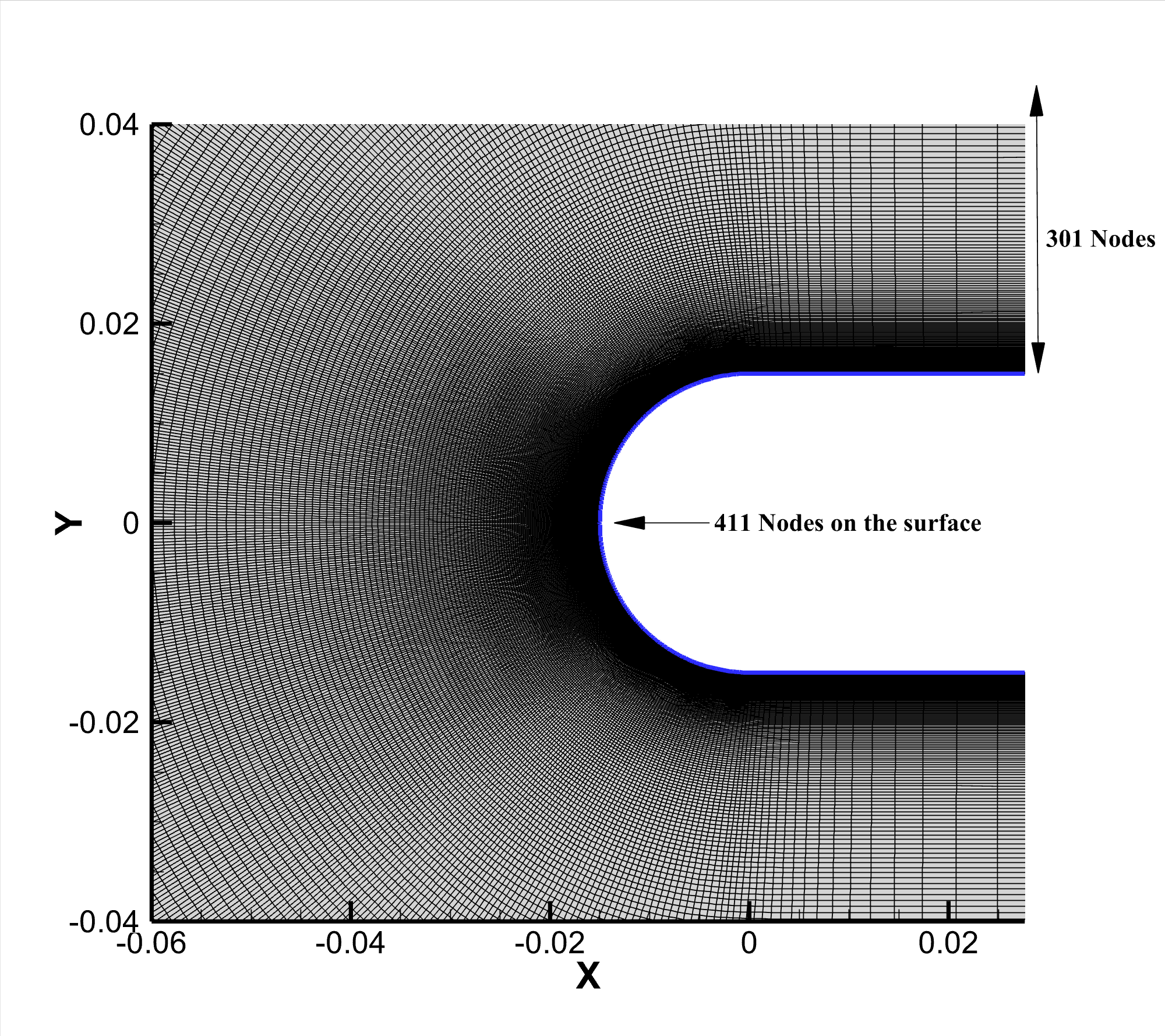}}\\
    \vspace{-4mm}
    \subfigure[]{\includegraphics[width=0.49\textwidth,trim=10pt 10pt 10pt 10pt,clip]{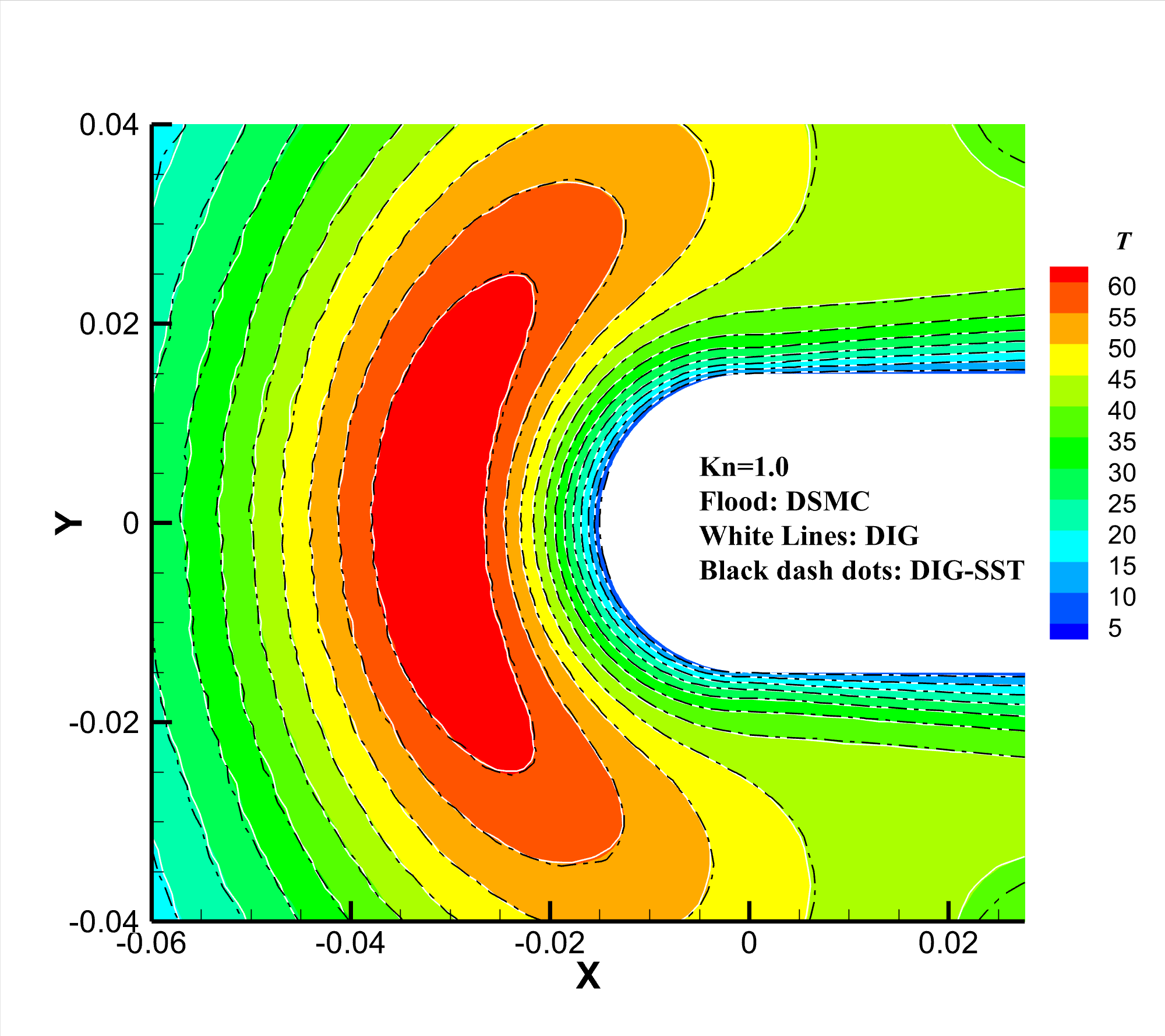}}
    \subfigure[]{\includegraphics[width=0.49\textwidth]{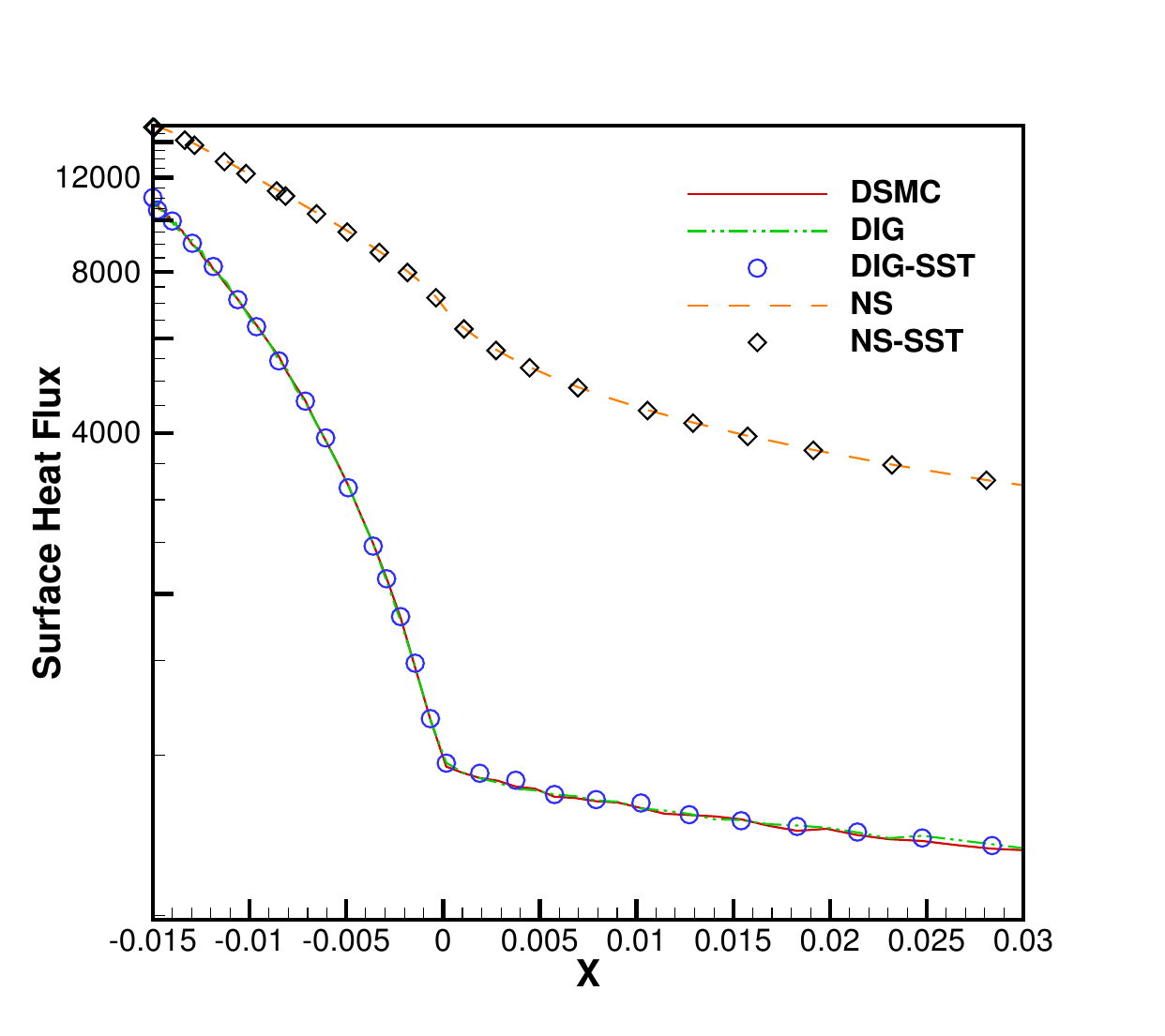}}
    \vspace{-5mm}
    \caption{(a) Schematic illustration of the blunt body configuration employed in this work, where the jet boundary is located at the leading edge. (b) The computational grid, shown with units in meters. (c) The contour of the total temperature obtained by different methods when $\text{Ma}=25$ and $\text{Kn}=1$. (d) Distributions of the surface heat flux.  }
    \label{fig:blunt_body_largekn}
\end{figure}

The hypersonic flow over a blunt body at a high Knudsen number is validated using the DSMC, DIG, and DIG-SST methods. The free-stream Mach number is $\text{Ma}=25$, with the temperatures of the solid wall and free stream set to be identical. The reference length, $L$, is defined as the diameter of the blunt body's leading edge, $D_1=30~$mm. Under these conditions, the Knudsen number is $\text{Kn}=1$, and the Reynolds number is $\text{Re}\approx25$. The computational domain and mesh are shown in Fig.~\ref{fig:blunt_body_largekn}, consisting of 123,000 cells. The time step is chosen such that the product of the time step and the most probable speed does not exceed the size of the smallest grid cell.

As shown in Fig.~\ref{fig:blunt_body_largekn}(c), the results from DSMC, DIG, and DIG-SST methods aligns perfectly, indicating that the turbulence model becomes negligible in highly rarefied environments. Figure~\ref{fig:blunt_body_largekn}(d) shows the surface heat flux predicted by different methods. The conventional NS equations, which fail to incorporate rarefied effects, exhibit significant deviations from DSMC results. In contrast, the predictions from both DIG and DIG-SST methods align closely with those of the DSMC approach, highlighting their capability to accurately capture rarefied flow phenomena. 

\section{Application of the DIG-SST method in Opposing jet}\label{sec:5}

Opposing jet technology has been introduced directly ahead of the aircraft surface to enhance the thermal protection and maneuverability of the aircraft. Numerous experiments have demonstrated that opposing jets can significantly reduce heat loads and drag in supersonic flows~\cite{aso-2003,daso-2009}. %Furthermore, several numerical studies employing turbulence models, such as the $k$-$\omega$ SST model, have further validated their heat and drag reduction capabilities~\cite{ou-2018}. 
%Additionally, the behavior of opposing jets in rarefied environments has been analyzed using the DSMC method~\cite{raeisi-2019}, with the study domain covering altitudes ranging from 60 km to 90 km~\cite{guo-2024}. Microscale jets have also been simulated and validated using hybrid NS-DSMC solvers~\cite{virgile-2022,liuw-2024}; however, these studies focus only on laminar flows and neglect the effects of turbulence. 
Here, the DIG-SST method is employed to comprehensively examine the interaction between turbulent and rarefied flows.

The computational domain is identical to that depicted in Fig.~\ref{fig:blunt_body_largekn}, with the distinction that a 4~mm wide wall boundary located at upstream of the blunt body surface has been replaced by a jet flow boundary. The characteristics of the jet flow are defined based on the pressure ratio, $P_{\text{ratio}}$, which represents the ratio of the jet flow pressure to the post-shock pressure of the free stream, followed by,
\begin{equation}
    P_{\text{ratio}}=\frac{p_{j,0}}{p_{f,s,0}},
\end{equation}
where $p_{j,0}$ is the stagnation pressure of the jet and $p_{f,s,0}$ is the stagnation pressure of the downstream shock, which is calculated based on the relation of the normal shock for free stream properties. Different pressure ratios have been employed to validate the turbulent effect.
The jet has a constant temperature of 250~K and an exit Mach number of 1.

In the DIG-SST method, which incorporates the $k$-$\omega$ SST turbulence model, the initial values of turbulent kinetic energy $k$ and dissipation frequency $\omega$ must be specified. For the incoming free-flow boundary, a turbulence intensity of $I_t = 0.3\%$ and a turbulent to laminar viscosity ratio $\mu_r = 15$ are initialized. At the jet flow boundary, the turbulence intensity is set to $I_t = 3\%$, with a non-dimensional characteristic length $l_t = 0.118$. The values of $k$ and $\omega$ are calculated as $k=1.5I_t(\|\bm{u}\|)^2$ and $\omega=\sqrt{k}/l_t$, respectively. Additionally, the number of particles initially employed in each cell as well as the sample size are the same as previous computational cases.

The opposing jet flow under different conditions at altitudes of 61.5 km and 80 km, corresponding to Knudsen numbers of 0.125 and 0.01, respectively, is analyzed, as summarized in Table~\ref{tab:jetcondition}. When the jet-to-free-stream number density ratio $n_0/n_\infty$ becomes excessively large, the DSMC method faces significant challenges due to its strict limitations on grid resolution and time step, rendering its results unsuitable as a reference. {For example, under a free-stream Mach number of 15 with $\text{Kn}=0.01$ and $P_\text{ratio}=2.5$, the density of the jet is about 370 times larger than the free stream, such that the mean free path is nearly 370 times smaller.} In this section, the results for $P_{\text{ratio}}=2.5$ and $\text{Ma}=5$ are first obtained and compared using the DSMC, DIG, and DIG-SST methods. Subsequently, the interaction of turbulence and rarefaction effects is investigated for supersonic jet flows at $\text{Ma}=15$ and for increased pressure ratio $P_{\text{ratio}}=5$, based on DIG and DIG-SST methods.

\begin{table}[t]
\centering
\footnotesize
\caption{The general setup for opposing jet flow simulations under various conditions is summarized. A \checkmark indicates both affordable computational cost and acceptable results, $\circ$ denotes affordable computational cost but with results lacking turbulence effects, and $\times$ signifies computational cost is prohibitive.}
\begin{tabular}{ccccccccccc}
\hline
\multirow{2}{*}{Height (km)} & \multirow{2}{*}{$\text{Kn} $} & \multirow{2}{*}{$T_\infty$} & \multirow{2}{*}{$T_{\text{jet}}/T_\infty$} & \multirow{2}{*}{Ma} & \multirow{2}{*}{$P_{\text{ratio}}$}&\multirow{2}{*}{$n_{\text{jet}}/n_\infty$} & \multirow{2}{*}{$\text{Re}_\text{jet}$} & \multicolumn{3}{c}{Methodology}                                                   \\ \cline{9-11} 
                        &                               & &                            &                                            &                     &                                            &                                         & DSMC                      & DIG                       & DIG-SST                   \\ \hline
\multirow{3}{*}{61.8}   & \multirow{3}{*}{0.01}         & \multirow{3}{*}{242.1}    & \multirow{3}{*}{1.03}                    & 5                   & 2.5 & 41.76                                      & 925.4                                   & $\circ$ & $\circ$ & \checkmark \\
                        &                               &                             &                                            & 15 &2.5                  & 371.07                                     & 6425.3                                  & $\bm{\times}$             & $\circ$                   & \checkmark \\
                        &                               &                             &                                            & 15  &5                & 742.1                                    & 12850.6                              & $\bm{\times}$             & $\circ$                   & \checkmark \\ \hline
\multirow{3}{*}{80}     & \multirow{3}{*}{0.125}        & \multirow{3}{*}{198.6}    & \multirow{3}{*}{1.26}                    & 5    &2.5               & 34.27                                      & 44.4                                   & \checkmark & \checkmark & \checkmark \\
                        &                               &                             &                                            & 15   &2.5               & 304.5                                     & 394.1                                & $\bm{\times}$             & \checkmark & \checkmark \\
                        &                               &                             &                                            & 15   &5               & 609.0                                     & 788.3                                & $\bm{\times}$             & $\circ$                   & \checkmark \\ \hline
\end{tabular}
\label{tab:jetcondition}
\end{table}

\subsection{Jet flow in small Mach number environment}

\begin{figure}[!t]
    \centering
\includegraphics[width=0.49\textwidth,trim=10pt 10pt 10pt 10pt,clip]{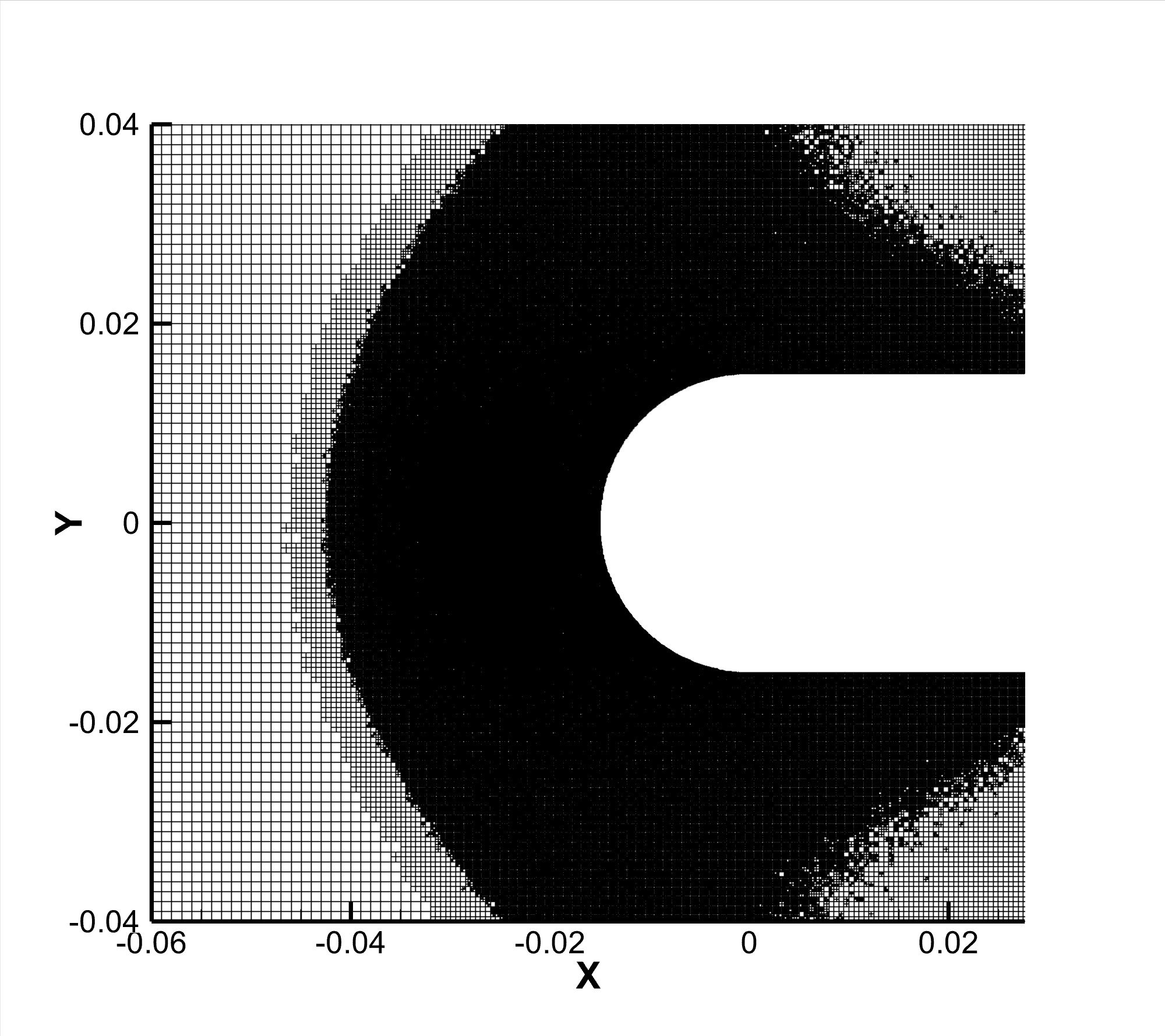}
    \includegraphics[width=0.49\textwidth,trim=10pt 10pt 10pt 10pt,clip]{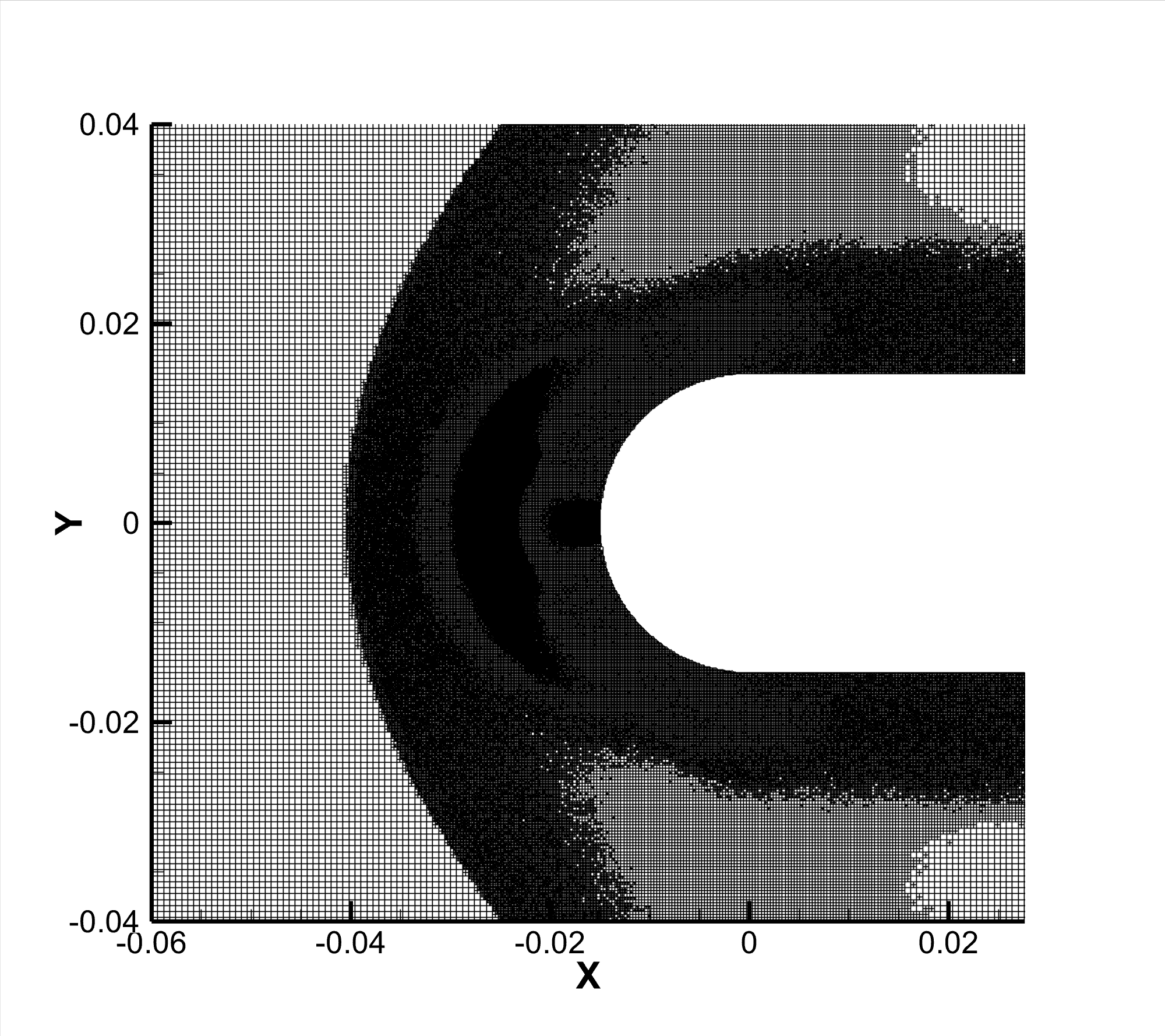}\\
    \caption{The adaptive grids of (left) $\text{Kn}=0.125$ and (right) $\text{Kn}=0.01$ used to generate the reference DSMC data in SPARTA, are applied within a computational domain same as that of the previous blunt body flow case. The time step $\Delta t$, multiplied by the most probable speed, is carefully constrained to remain smaller than the length of the smallest grid cell. }
    \label{fig:DSMC_jetmesh}
\end{figure}

\begin{figure}[p]
    \centering
\includegraphics[width=0.49\textwidth,trim=10pt 10pt 10pt 10pt,clip]{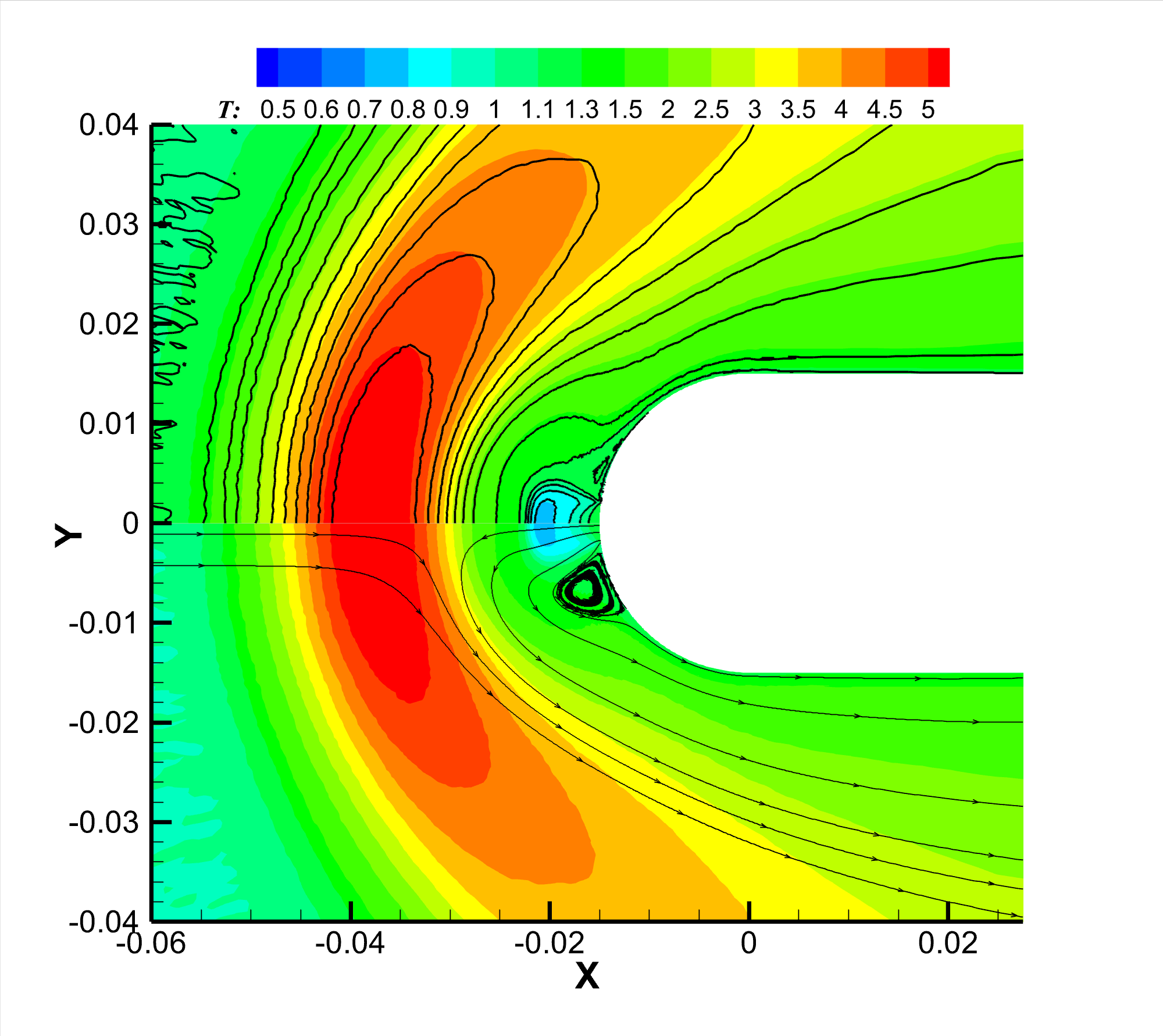}
    \includegraphics[width=0.49\textwidth,trim=10pt 10pt 10pt 10pt,clip]{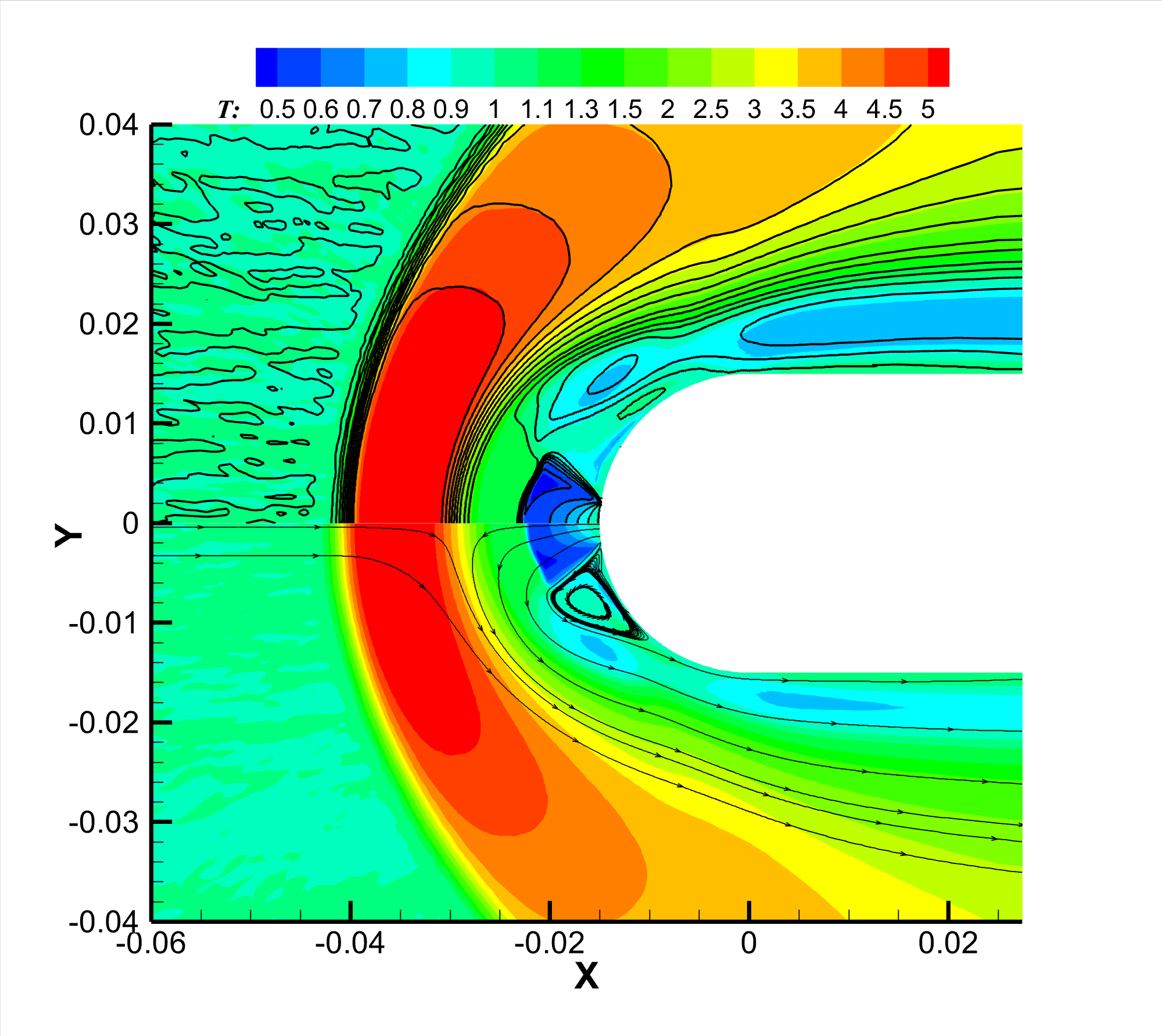}\\
    \vspace{-8mm}
    \includegraphics[width=0.49\textwidth,trim=10pt 10pt 10pt 10pt,clip]{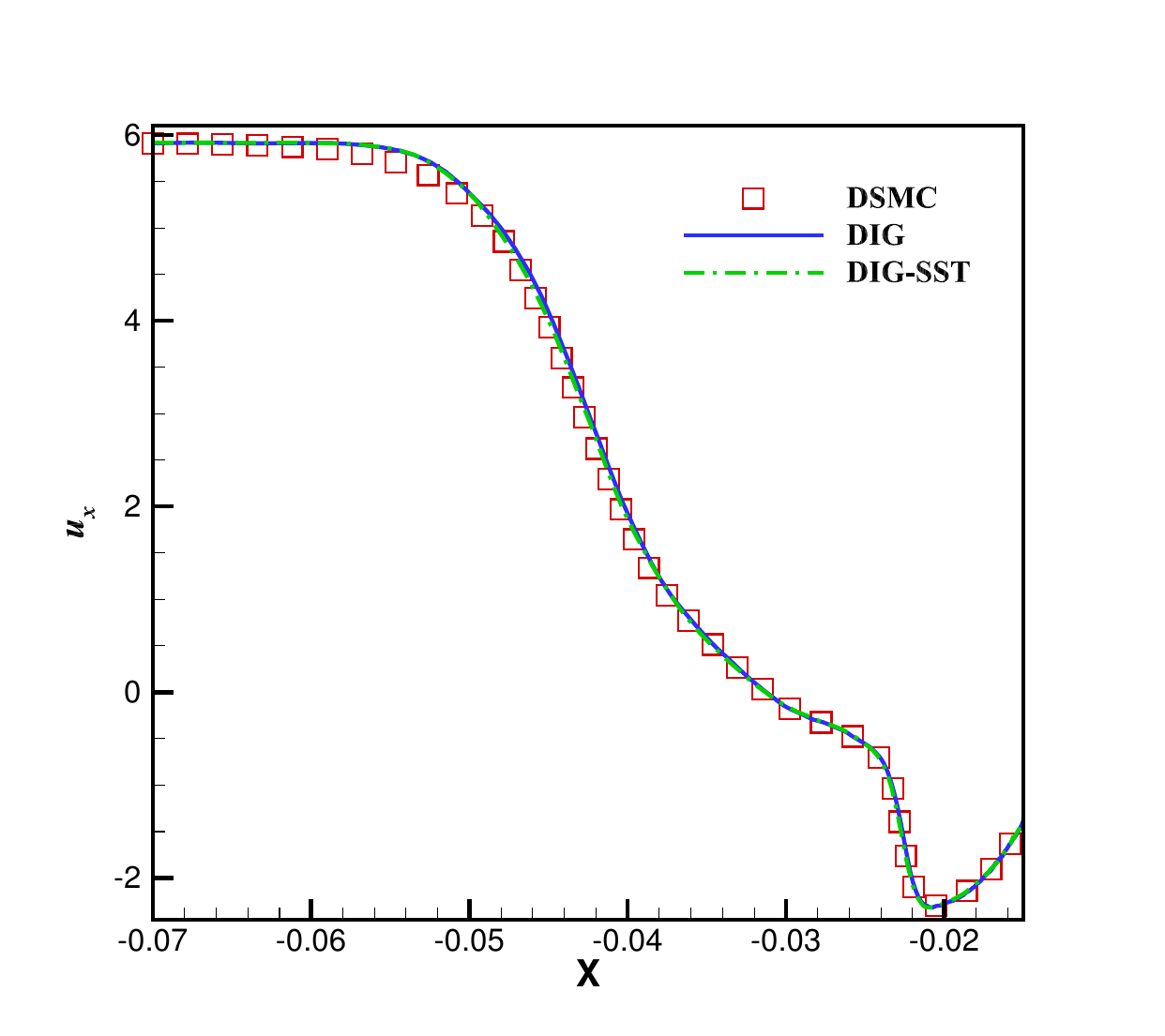}
    \includegraphics[width=0.49\textwidth,trim=10pt 10pt 10pt 10pt,clip]{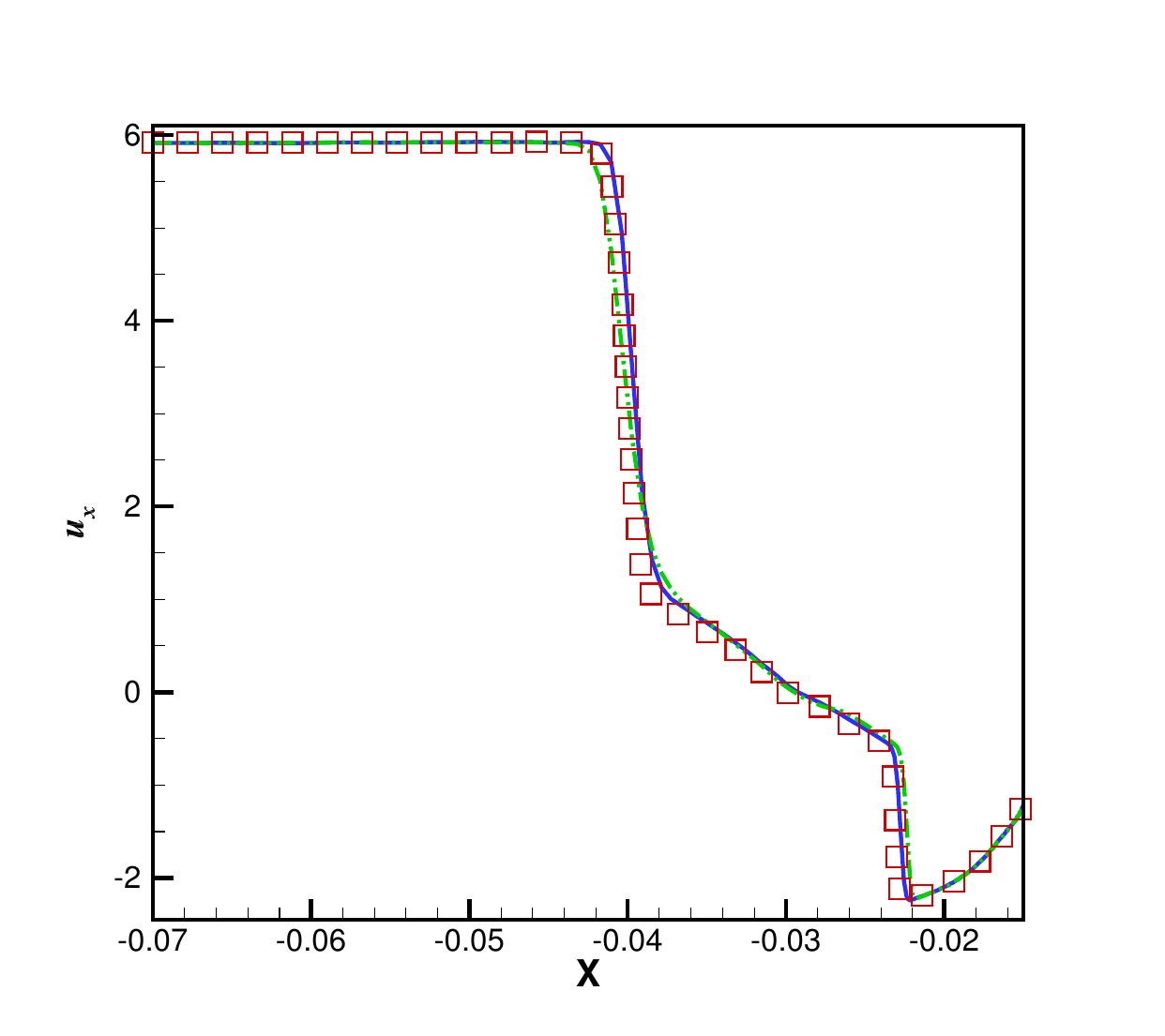}\\
    \vspace{-8mm}
    \includegraphics[width=0.49\textwidth,trim=10pt 10pt 10pt 10pt,clip]{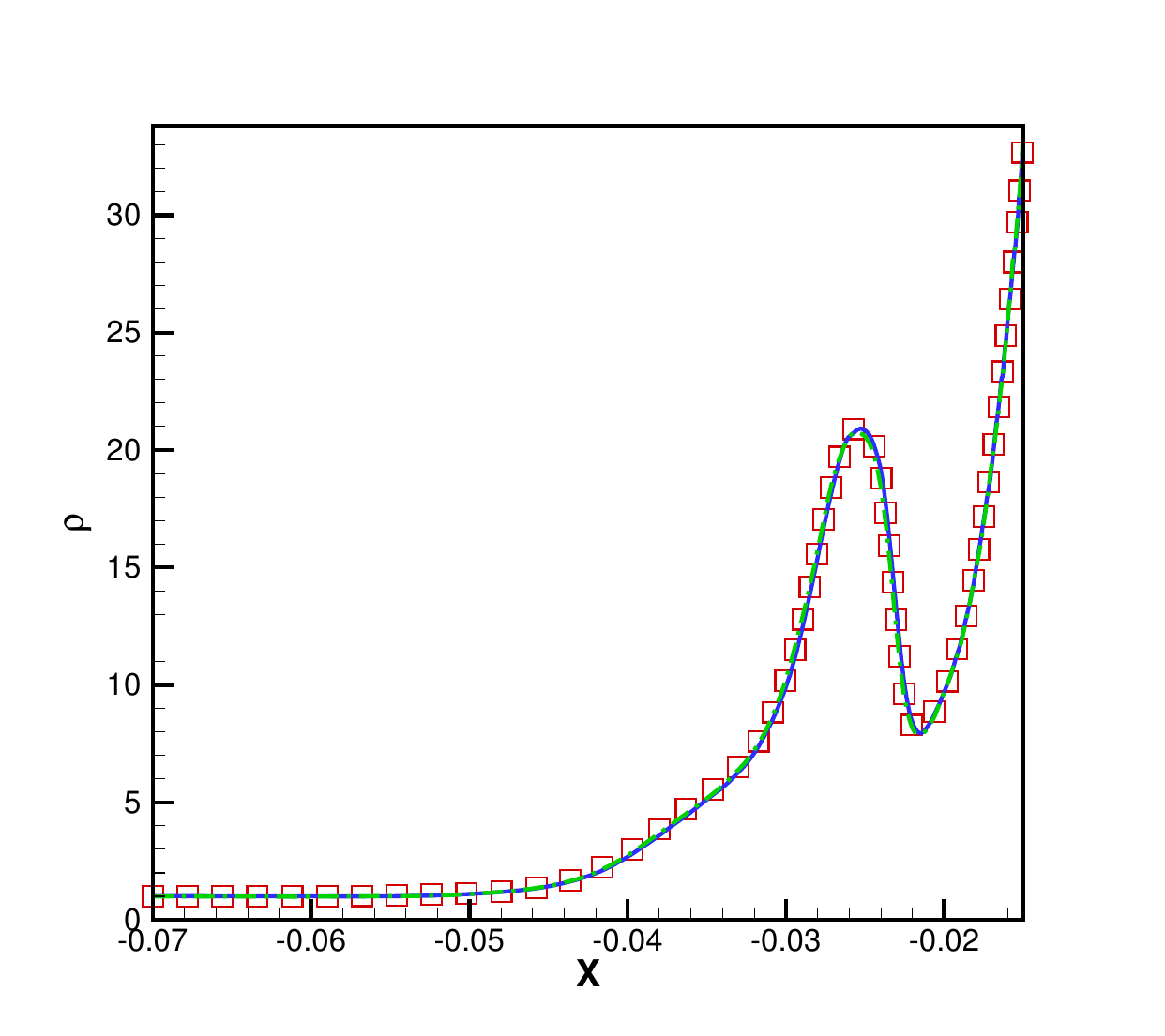}
    \includegraphics[width=0.49\textwidth,trim=10pt 10pt 10pt 10pt,clip]{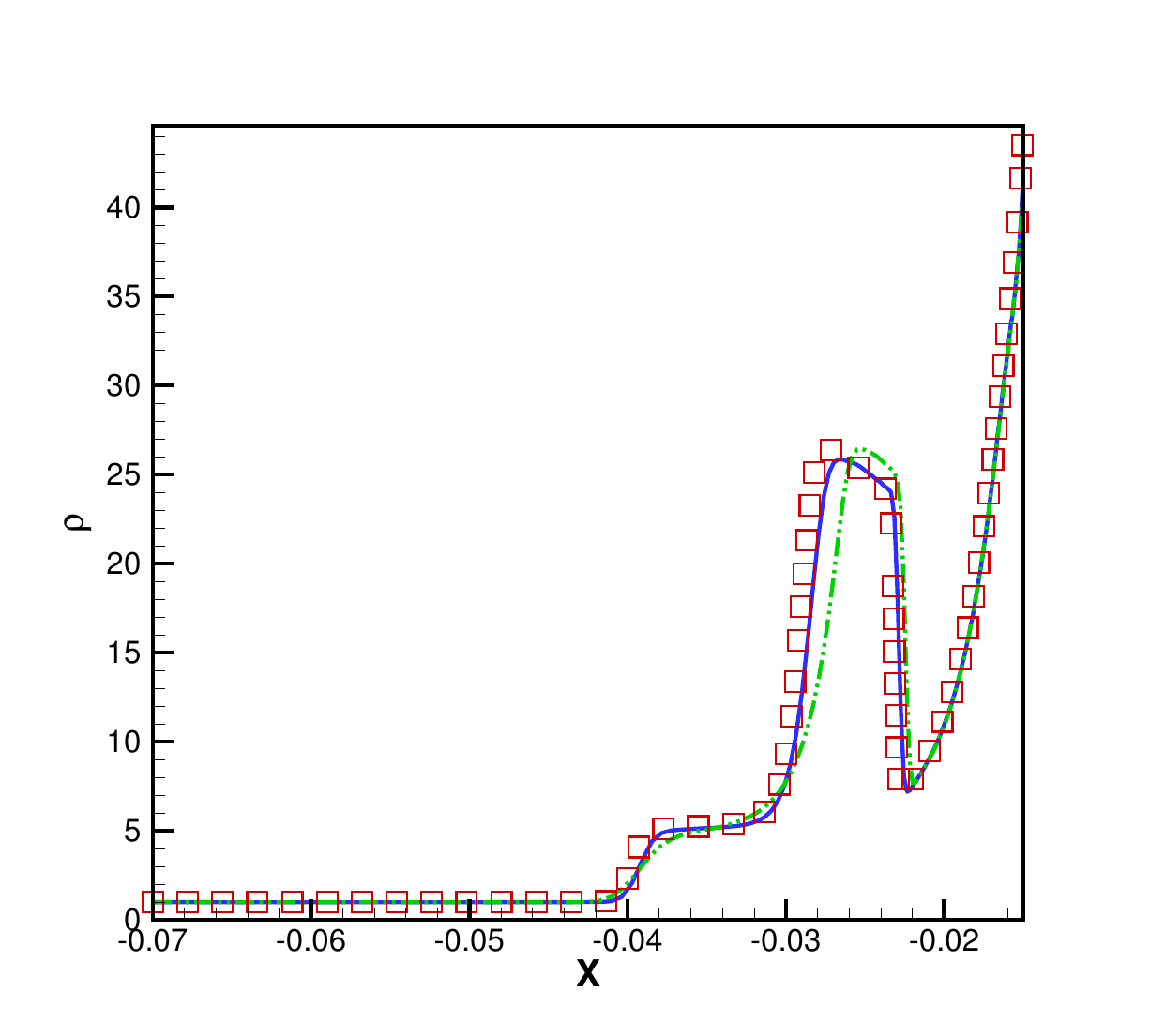}
    \vspace{-5mm}
    \caption{The temperature contour as well as the macroscopic properties extracted along the horizontal central line when $\text{Ma}=5$, $\text{Kn}=0.125$ (left) and $\text{Kn}=0.01$ (right). The temperature contours are obtained and compared by different methods: DSMC (contour in upper half), DIG (lines in upper half) and DIG-SST (contour in lower half).   }
    \label{fig:jetMa5}
\end{figure}

As discussed in the previous section, the cell size near the jet must be significantly smaller than that in the far field to satisfy the grid requirements of the DSMC method. When $\text{Ma}=5$ with $\text{Kn}=0.125$ and $\text{Kn}=0.01$ , given that the local mean free path is inversely proportional to the local number density, the cell size near the jet is approximately 40 times smaller than that in the far field according to Table \ref{tab:jetcondition}. As shown in Fig.~\ref{fig:DSMC_jetmesh}, to efficiently obtain the reference DSMC data, the Cartesian grids in SPARTA are adaptively refined and coarsened based on the local mean free path, and the grid size is strictly maintained below 1/3 of the local mean free path. Note that without the application of adaptive grid techniques, the computational cost for both cases would become prohibitively high. Additionally, the computational domain as well as the grid size applied in DIG and DIG-SST are the same as previous blunt body flow case.

\begin{figure}[!t]
    \centering
    \includegraphics[width=0.49\textwidth]{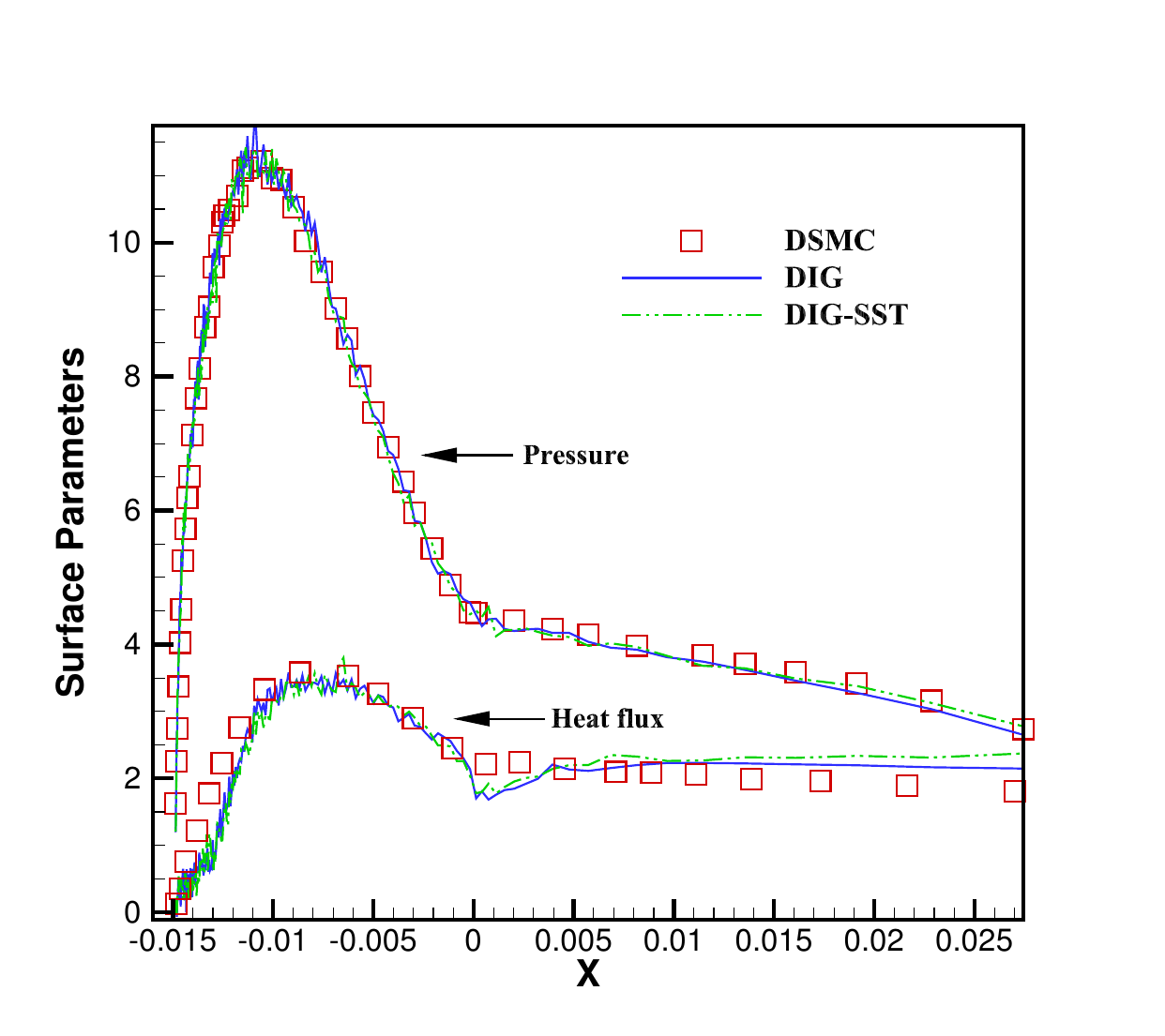}
    \includegraphics[width=0.49\textwidth]{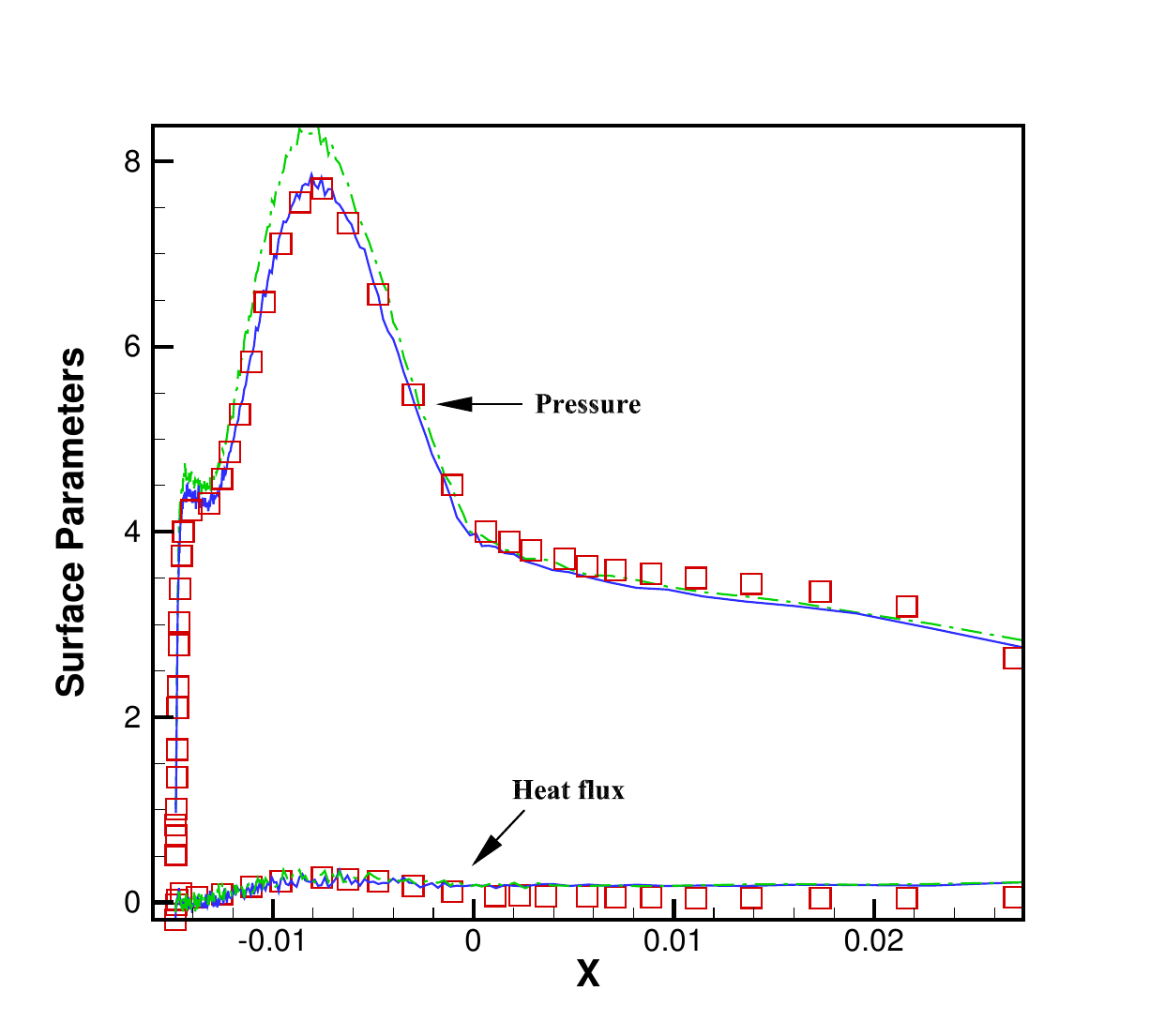}\\
    \caption{Surface properties predicted by different methods when $\text{Ma}=5$, $\text{Kn}=0.125$ (left) and $\text{Kn}=0.01$ (right).}
    \label{fig:Ma5surfacecompare}
\end{figure}
Figure \ref{fig:jetMa5} compares the macroscopic field obtained by the DSMC, DIG and DIG-SST methods. When $\text{Kn} = 0.125$, the Reynolds number is relatively low, the turbulent viscosity predicted by the $k$-$\omega$ SST turbulence model is negligible. Under this condition, the numerical results from the DIG-SST method closely match those of the DIG method, which itself recovers to the original DSMC method due to the relatively high Knudsen number. However, as the Knudsen number decreases to $\text{Kn} = 0.01$, the Reynolds number increases, and the turbulent viscosity predicted by the $k$-$\omega$ SST model becomes significant. In this case, the macroscopic field predicted by the DIG-SST method deviates notably from the DIG and DSMC results, while the DIG and DSMC methods remain in close agreement. The same phenomenon occurs in the prediction of the surface quantities. Figure \ref{fig:Ma5surfacecompare} illustrates the surface pressure and heat flux distributions for different Knudsen numbers. At $\text{Kn} = 0.125$, the predictions from all methods are in good agreement, primarily due to the dominant rarefaction effects. However, at $\text{Kn} = 0.01$, the influence of turbulence becomes significant, leading to slight deviations in surface pressure predicted by the DIG-SST method compared to the other two methods. As the surface heat flux in this case is relatively low, the differences in predictions by the DIG-SST method, while present, remain less pronounced.

\begin{figure}[!t]
    \centering
    \includegraphics[width=0.49\textwidth,trim=10pt 10pt 10pt 10pt,clip]{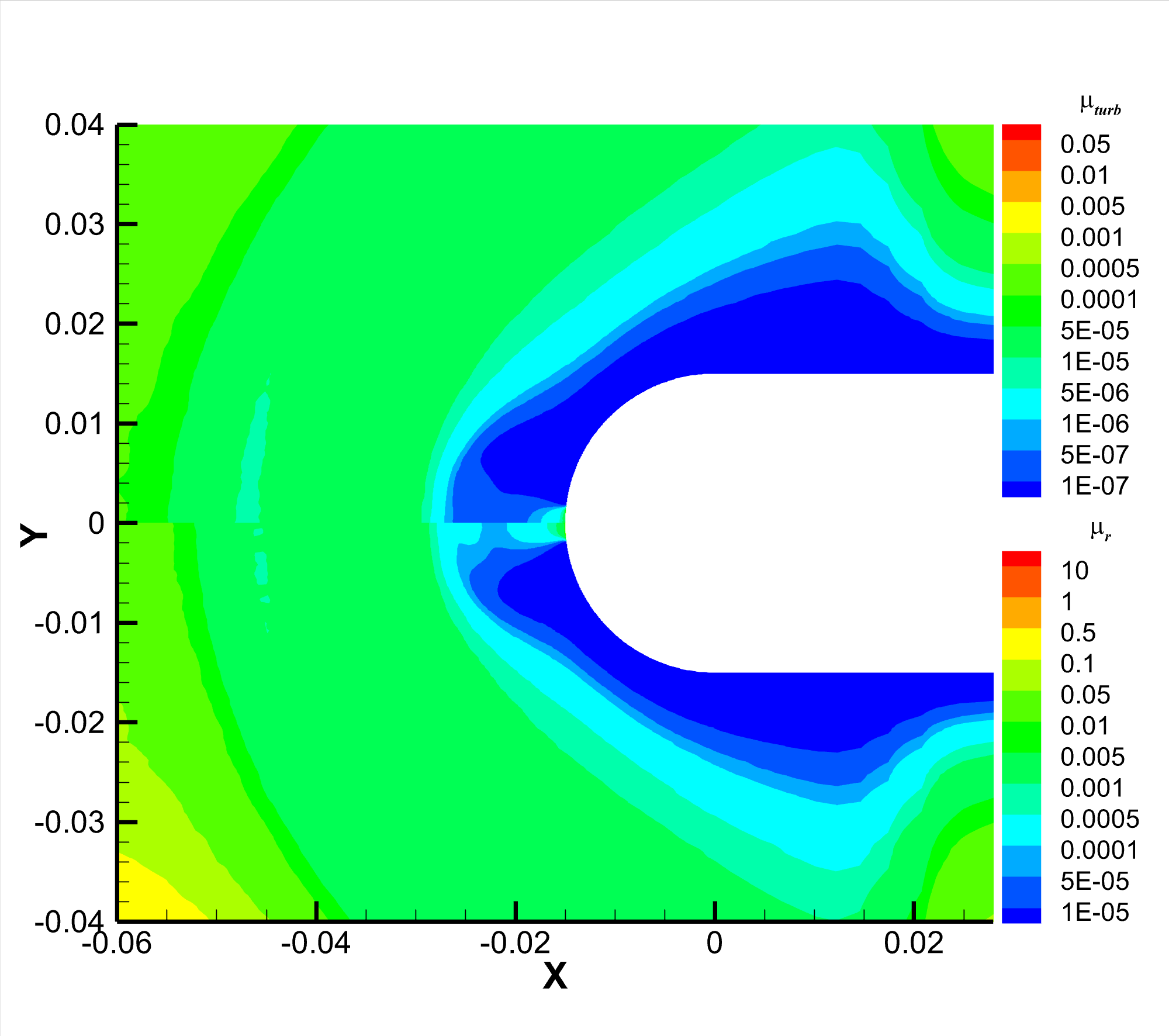}
    \includegraphics[width=0.49\textwidth,trim=10pt 10pt 10pt 10pt,clip]{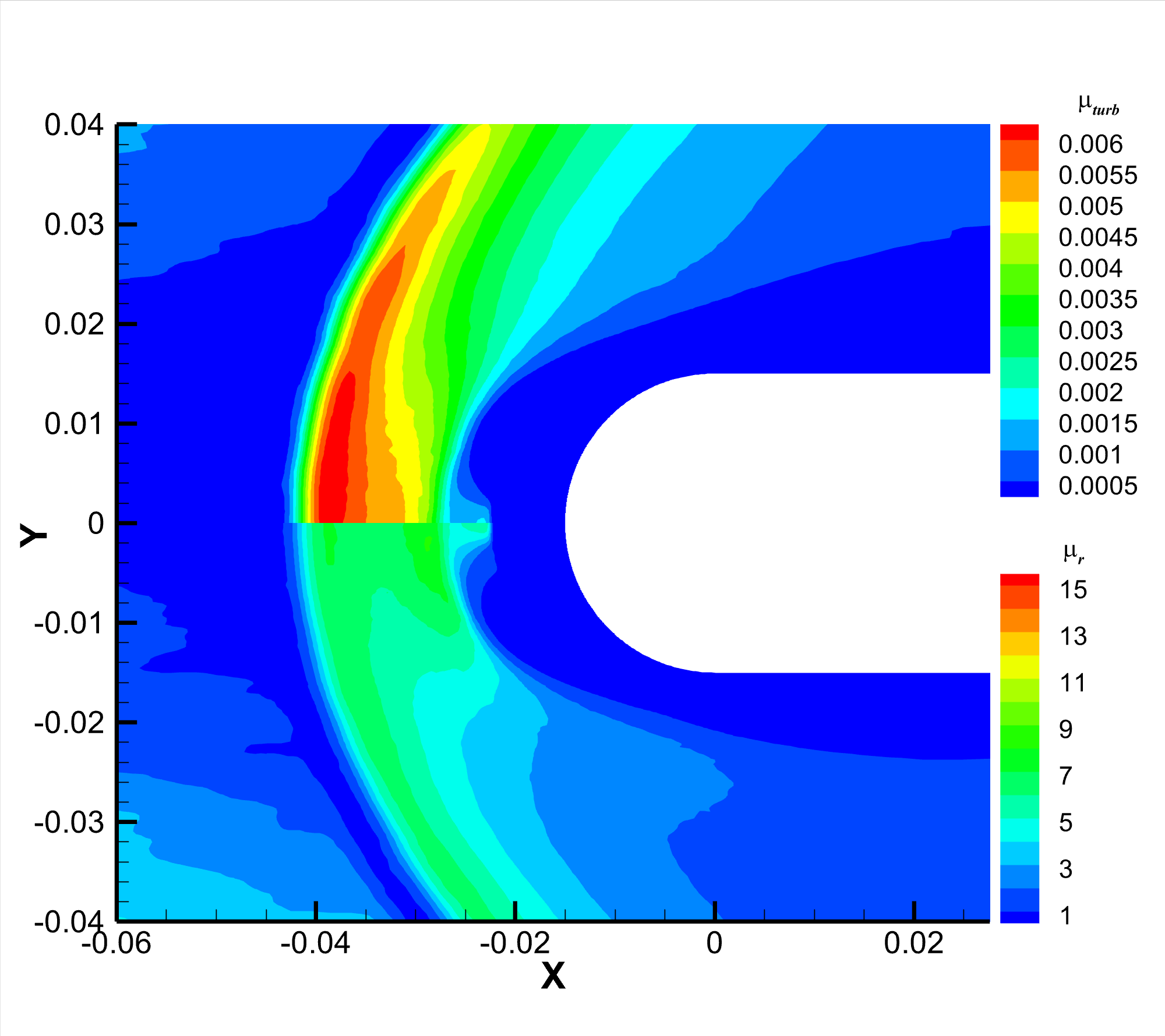}\\
    \caption{Contour of turbulent viscosity $\mu_{turb}$ (top half) and the ratio of the turbulent to laminar viscosity $\mu_{r}$ from DIG-SST method (lower half) when $\text{Ma}=5$. For $\text{Kn}=0.125$ (left), the turbulent effect is negligible, whereas for $\text{Kn}=0.01$ (right), the primary turbulence arises in the free stream and outer shock regions. }
    \label{fig:Ma5mutur}
\end{figure}
Figure $\ref{fig:Ma5mutur}$ shows the ratio between the turbulent and physical viscosities predicted by DIG-SST when $\text{Ma}=5$. At $\text{Kn}=0.125$,  $\mu_r$ in the shock and jet flow regions ranges from $10^{-4}$ to $10^{-5}$, indicating that the turbulent viscosity $\mu_{turb}$ is negligible compared to the laminar viscosity $\mu_{lam}$ in the rarefied regime. In contrast, at $\text{Kn}=0.01$, the distribution of $\mu_{turb}$ shows a pronounced concentration near the outer shock region, leading to noticeable turbulence effects on the shock position and macroscopic property fields.

\subsection{Jet flow in large Mach number environment}

% The Mach numbers of the free stream are increased to 15 and 30 to investigate the turbulent jet in rarefied environment with the Knudsen numbers remained unchanged. As shown in Table~\ref{tab:jetcondition}, the number density of the jet is very large, leading to the unacceptable computational cost for DSMC method. In this section, the results obtained by DIG and DIG-SST method are compared to investigate the coexist of turbulent and rarefied effects.
To examine the behavior of turbulent jets in rarefied environments, the free-stream Mach numbers are increased to 15, while the Knudsen numbers remain unchanged. As indicated in Table~\ref{tab:jetcondition}, the high number density of the jet results in prohibitive computational costs for the DSMC method. Thus in this section, only the results obtained using the DIG and DIG-SST methods are compared to analyze the interaction between turbulence and rarefaction effects.

\begin{figure}[p]
    \centering
    \hspace{-8mm}
    \includegraphics[width=0.38\textwidth,trim=10pt 10pt 10pt 10pt,clip]{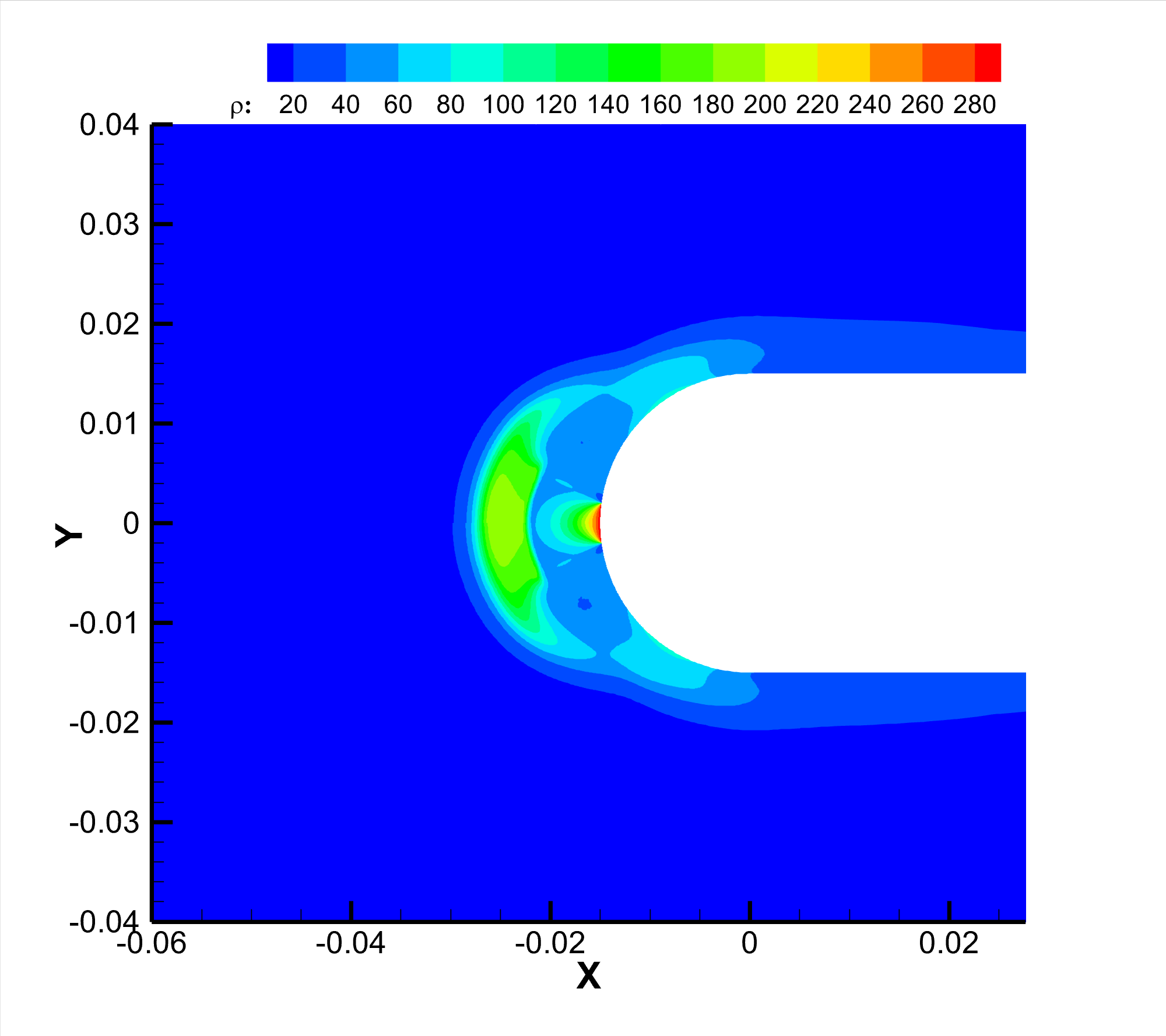}
    \hspace{-11mm}
    \includegraphics[width=0.38\textwidth,trim=10pt 10pt 10pt 10pt,clip]{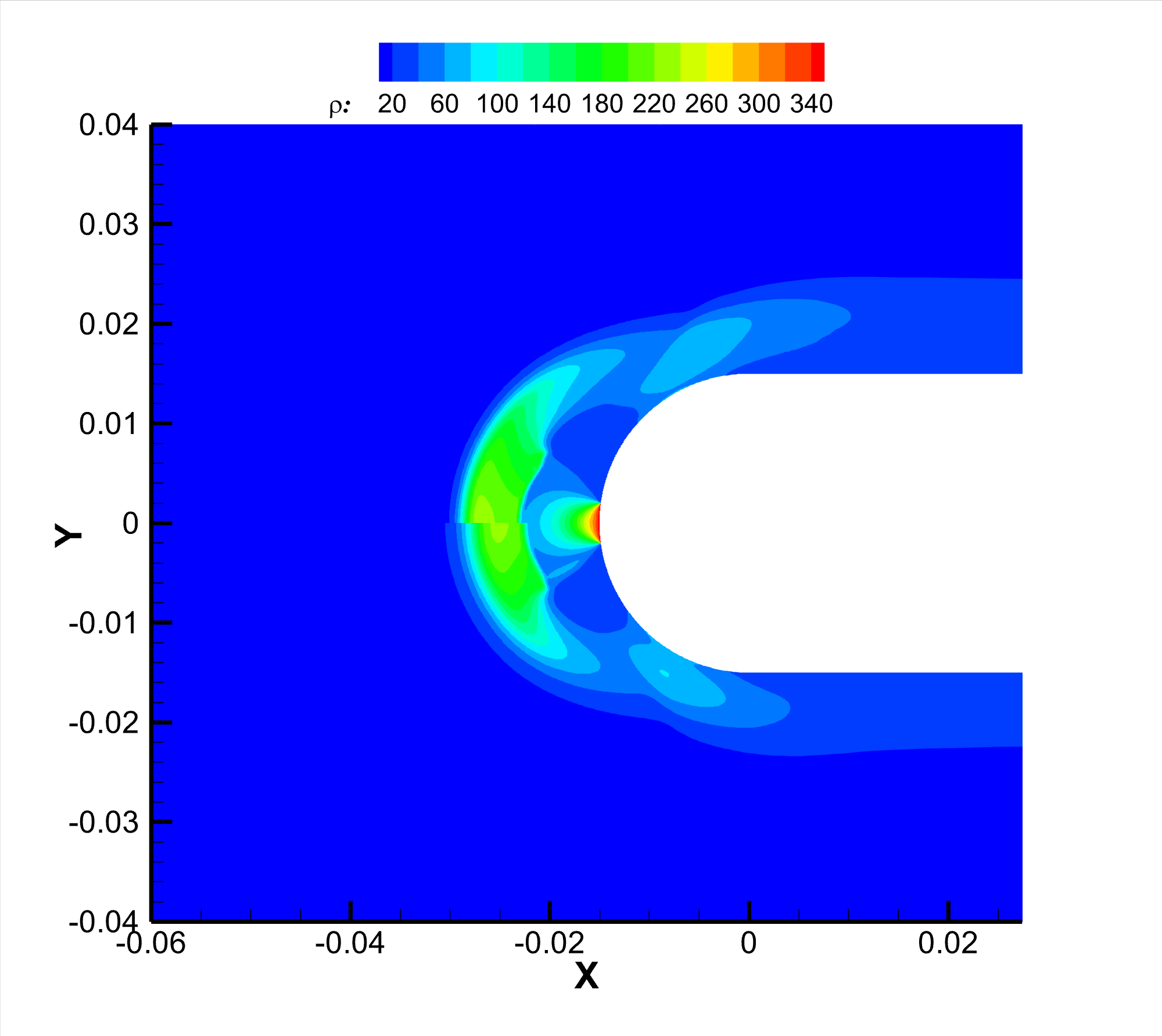}
    \hspace{-11mm}
    \includegraphics[width=0.38\textwidth]{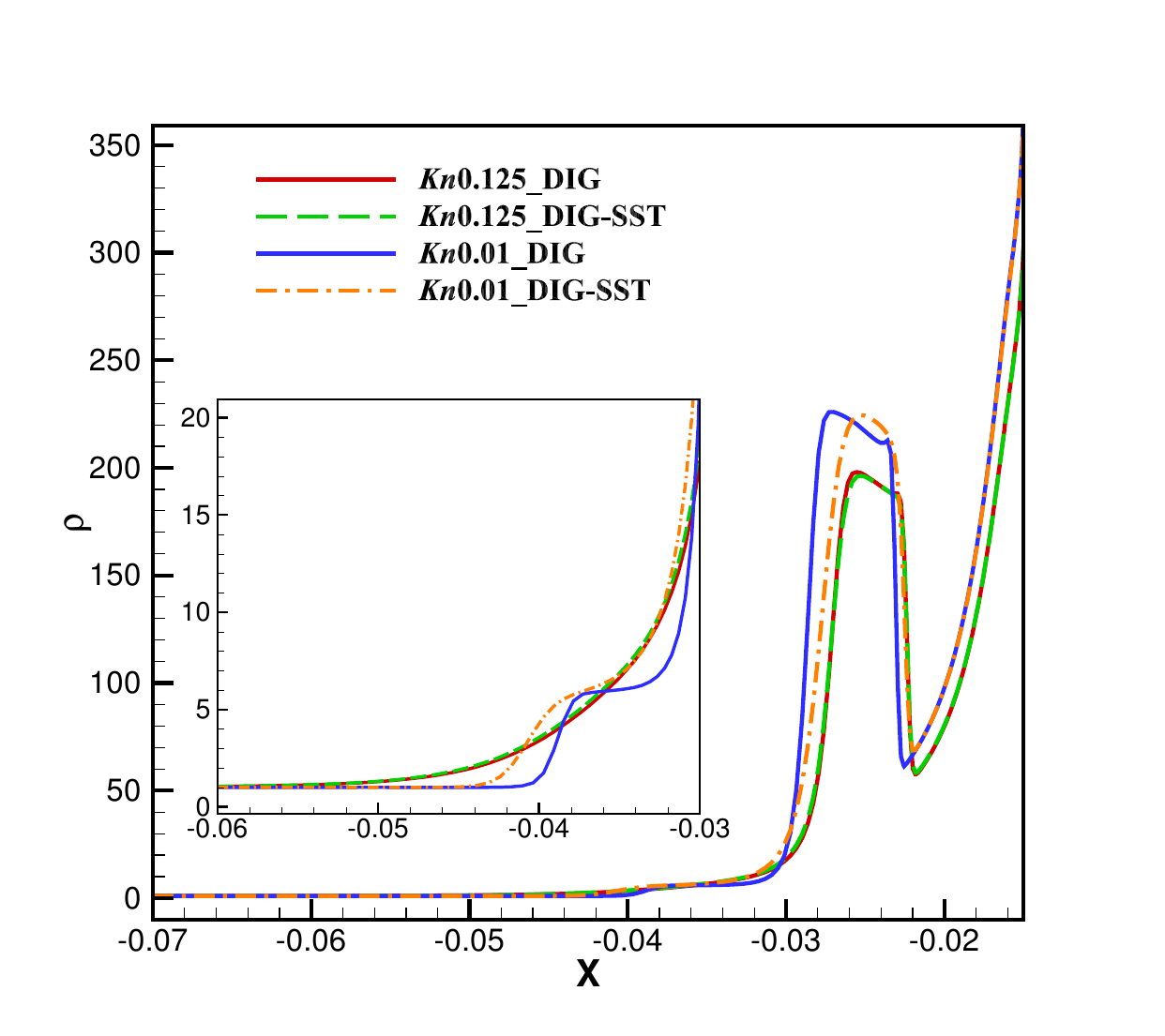}	\\
    \vspace{-2.4mm}
     \hspace{-8mm}
    \includegraphics[width=0.38\textwidth,trim=10pt 10pt 10pt 10pt,clip]{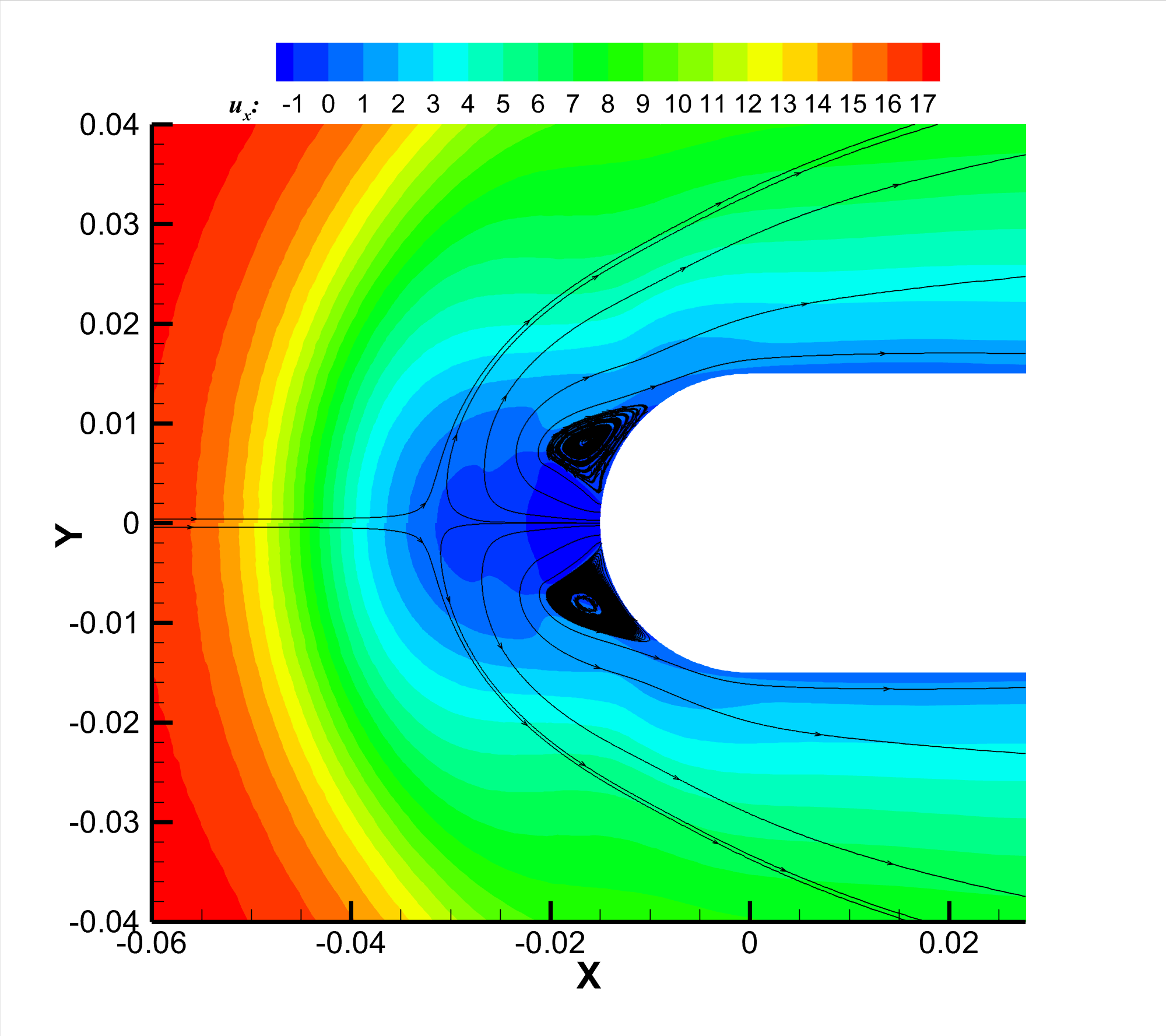}
    \hspace{-11mm}
    \includegraphics[width=0.38\textwidth,trim=10pt 10pt 10pt 10pt,clip]{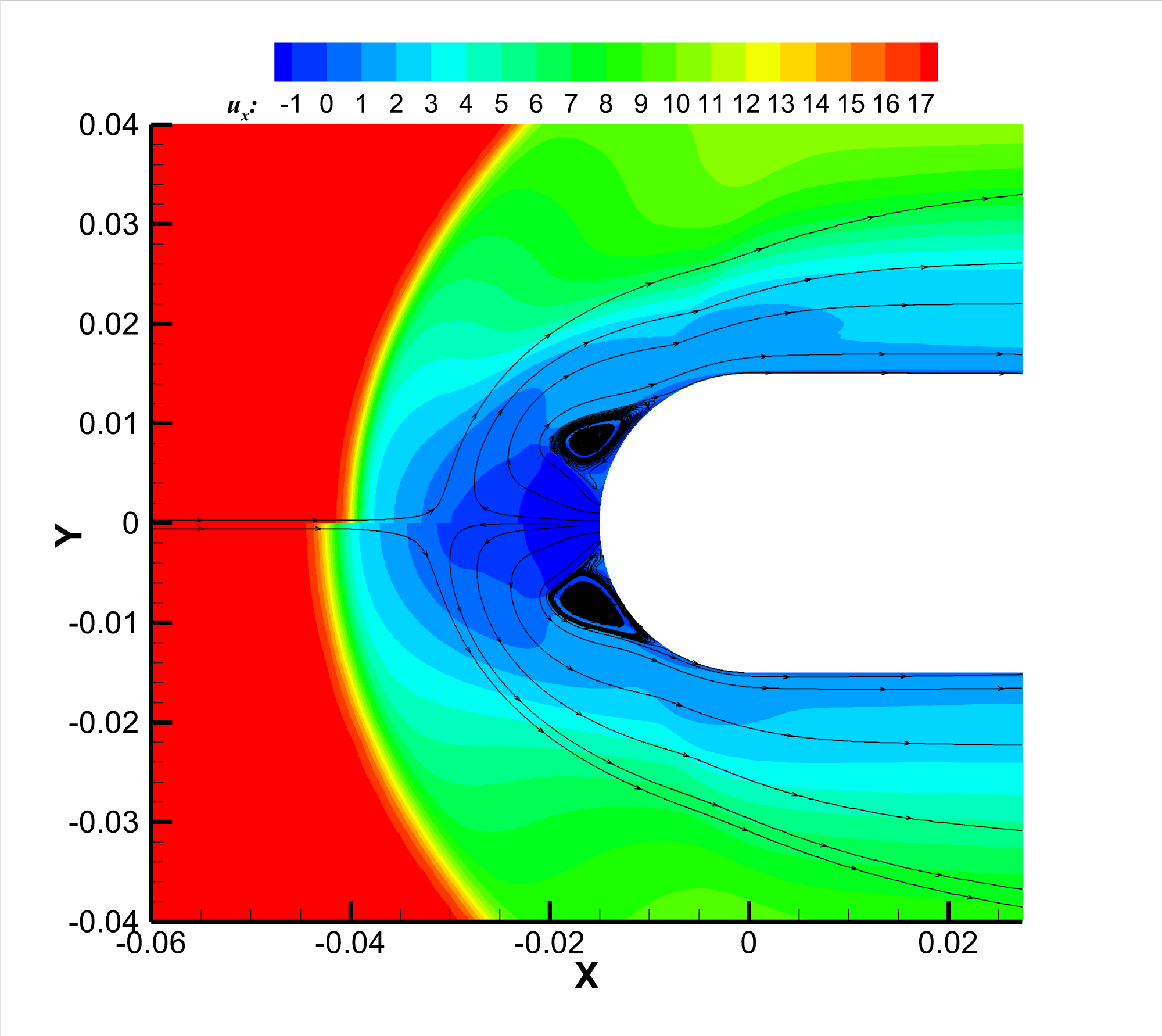}
    \hspace{-11mm}
    \includegraphics[width=0.38\textwidth]{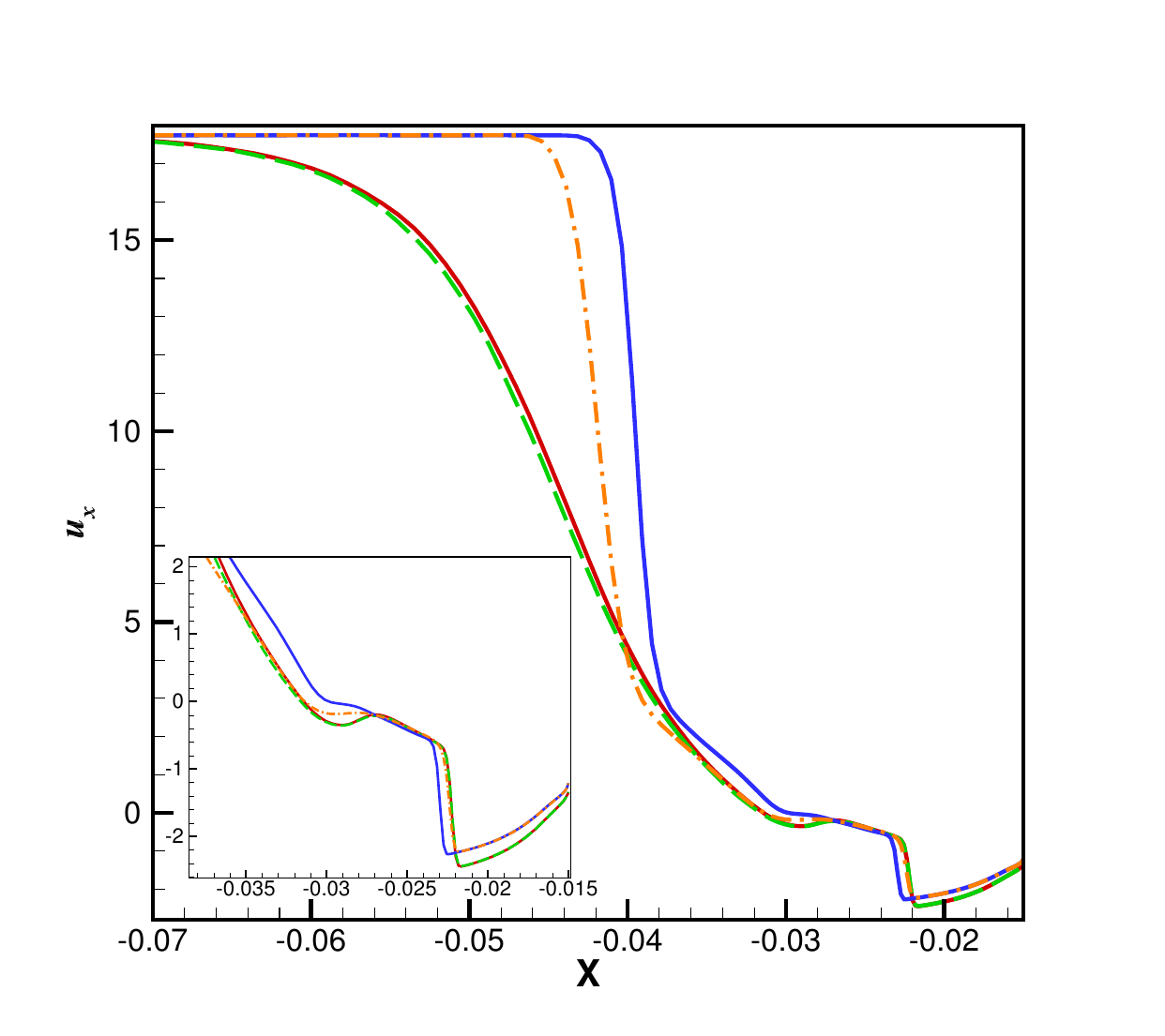}   \\
    \vspace{-2.4mm}
     \hspace{-8mm}
    \includegraphics[width=0.38\textwidth,trim=10pt 10pt 10pt 10pt,clip]{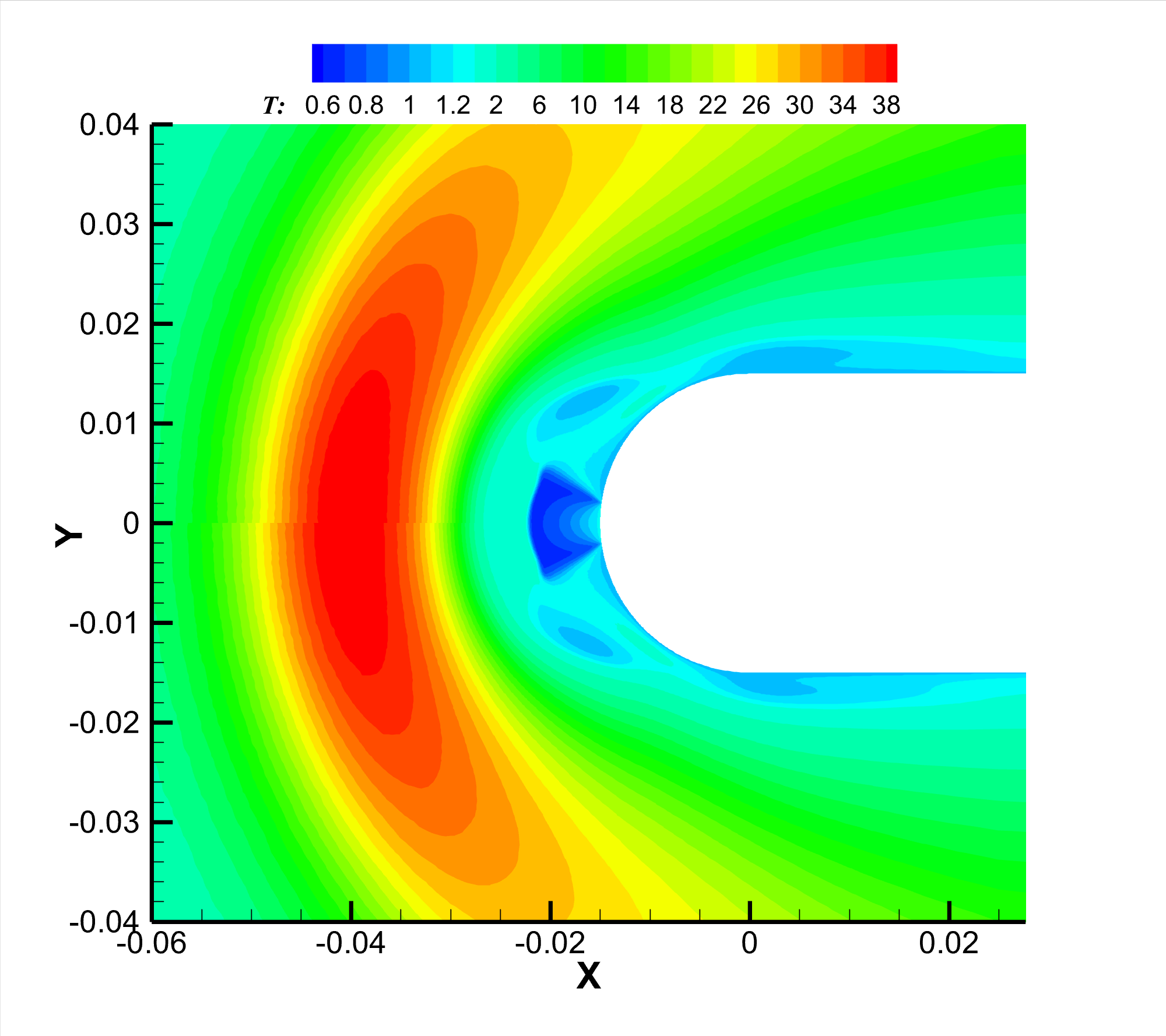}
    \hspace{-11mm}
    \includegraphics[width=0.38\textwidth,trim=10pt 10pt 10pt 10pt,clip]{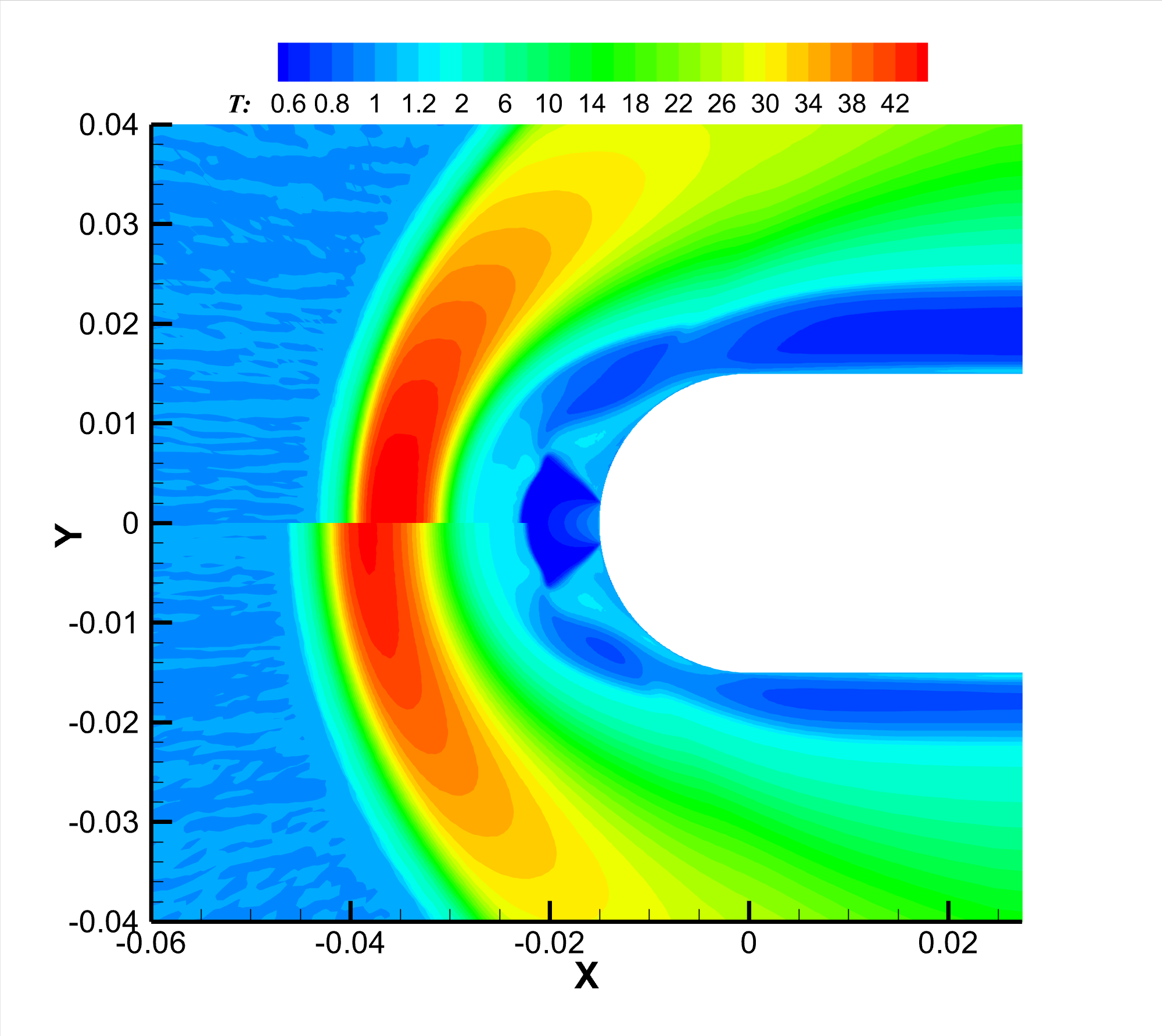}
    \hspace{-11mm}
    \includegraphics[width=0.38\textwidth]{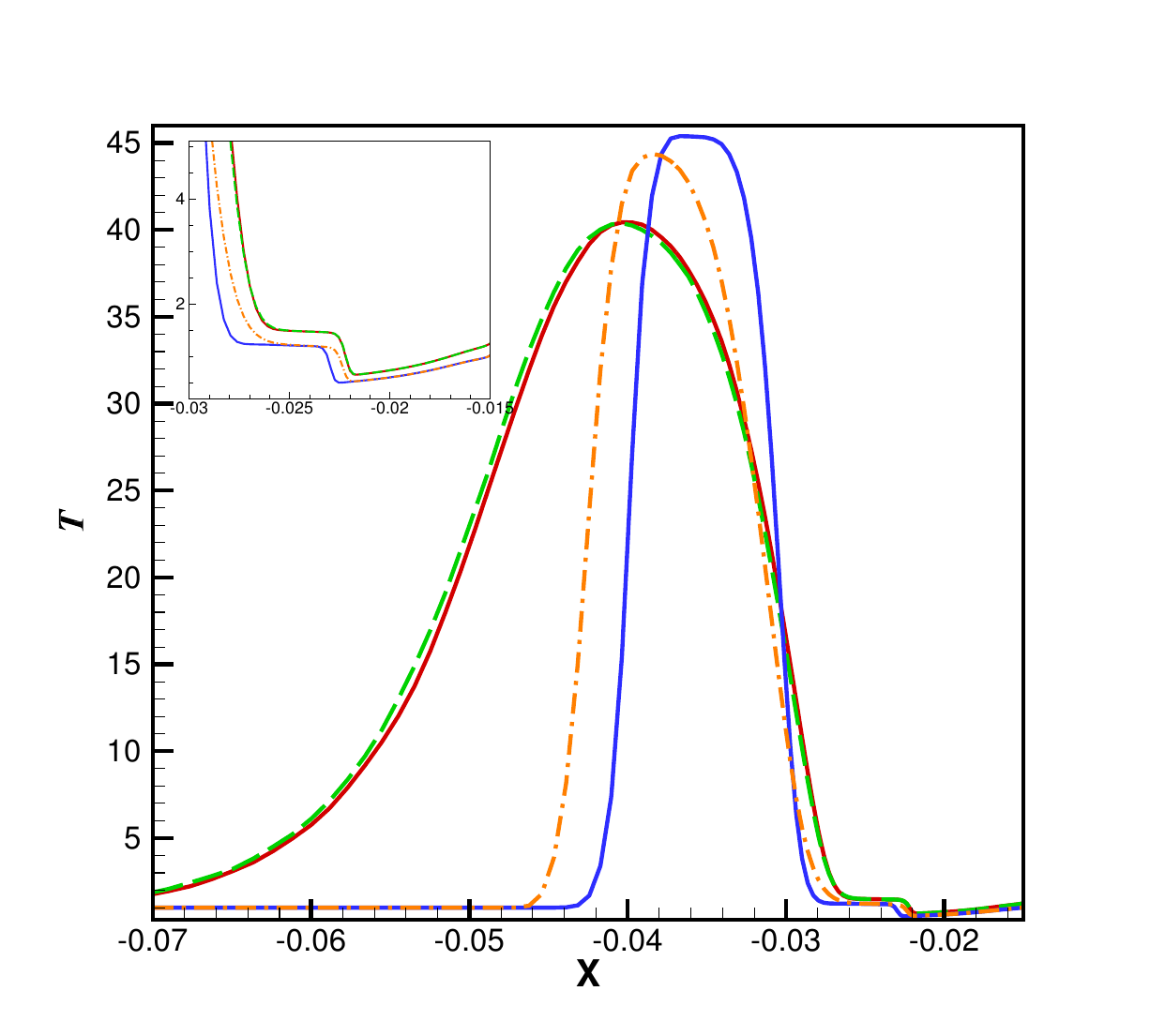}  
    \vspace{-8mm}
    \caption{Contours of macroscopic properties for $\text{Ma}=15$ at $\text{Kn}=0.125$ (left column) and $\text{Kn}=0.01$ (middle column) for $P_{\text{ratio}}=2.5$, with results from the DIG method shown in the upper half and DIG-SST in the lower half of each contour. (right column) the corresponding profiles extracted along the horizontal central line. Inset: zoomed-in figures to show the difference in shock and jet region.}
    \label{fig:Ma15macroscopicproperties}
\end{figure}

\begin{figure}[p]
    \centering
    \hspace{-8mm}
    \includegraphics[width=0.38\textwidth,trim=10pt 10pt 10pt 10pt,clip]{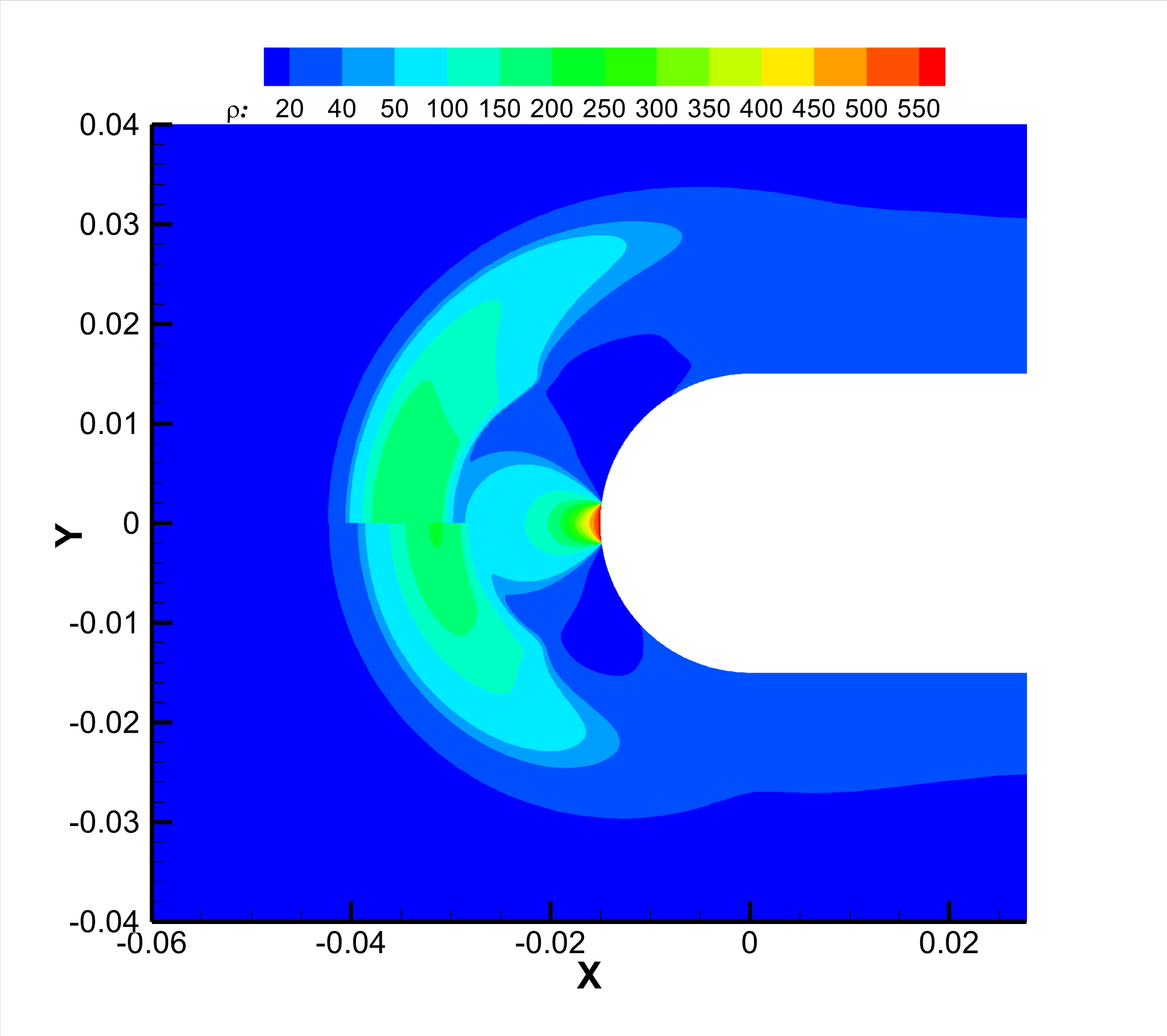}
    \hspace{-11mm}
    \includegraphics[width=0.38\textwidth,trim=10pt 10pt 10pt 10pt,clip]{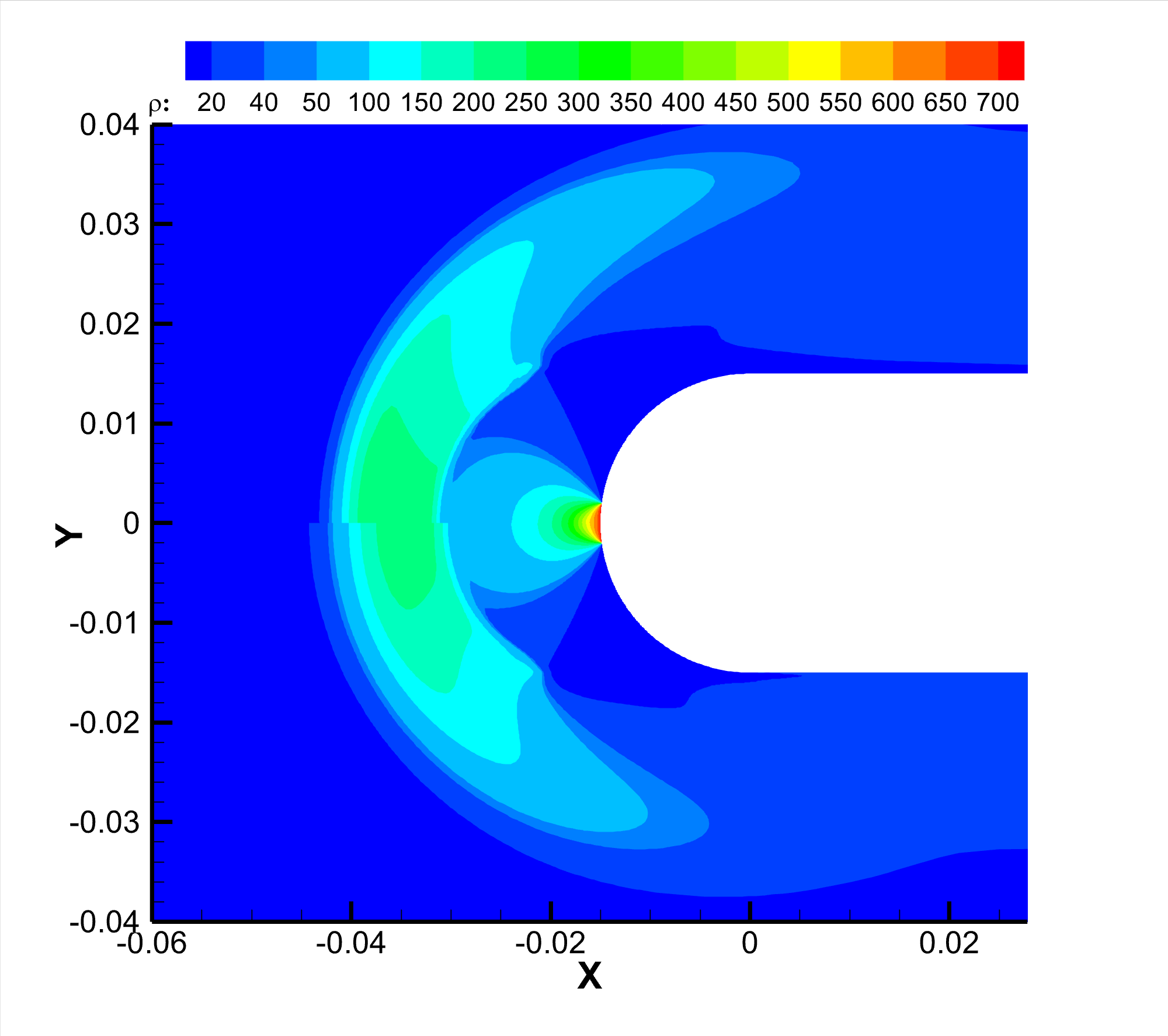}
    \hspace{-11mm}
    \includegraphics[width=0.38\textwidth]{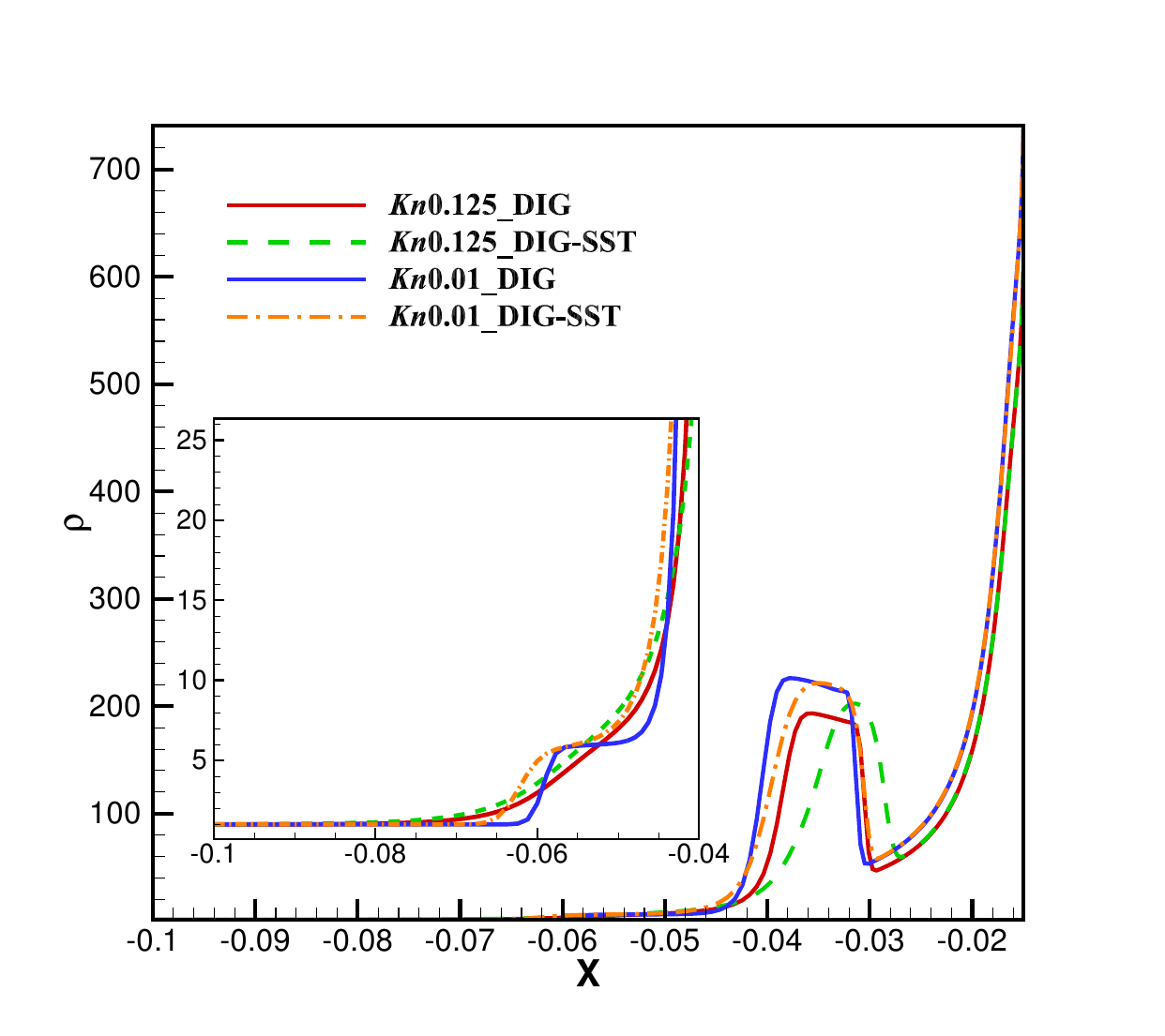}	\\
    \vspace{-1mm}
     \hspace{-8mm}
    \includegraphics[width=0.38\textwidth,trim=10pt 20pt 10pt 50pt,clip]{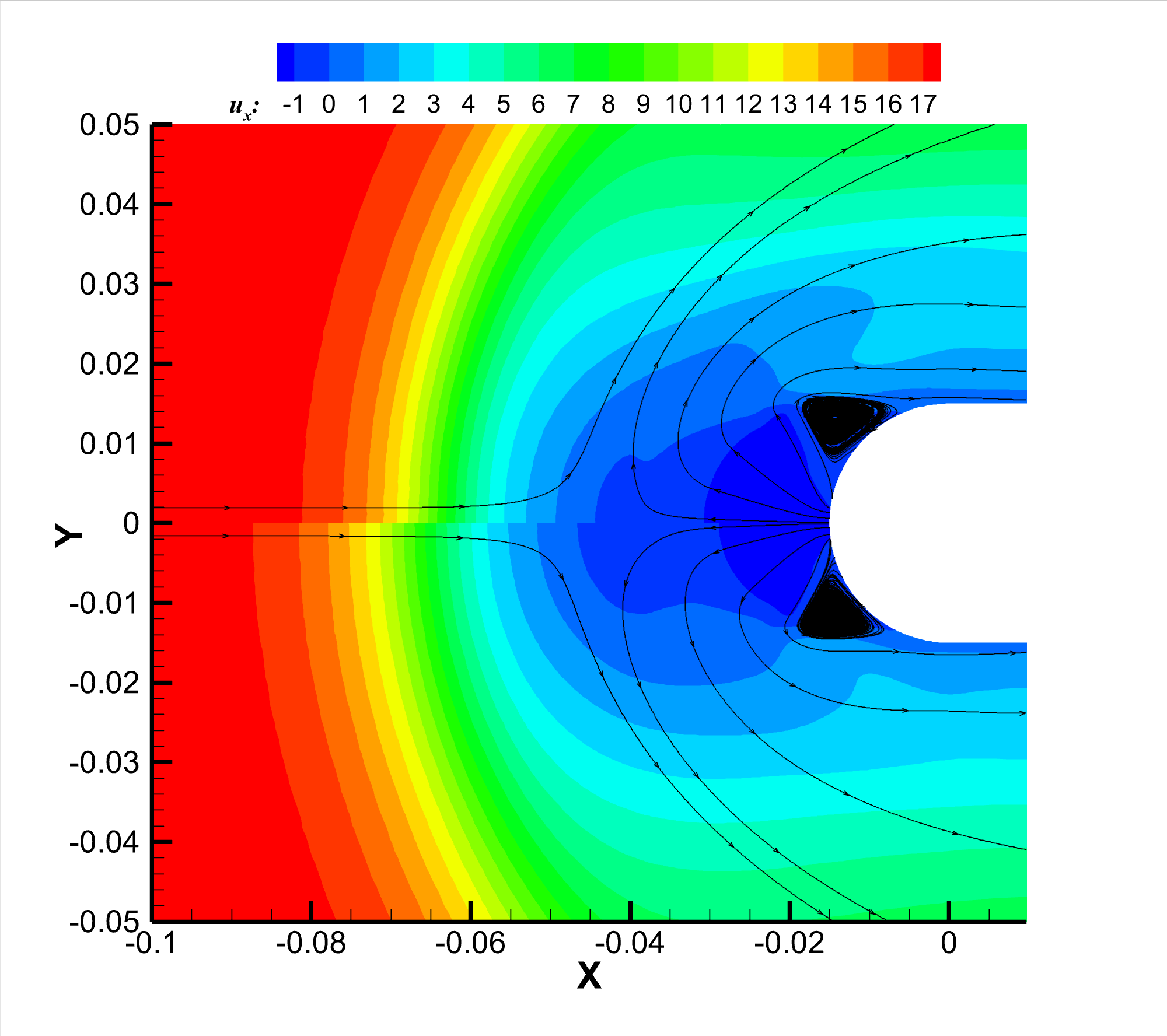}
    \hspace{-11mm}
    \includegraphics[width=0.38\textwidth,trim=10pt 20pt 10pt 50pt,clip]{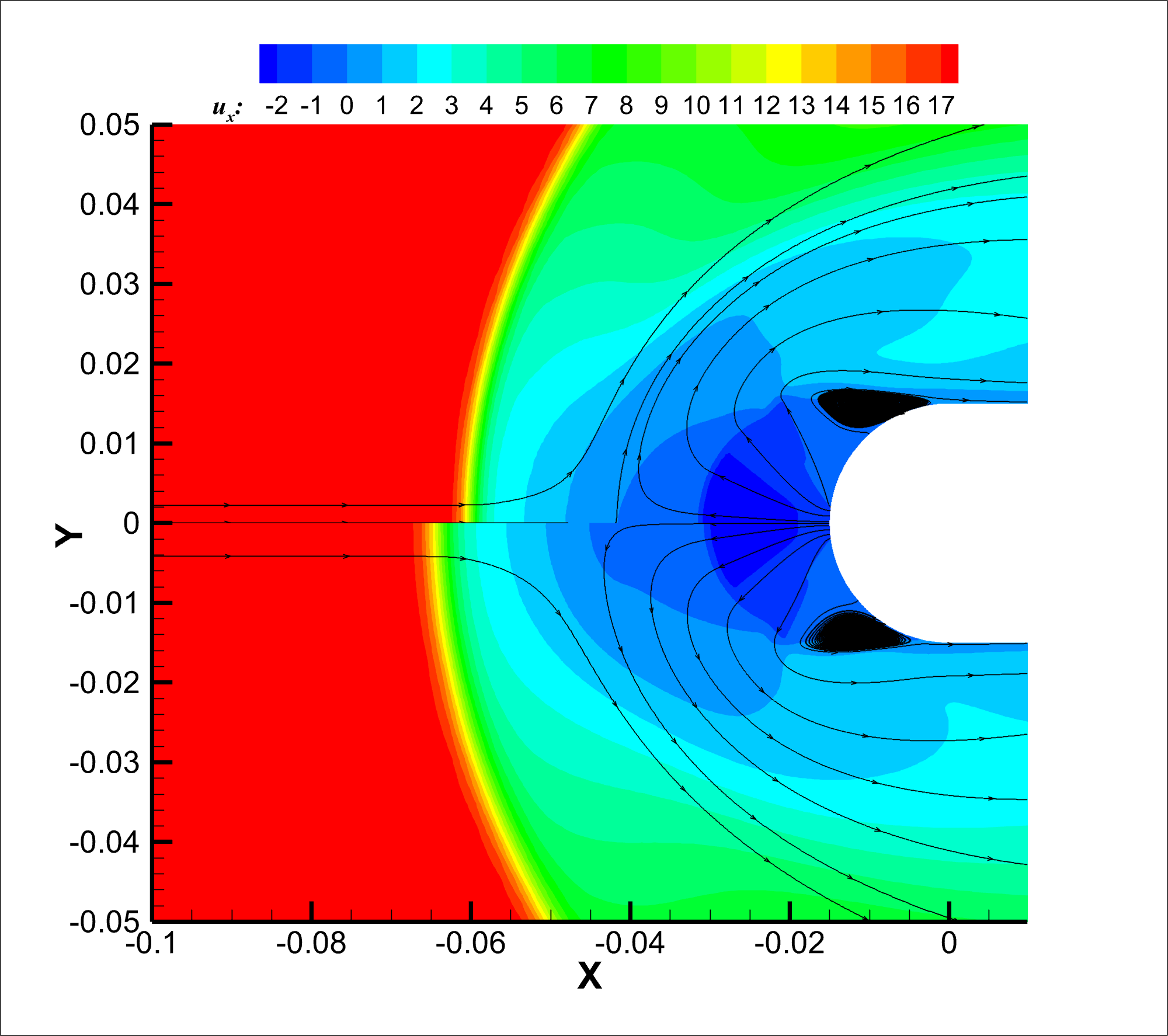}
    \hspace{-11mm}
    \includegraphics[width=0.38\textwidth]{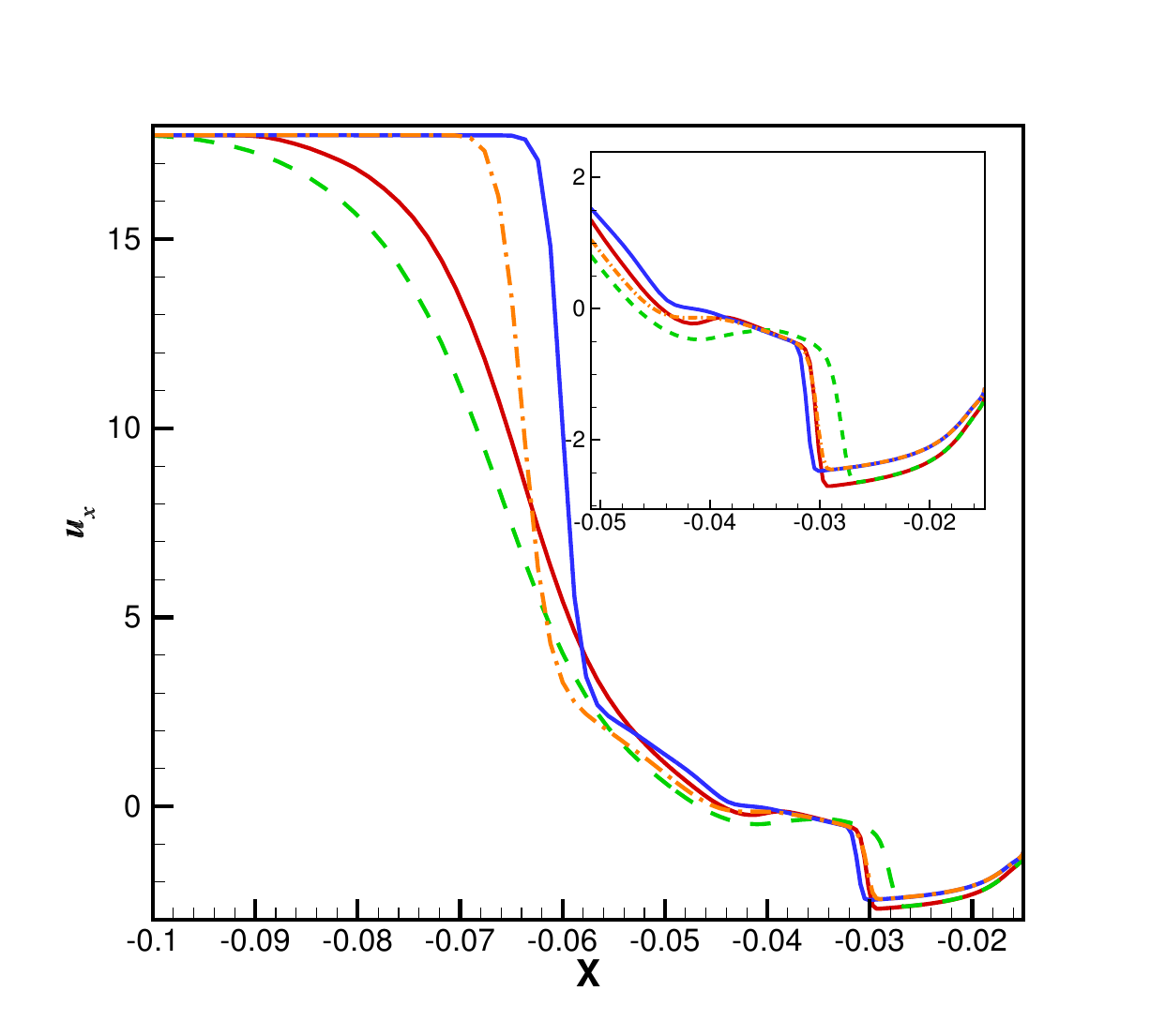}   \\
    \vspace{-2mm}
     \hspace{-8mm}
    \includegraphics[width=0.38\textwidth,trim=10pt 20pt 10pt 50pt,clip]{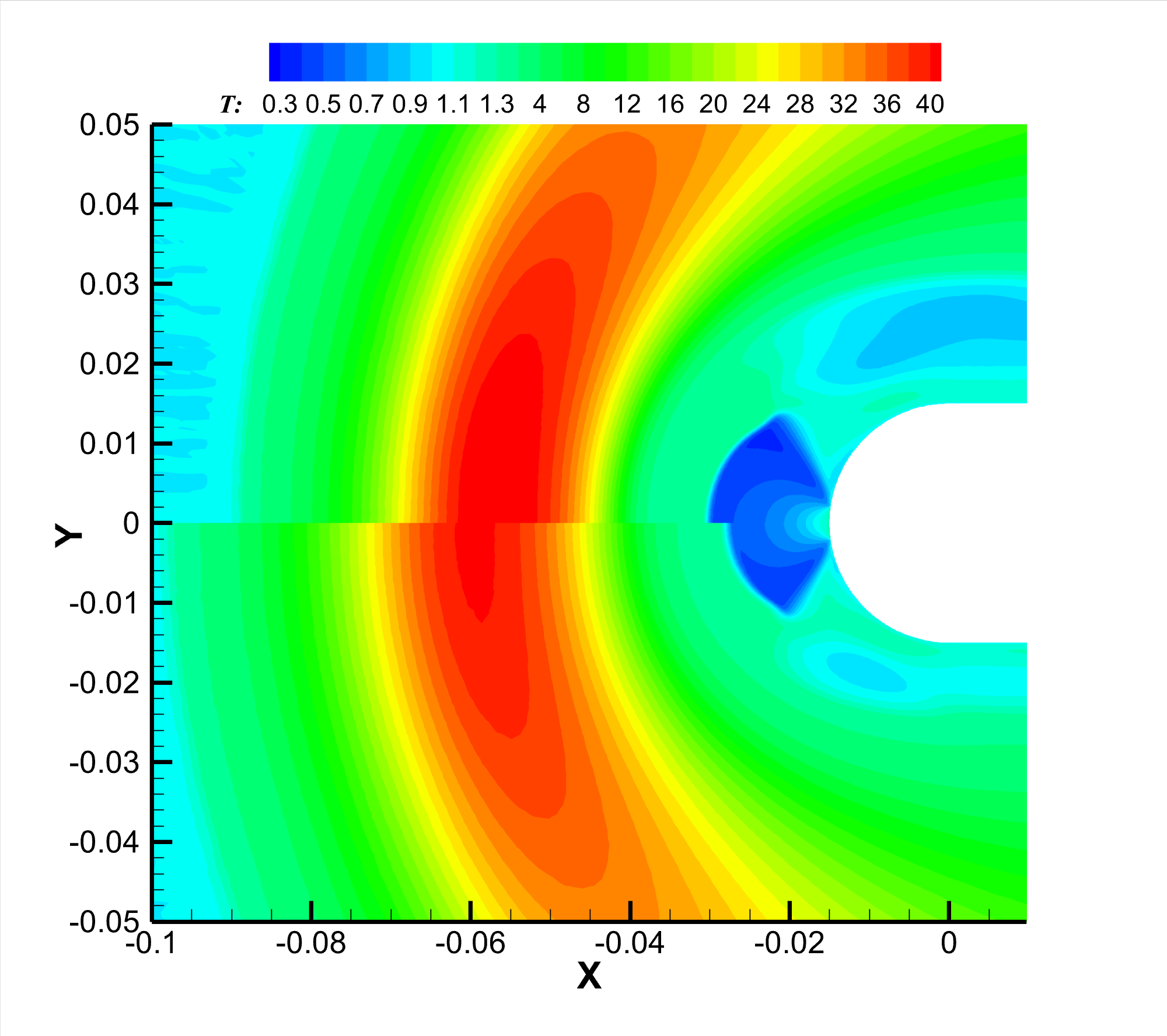}
    \hspace{-11mm}
    \includegraphics[width=0.38\textwidth,trim=10pt 20pt 10pt 50pt,clip]{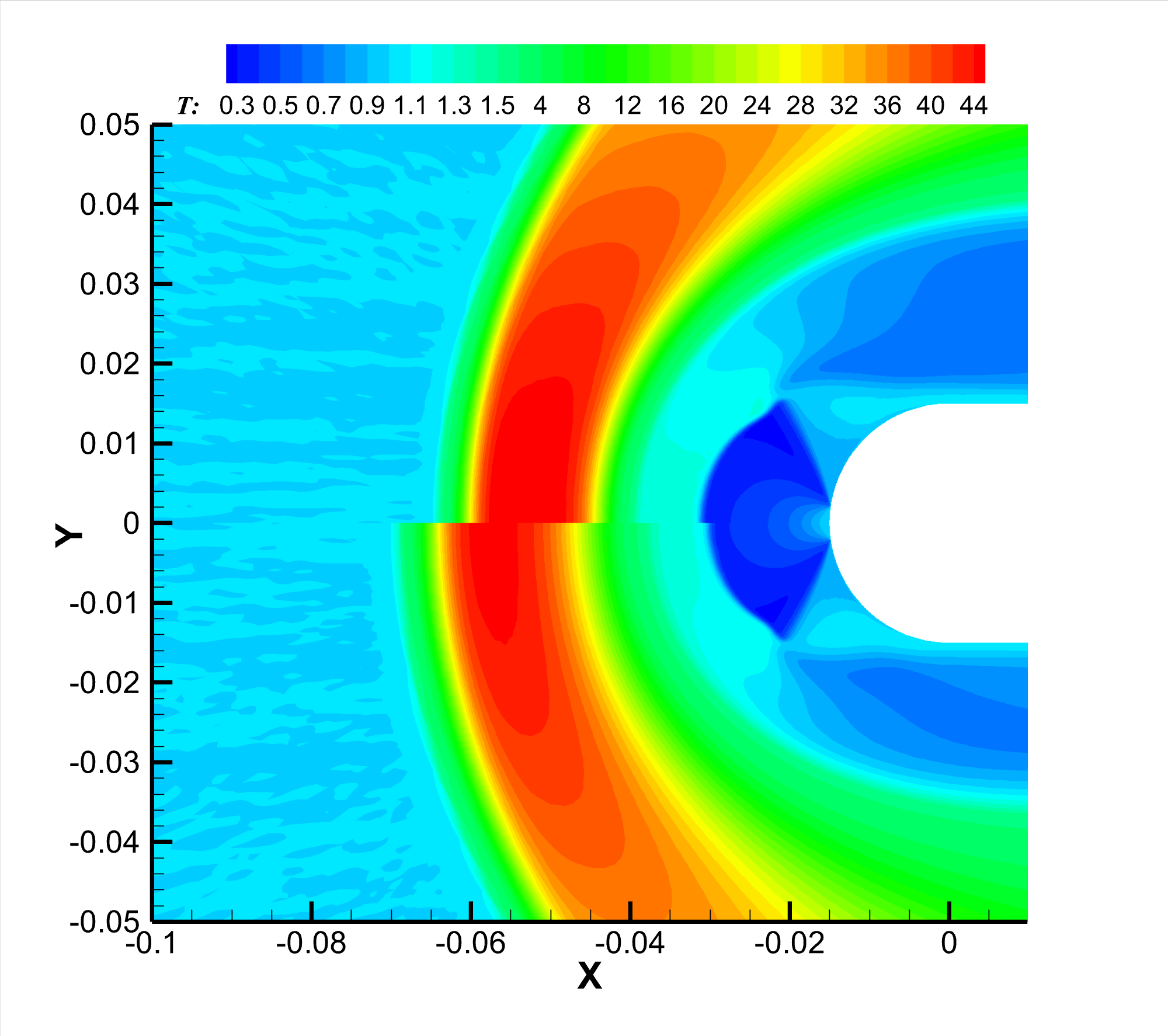}
    \hspace{-11mm}
    \includegraphics[width=0.38\textwidth]{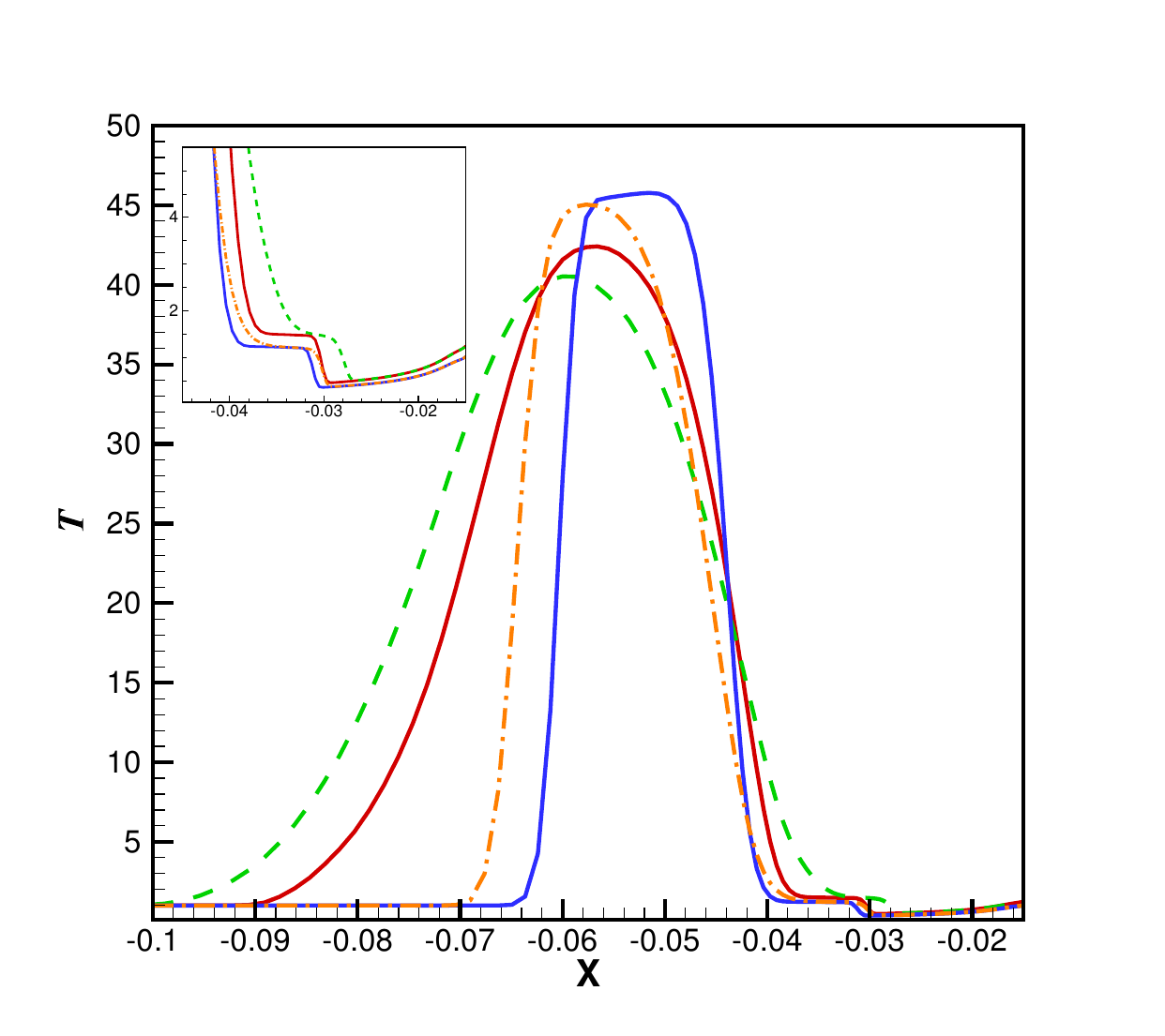}  
    \vspace{-6mm}
    \caption{Contours of macroscopic properties for $\text{Ma}=15$ at $\text{Kn}=0.125$ (left column) and $\text{Kn}=0.01$ (middle column) for $P_{\text{ratio}}$=5, with results from the DIG method shown in the upper half and DIG-SST in the lower half of each contour. (right column) the corresponding profiles extracted along the horizontal central line. Inset: zoomed-in figures to show the difference in shock and jet region.}
    \label{fig:Ma15PR5macroscopicproperties}
\end{figure}

Figures~\ref{fig:Ma15macroscopicproperties} presents a comparison of the macroscopic properties field obtained by DIG and DIG-SST methods. As the jet propagates into the flow field, it undergoes rapid expansion and acceleration. This expansion continues until the jet interacts with the free stream, leading to the formation of a distinct Mach disk. The jet flow is compressed and decelerated by the Mach disk, and then adjusts to a state of local equilibrium with the post-shock free stream. This equilibrium results in a lateral deflection of the jet, accompanied by a reverse flow component that reattaches to the surface of the model, completing the flow structure. For $P_\text{ratio}=2.5$, the jet flow is fully turbulent when $\text{Ma}=15$ and $\text{Kn}=0.01$. As shown in Fig.~\ref{fig:Ma15macroscopicproperties}, the Mach disk predicted by DIG-SST shifts slightly closer to the jet region compared to that from DIG. Additionally, the outer shock in the DIG-SST results moves marginally towards the free-stream region compared to the DIG results.  As the Knudsen number increases to 0.125, the enhanced rarefaction effect weakens the outer shock, causing it to become thicker. And in the jet-controlled region, the shape of the Mach disk becomes flatter, as the relatively low Reynolds number of the jet flow diminishes the turbulent effects. Consequently, the macroscopic properties obtained from both the DIG and DIG-SST methods are nearly identical. 

As shown in Fig.~\ref{fig:Ma15PR5macroscopicproperties}, when the pressure ratio $P_\text{ratio}$ increases to 5, compared to $P_\text{ratio}=2.5$,  due to the larger expansion of the jet flow, the Mach disk as well as the outer shock moves towards the free-stream region, while the reattachment point for velocity along the model surface moves towards the outflow region. Since the jet flow is totally turbulent for both $\text{Kn}=0.125$ and $\text{Kn}=0.01$, larger difference occurs for macroscopic properties field obtained by DIG and DIG-SST methods. Moreover, the macroscopic properties extracted along the horizontal centerline are compared in the right column of Fig.~\ref{fig:Ma15PR5macroscopicproperties}. Similarly, the results reveal that coupling with the $k$-$\omega$ SST turbulence model causes the macroscopic properties to change earlier compared to the original DIG method. For instance, compared to the profile obtained using DIG, the temperature profile from DIG-SST starts to rise earlier in the upstream region. Thus, these phenomena demonstrate that the turbulent effects become more pronounced when $P_\text{ratio}=5$, regardless of whether $\text{Kn}=0.125$ or $\text{Kn}=0.01$.

% When $\text{Ma}=15$ and $\text{Kn}=0.01$, where the jet flow is fully turbulent and $P_\text{ratio}=2.5$, the Mach disk predicted by DIG-SST shifts slightly closer to the jet exit compared to that from DIG. Additionally, the outer shock in the DIG-SST results moves marginally towards the free-stream region compared to the DIG results.  As the Knudsen number increases to 0.125, the enhanced rarefaction effect weakens the outer shock, causing it to become thicker. And in the jet-controlled region, the shape of the Mach disk becomes flatter, as the relatively low Reynolds number of the jet flow diminishes the turbulent effects. Consequently, the macroscopic properties obtained from both the DIG and DIG-SST methods are nearly identical. Compared to $P_\text{ratio}=2.5$, when pressure ratio $P_\text{ratio}$ increases to 5, the Mach disk as well as the outer shock moves towards the free-stream region, while the reattachment point for vel

\begin{figure}[!t]
    \centering
    \subfigure[$\text{Kn}=0.125$, $P_{\text{ratio}}=2.5$]{\includegraphics[width=0.49\textwidth,trim=10pt 60pt 10pt 150pt,clip]{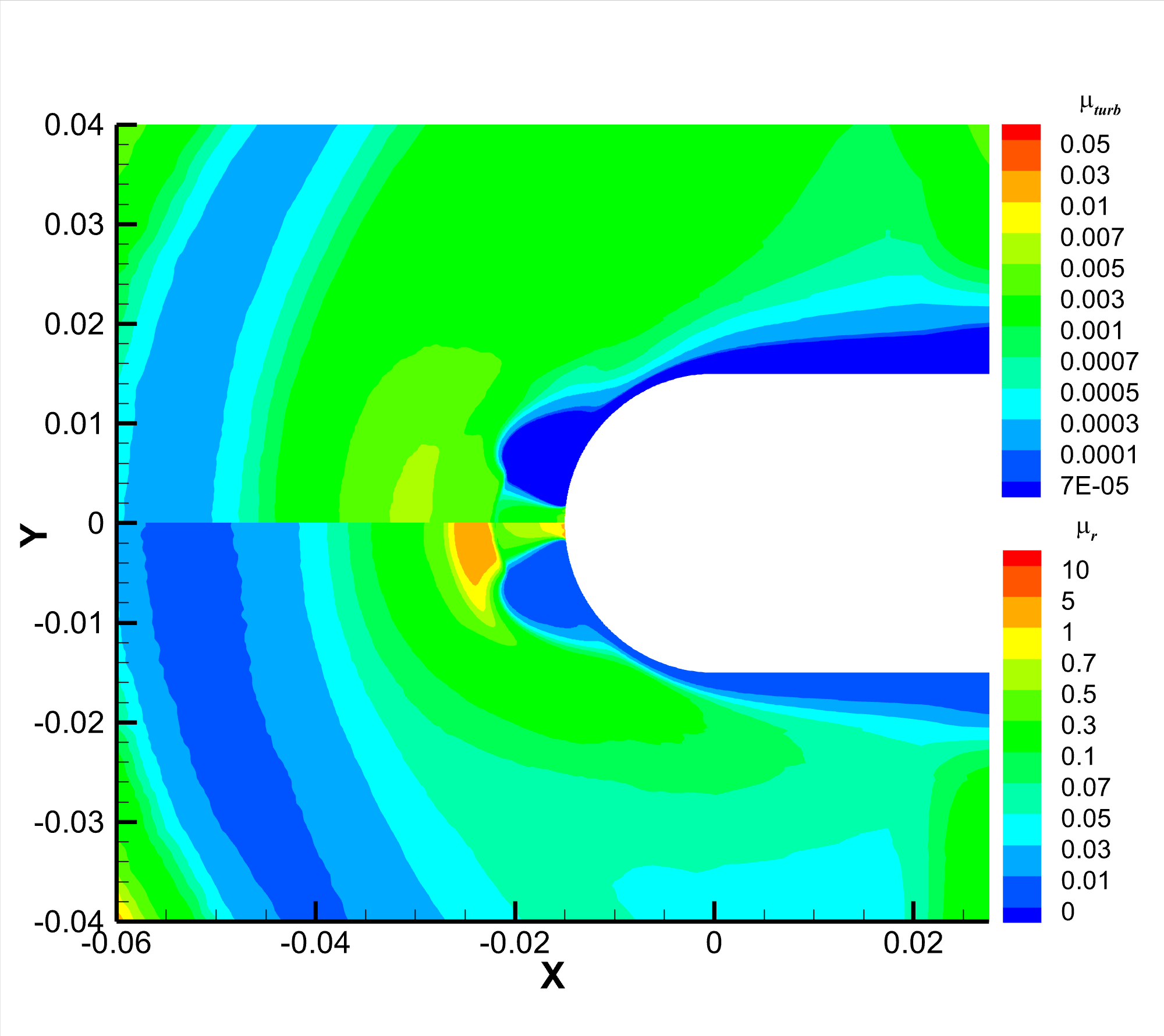}}
    \subfigure[$\text{Kn}=0.01$, $P_{\text{ratio}}=2.5$]{\includegraphics[width=0.49\textwidth,trim=10pt 60pt 10pt 150pt,clip]{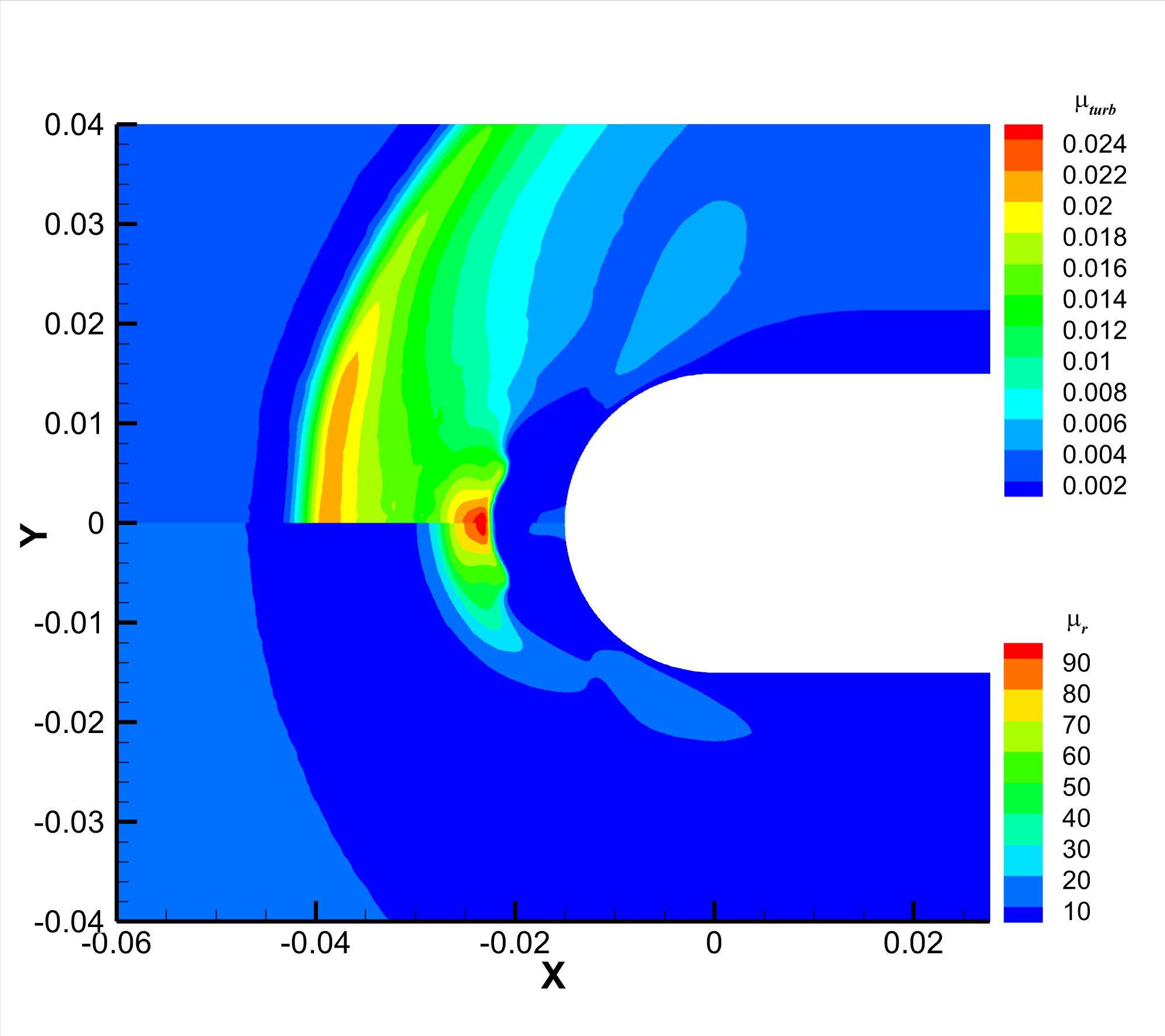}}\\
    \vspace{-3mm}
    \subfigure[$\text{Kn}=0.125$, $P_{\text{ratio}}=5$]{\includegraphics[width=0.49\textwidth,trim=10pt 60pt 10pt 150pt,clip]{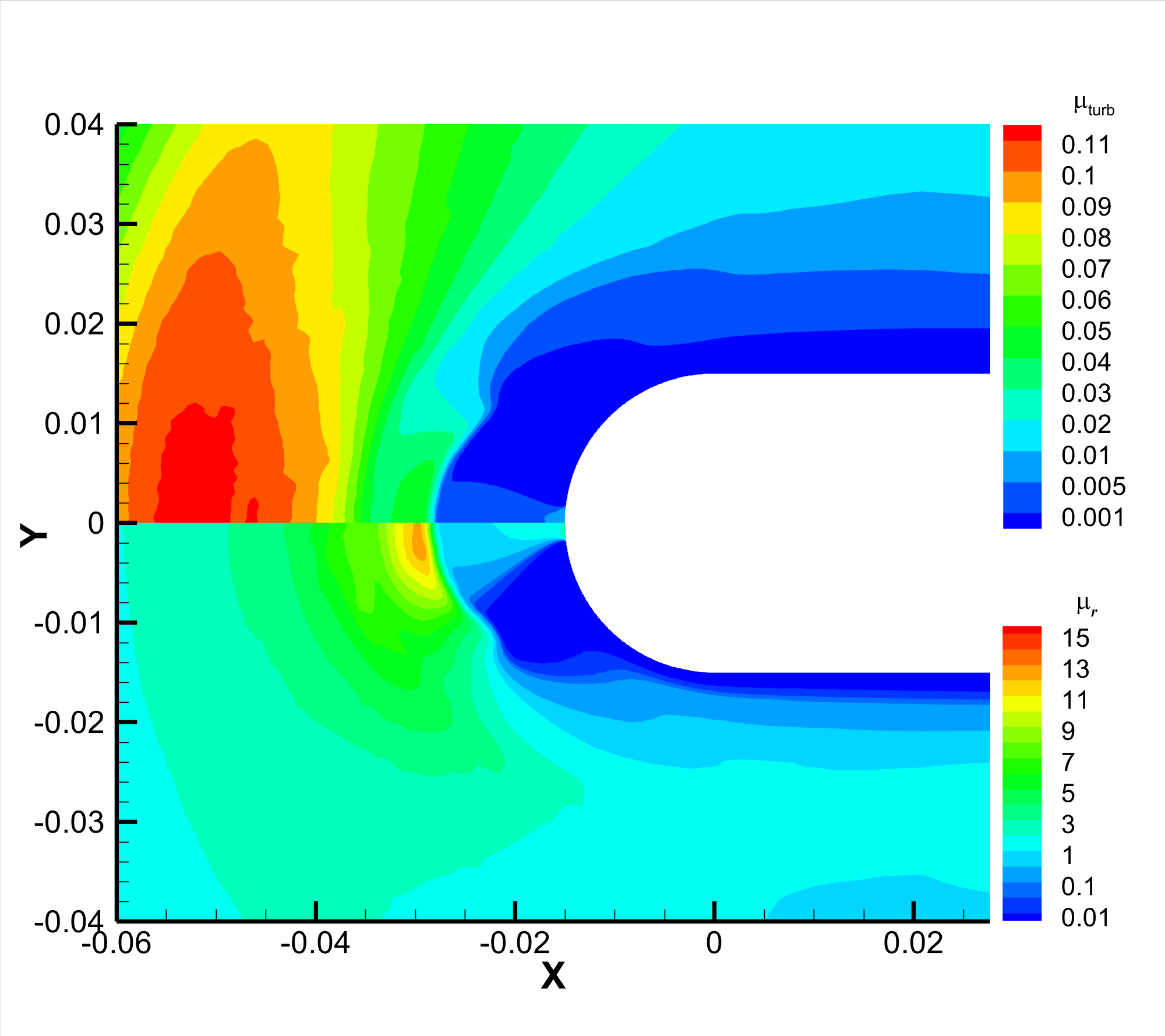}}
    \subfigure[$\text{Kn}=0.01$, $P_{\text{ratio}}=5$]{\includegraphics[width=0.49\textwidth,trim=10pt 60pt 10pt 150pt,clip]{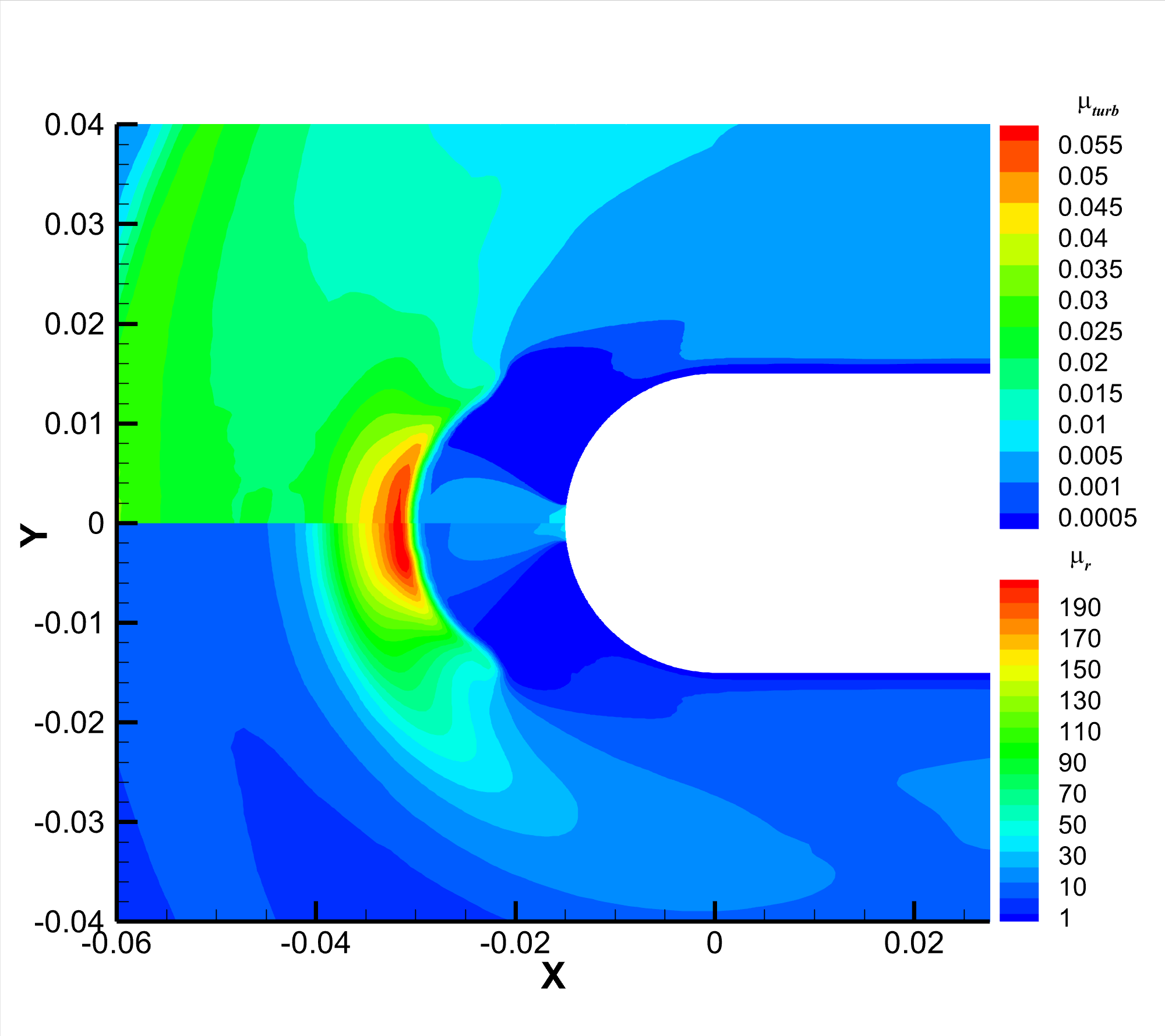}}\\
    \vspace{-3mm}
    \caption{Contours of turbulent viscosity $\mu_{turb}$ (top half) and the ratio of the turbulent to laminar viscosity $\mu_r$ (lower half) obtained by DIG-SST method when $\text{Ma}=15$ under various conditions. The primary turbulence arises between the outer shock the Mach disk.}
    \label{fig:Ma15murmuturb}
\end{figure}

Figure~\ref{fig:Ma15murmuturb} presents the contours of turbulent viscosity $\mu_{turb}$ and the ratio of turbulent to laminar viscosity $\mu_{r}$ obtained using the DIG-SST method for $\text{Ma}=15$ under different conditions. When $\text{Kn}=0.125$ and $P_{\text{ratio}}=2.5$, $\mu_r$ exhibits a small increase in the region between the outer shock and the Mach disk (around $\text{X}=-0.025$ m), while remaining relatively low elsewhere. Consequently, the influence of turbulence in most regions is negligible, and the results closely resemble those obtained with the purely laminar DIG method. However, as the Knudsen number decreases to 0.01, $\mu_r$ becomes significantly larger in the same region, indicating a pronounced turbulent effect, which substantially alters the flow characteristics compared to the laminar case. As the pressure ratio $P_{\text{ratio}}$ increases to 5, the region with the highest $\mu_r$ (around $\text{X}=-0.03$ m) shifts further away from the jet flow region.
However, in both the jet flow and outer shock regions, $\mu_r$ remains significantly elevated, indicating that turbulence plays a more dominant role in these regions compared to lower pressure ratios.

% where the turbulent effect dominates and the ratio $\mu_r$ is extremely large when the Knudsen number decreases to 0.01.

\begin{figure}[!t]
    \centering
    \subfigure[]{\includegraphics[width=0.49\textwidth,trim=10pt 20pt 10pt 50pt,clip]{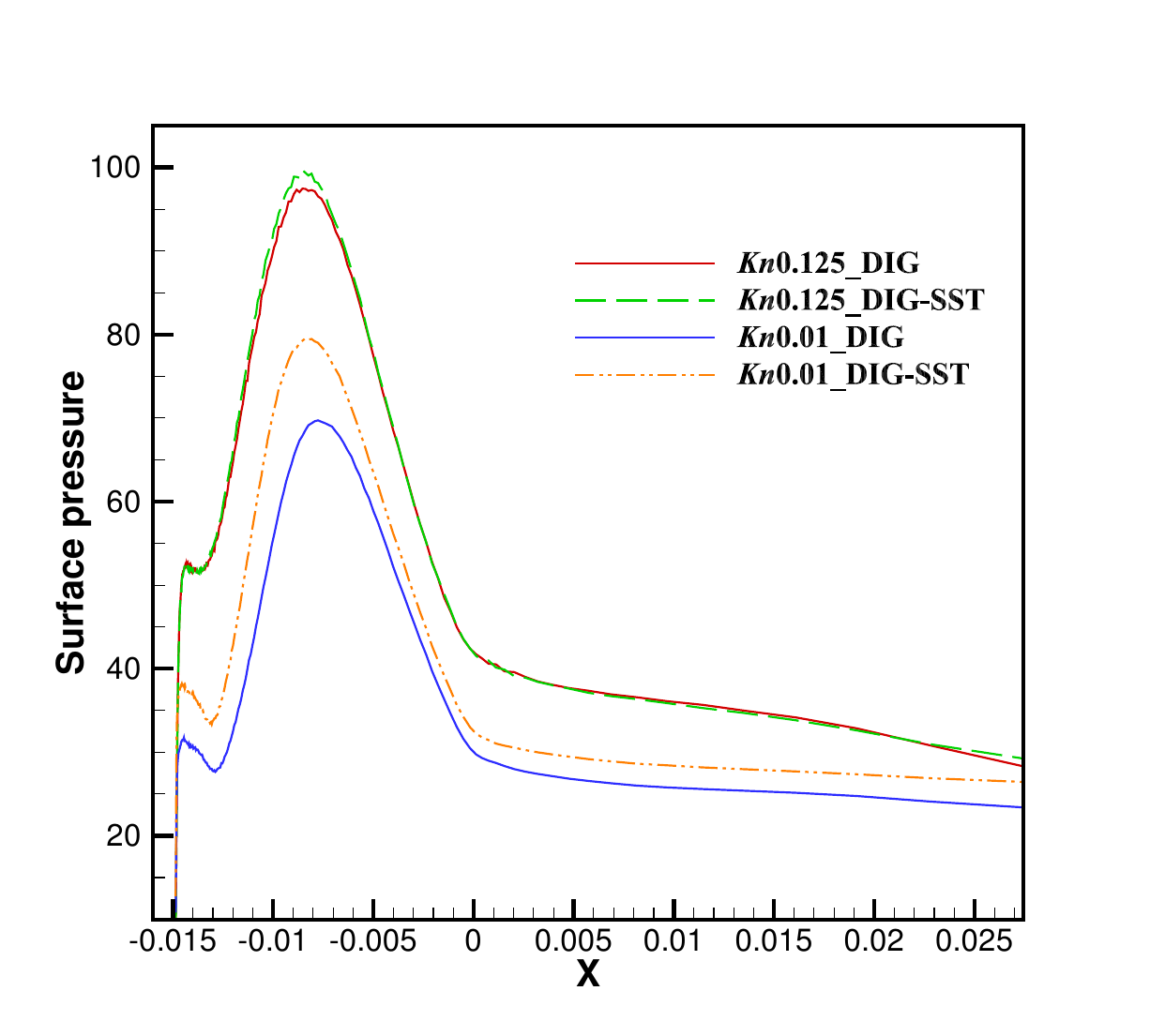}}
    \subfigure[]{\includegraphics[width=0.49\textwidth,trim=10pt 20pt 10pt 50pt,clip]{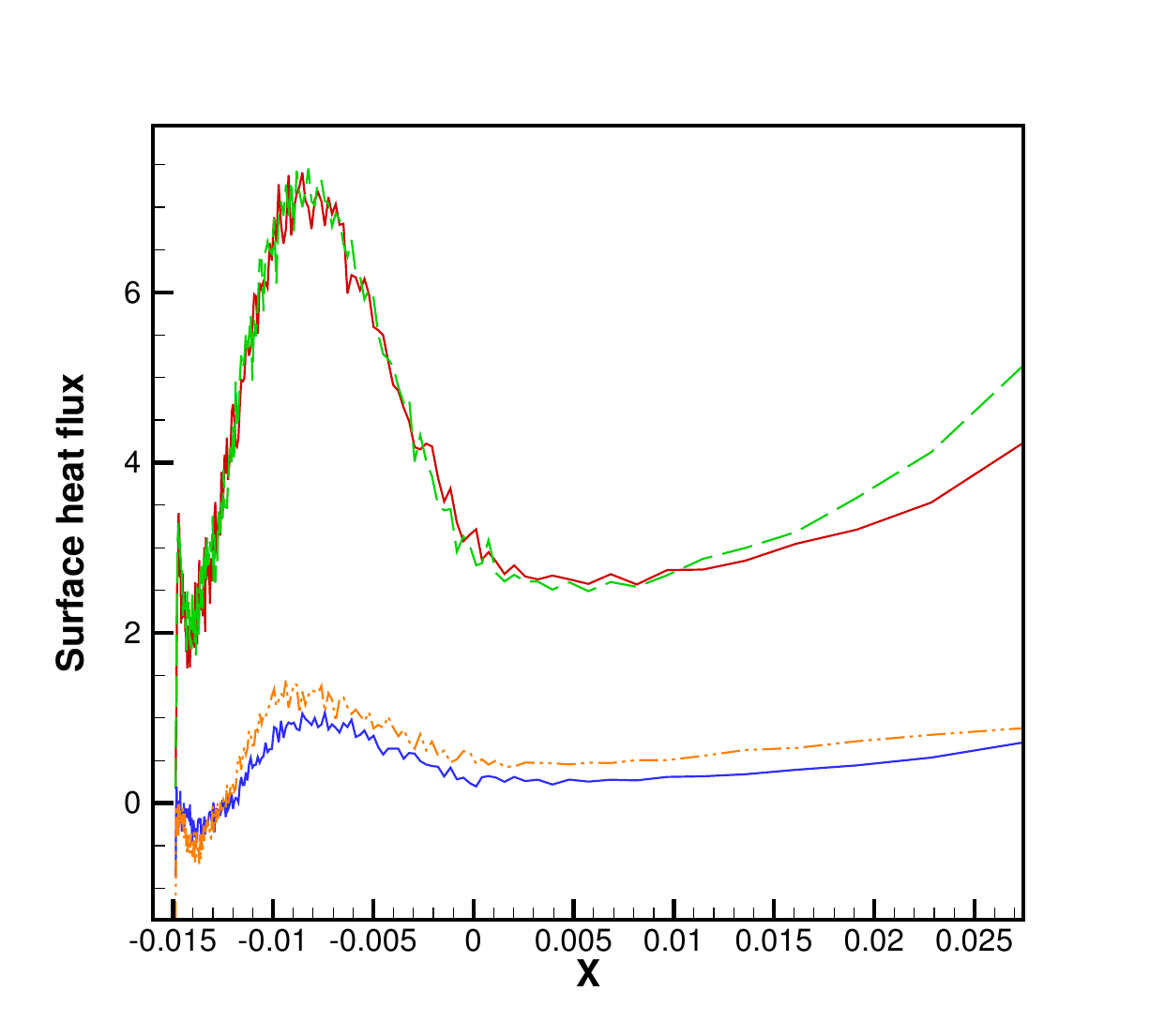}}\\
    \vspace{-2mm}
    \subfigure[]{\includegraphics[width=0.49\textwidth,trim=10pt 20pt 10pt 50pt,clip]{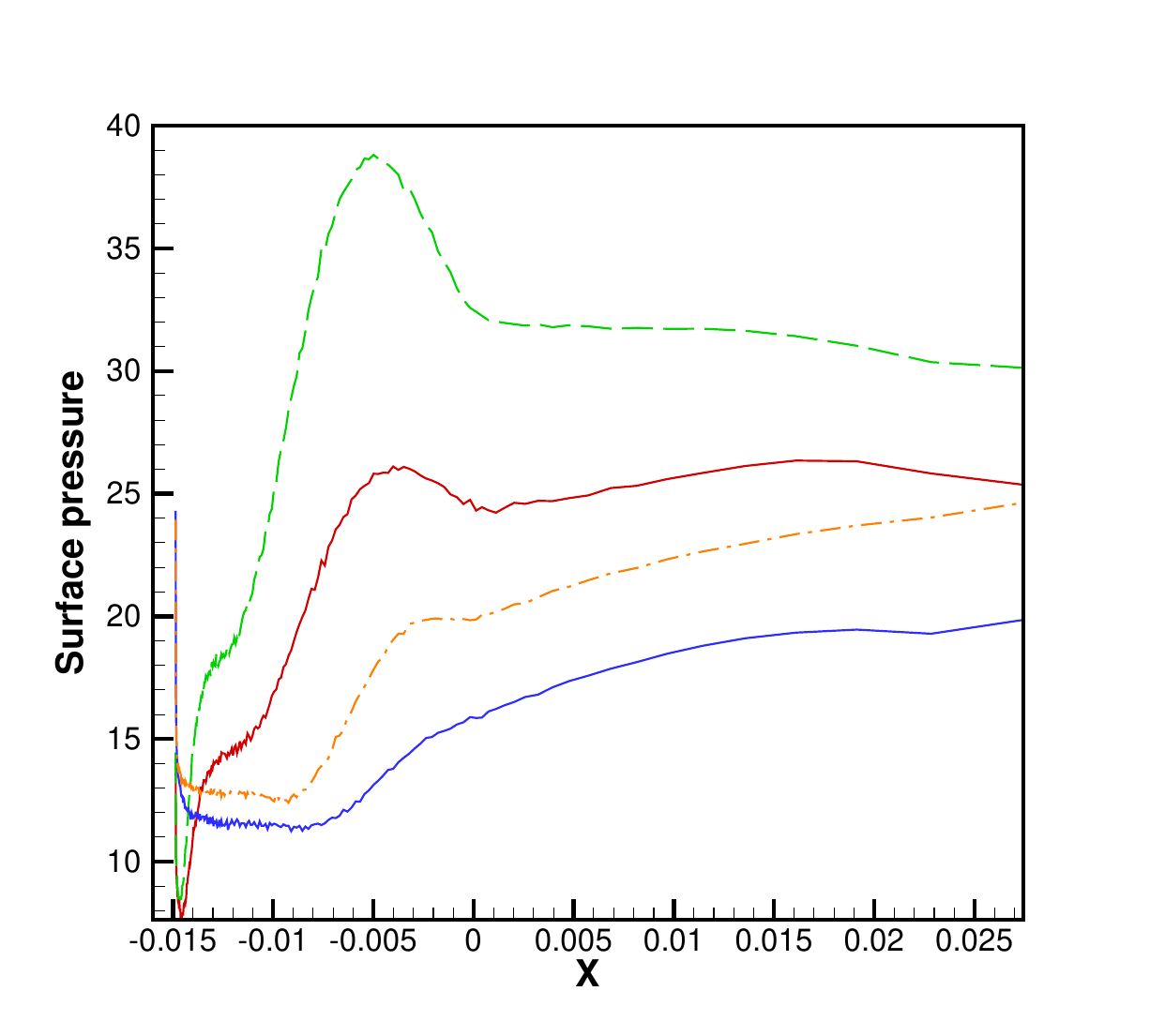}}
    \subfigure[]{\includegraphics[width=0.49\textwidth,trim=10pt 20pt 10pt 50pt,clip]{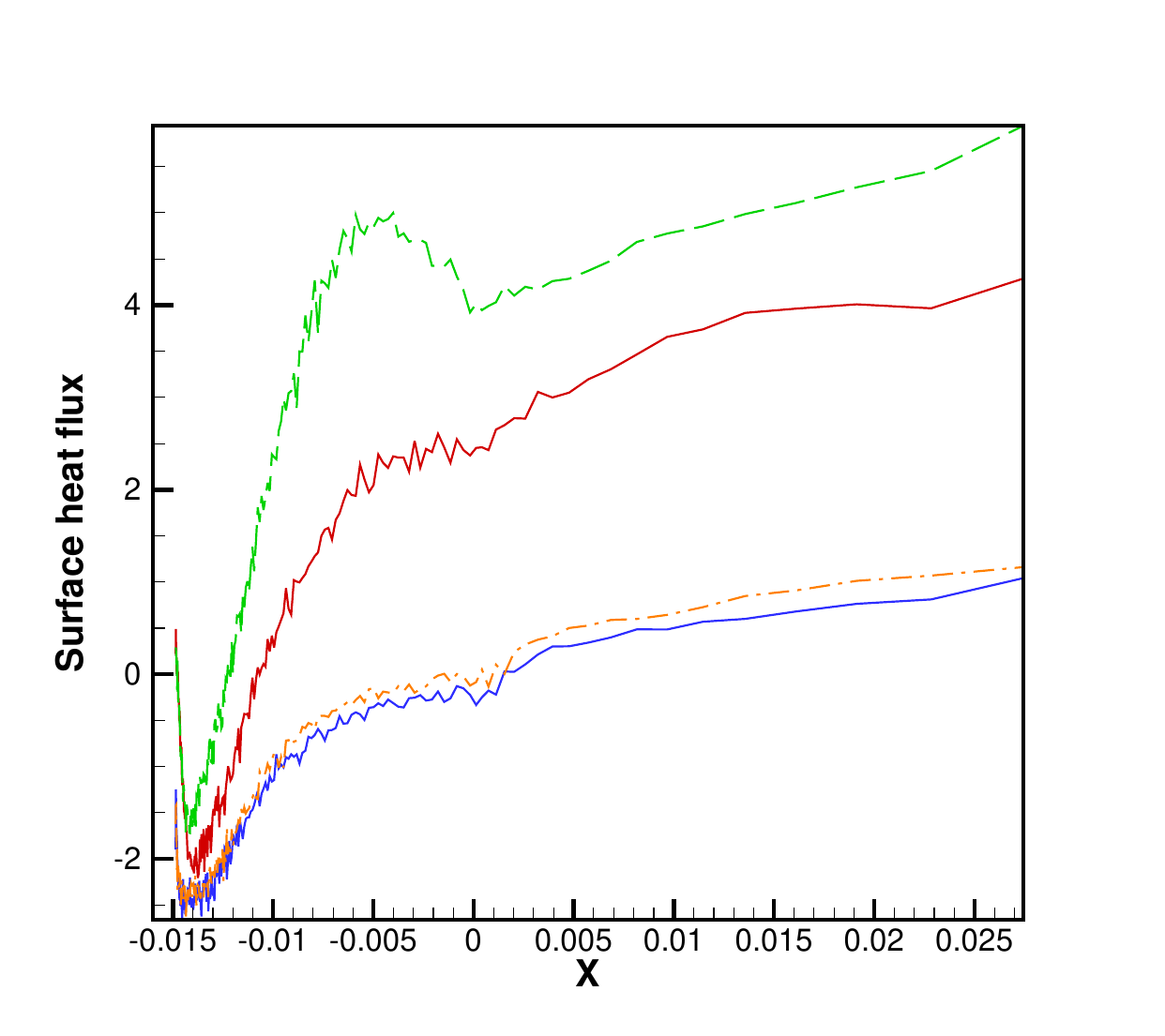}}\\
    \vspace{-2mm}
    \caption{The surface properties obtained by different methods for $P_{\text{ratio}}=2.5$ (upper row) and $P_{\text{ratio}}=5$ (lower row) under different conditions.}
    \label{fig:Ma15surface}
\end{figure}

Figure~\ref{fig:Ma15surface} illustrates the distributions of pressure and heat flux along the blunt body surface. When $\text{Kn}=0.125$ and $P_{\text{ratio}}=2.5$, turbulence effects are relatively weak, and the surface properties obtained using DIG-SST closely resemble those from the DIG method. However, as the Knudsen number decreases to 0.01, the significantly increased jet Reynolds number amplifies the turbulent effect, resulting in an approximately 10\% rise in the peak of surface pressure and heat flux (around $\text{X}=-0.008$ m). With $P_{\text{ratio}}$ increased to 5, the reattachment point shifts toward the outflow region, and accordingly, the peaks of the surface properties also migrate in that direction (around $\text{X}=-0.003$ m). Furthermore, under $\text{Kn}=0.125$, the coexistence of turbulent and rarefied effects leads to surface property values from DIG-SST that are nearly double those predicted by DIG. A similar trend is observed for surface pressure when $\text{Kn}$ decreases to 0.01, with DIG-SST yielding approximately 30\% higher values at the reattachment point. In contrast, at the reattachment point, the angle between the velocity streamlines and the model surface is exceedingly small, indicating that the streamlines are nearly parallel to the surface. 
This phenomena produces an almost zero heat flux at the reattachment point, rendering the predictions from both methods essentially indistinguishable.

\section{Conclusions and outlooks}\label{sec:6}

In summary, we have developed a multiscale stochastic-deterministic coupling method DIG-SST to model flows spanning from the rarefied regime to the turbulent flow region, where the constitutive relations contain the Newton and Fourier laws from the molecular motion, the high-order terms for the  rarefaction effects, and the turbulent transport coefficients from the SST model. Numerical simulations have confirmed the asymptotic-preserving properties of the proposed method, that is, when the Reynolds number is high, it turns to the pure $k$-$\omega$ SST turbulence model; when the Knudsen number is high,  it turns to the Boltzmann solver. When the Reynolds number and Knudsen number are both small, it turns the to laminar NS equations.

%This is achieved by intermittently incorporating the macroscopic synthetic equations and the $k$-$\omega$ SST turbulence model into the standard DSMC framework. After a specific time-stepping interval in DSMC, macroscopic properties are statistically sampled and employed as the initial input for solving macroscopic synthetic equations. These equations, coupled with the $k$-$\omega$ SST model, govern the interaction between turbulence and rarefaction, primarily through the reconstruction of constitutive relations for shear stress and heat flux. This reconstruction incorporates not only turbulent and laminar transport coefficients but also higher-order terms that capture rarefaction effects. Furthermore, the particle information is directly modified to align with the solutions of the synthetic equations. Moreover, since the $k$-$\omega$ SST model is formulated within the RANS framework, coarser grid can be employed to capture the turbulent flow pattern in the proposed algorithm compared to the DNS method.
% Overall, the accuracy of this coupling method is primary affected by the SST model.

With the DIG-SST method, we have investigated the opposing jet flow problems under varying free-stream Knudsen numbers and jet pressure ratios. It is found that, compared to the original DIG method, 
incorporating the turbulence model leads to a notable increase in surface pressure and heat flux at the reattachment region. This effect becomes more pronounced when stronger interactions between turbulence and rarefaction occur. For instance, under conditions of a higher global Knudsen number and an increased jet pressure ratio, the DIG-SST method predicts nearly twice the surface pressure and heat flux compared to the DIG method. These findings highlight the crucial role of turbulence effects in opposing jet flow problems even under rarefied free-stream conditions. The interaction between turbulence and rarefaction plays a critical role, as it is essential for optimizing thermal protection systems and aerodynamic control strategies in hypersonic vehicles.

We believe that, with small modifications, the proposed coupling technique can be readily extended to incorporate other turbulence models (i.e., Spalart-Allmaras model~\cite{spalart-1922} and $k$-$\epsilon$ model~\cite{wilcox-1993}), allowing for the selection of an appropriate turbulence model based on specific industrial requirements and application scenarios. Moreover, given that the DSMC method is more adept at incorporating complex physicochemical processes, the proposed approach establishes a foundational bridge that can be extended to simulate multi-component flows and incorporate intricate chemical reactions. This extension will enable the study of the interplay between rarefied and turbulent effects in chemically reacting flows, thereby more closely reflecting the conditions encountered in real-world aerospace applications.

\section*{Acknowledgments}
This work is supported by the National Natural Science Foundation of China (12450002). Special thanks are given to the Center for Computational Science and Engineering at the Southern University of Science and Technology.

\bibliographystyle{elsarticle-num}
\bibliography{ref}
\end{document}